\documentclass[aps,twocolumn,pra,superscriptaddress,articlepaper,nofootinbib,longbibliography]{revtex4-2}
\usepackage{amsmath, amsthm, amssymb, amsfonts}
\usepackage{physics}
\usepackage{mathtools}
\usepackage{upgreek}
\usepackage{booktabs}
\usepackage{url}
\usepackage{float}
\usepackage{enumitem}
\AtBeginDocument{
\heavyrulewidth=.08em
\lightrulewidth=.05em
\cmidrulewidth=.03em
\belowrulesep=.65ex
\belowbottomsep=0pt
\aboverulesep=.4ex
\abovetopsep=0pt
\cmidrulesep=\doublerulesep
\cmidrulekern=.5em
\defaultaddspace=.5em
}

\usepackage{txfonts}
\usepackage[pdftex]{graphicx}
\usepackage[usenames, dvipsnames, svgnames, table]{xcolor}
\usepackage[colorlinks=true, citecolor=Blue,linkcolor=BrickRed, urlcolor=magenta]{hyperref}
\usepackage{bm}
\usepackage{placeins} 
\usepackage{setspace}
\usepackage[final]{changes}

\usepackage{soul} 

\usepackage{cleveref}
\crefname{section}{Section}{Sections}
\crefname{appendix}{Appendix}{Appendices}
\crefname{figure}{Figure}{Figures}
\crefname{table}{Table}{Tables}
\crefname{equation}{Eq.}{Eqs.}
\crefname{assumption}{Assumption}{Assumptions}

\usepackage{siunitx}

\newcommand{\OSI}[2]{$\mathcal{O}\left(\SI{#1}{#2}\right)$}

\usepackage{dsfont} 
\usepackage{mathtools}
\usepackage{empheq}
\newcommand{\mrm}[1]{\mathrm{#1}}

\newcommand{\prtHz}{\per\sqrt{\mrm\Hz}}

\newcommand{\F}{\mathbf{F}} 
\newcommand{\D}[2]{\mathbf{D}^{#1}_{#2}}

\newcommand{\Dc}[2]{\mathcal{D}^{#1}_{#2}}
\newcommand{\dotDc}[2]{\dot{\mathcal{D}}^{#1}_{#2}}

\newcommand{\A}[2]{\mathbf{A}^{#1}_{#2}}

\newcommand{\Ac}[2]{\mathcal{A}^{#1}_{#2}}

\newcommand{\T}[2]{\mathbf{T}^{#1}_{#2}}

\newcommand{\Tc}[2]{\mathcal{T}^{#1}_{#2}}

\renewcommand{\F}{\mathbf{F}}

\newcommand{\ifo}[1]{\mathrm{ifo}_{#1}}
\newcommand{\isi}{\mathrm{isi}}   
\newcommand{\rfi}{\mathrm{rfi}}   
\newcommand{\tmi}{\mathrm{tmi}}   

\begin{document}
\title{Observable-based reformulation of time-delay interferometry}

\author{Kohei Yamamoto}
\email{y9m9k0h@gmail.com}
\affiliation{Center for Space Sciences and Technology, University of Maryland, Baltimore County, 1000 Hilltop Circle, Baltimore, Maryland 21250, USA}
\affiliation{Gravitational Astrophysics Lab, NASA/GSFC, 8800 Greenbelt Road, Greenbelt, Maryland 20771, USA}
\affiliation{Center for Research and Exploration in Space Science and Technology, NASA/GSFC, 8800 Greenbelt Road, Greenbelt, Maryland 20771, USA}
\affiliation{Max-Planck-Institut f\"ur Gravitationsphysik (Albert-Einstein-Institut), Callinstra\ss e 38, 30167 Hannover, Germany}
\affiliation{Leibniz Universität Hannover, Institut für Gravitationsphysik, Callinstra\ss e 38, 30167 Hannover, Germany}

\author{Jan Niklas Reinhardt}
\affiliation{Max-Planck-Institut f\"ur Gravitationsphysik (Albert-Einstein-Institut), Callinstra\ss e 38, 30167 Hannover, Germany}
\affiliation{Leibniz Universität Hannover, Institut für Gravitationsphysik, Callinstra\ss e 38, 30167 Hannover, Germany}

\author{Olaf Hartwig}
\affiliation{Max-Planck-Institut f\"ur Gravitationsphysik (Albert-Einstein-Institut), Callinstra\ss e 38, 30167 Hannover, Germany}
\affiliation{Leibniz Universität Hannover, Institut für Gravitationsphysik, Callinstra\ss e 38, 30167 Hannover, Germany}

\begin{abstract}
Spaceborne gravitational-wave observatories utilize a postprocessing technique known as time-delay interferometry (TDI) to reduce the otherwise overwhelming laser frequency noise by around 8 orders of magnitude.
While, in its traditional form, TDI considers the spacecraft as point masses, recent studies have enhanced this simplified scenario by incorporating more realistic metrology chain models, which include onboard optical, electronic, and digital delays.
These studies have updated the TDI algorithm to include onboard delays obtained from prelaunch and in-flight calibrations.
Conversely, the processing scheme presented in this article treats onboard delays as an integral part of the TDI combinations:
instead of having separate calibration stages, it directly expresses all delays appearing in the algorithm in terms of onboard measurements, especially pseudo-random-noise ranging (PRNR) measurements.
The only onboard delays that need to be corrected in our processing scheme are PRNR delays in the digital domain, which are determined by commandable digital-signal-processing parameters; hence, they can be easily managed in postprocessing.
Furthermore, our processing scheme does not require a prior interspacecraft clock synchronization, and it automatically corrects for potential relative drifts between the clocks driving local phase measurement systems.
The proposed observable-based formulation closely relates TDI to the actual metrology system, and it clearly outlines how to manage onboard measurements in postprocessing.
Hence, it is expected to lead to fundamental advancements in TDI, providing both conceptual completeness and unique practical benefits.
\end{abstract}

\maketitle

\section{Introduction}\label{sec:intro}
On September 14, 2015, the Laser Interferometer Gravitational-Wave Observatory (LIGO) achieved the first direct gravitational-wave (GW) detection~\cite{GW150914}.
In the following years, many GW signals have been detected by ground-based observatories~\cite{LIGO:Catalog1,LIGO:Catalog2,LIGO:Catalog3}.
However, these detectors are limited by seismic and gravity gradient noise below \SI{1}{\Hz}.
Launching a detector into space will overcome this limit and open up channels into the millihertz regime, which is rich in GWs from systems involving heavier astronomical bodies, such as massive black hole binaries and extreme-mass-ratio inspirals~\cite{Seoane2023}.

There are several missions aiming for spaceborne GW detection~\cite{TianQin,Taiji,DECIGO}.
One of them is the Laser Interferometer Space Antenna (LISA), which is sensitive to GWs in the frequency band between \SI{0.1}{\milli\Hz} and \SI{1}{\Hz}~\cite{LisaRed}.
LISA comprises three spacecraft (SC) on heliocentric orbits that form a nearly equilateral triangle with an armlength of about \SI{2.5}{\giga\m}, corresponding to an interspacecraft light travel time (ILTT) of about \SI{8.3}{\second}.
GWs induce picometer armlength variations in this constellation.
To achieve the picometer-stable armlength measurements, LISA employs single-link interspacecraft laser interferometry using a split interferometry concept~\cite{LisaRed,Otto:2015erp}.

Each LISA SC houses two free-falling test masses (TMs) as local gravity field references~\cite{Armano:LPF-I,Armano:LPF-II}, which are oriented toward the respective distant SC.
An ultrastable optical bench (OB)~\cite{Brzozowski2022} is associated with each TM and placed between the TM and the distant SC (see \cref{fig:sc_notation}).
A \SI{1064}{\nano\m} laser~\cite{Numata2023} is fiber fed to the OB, from where it is directed to the distant SC via a telescope~\cite{Verlaan:Telescope} and to the adjacent OB through a fiber backlink~\cite{Fleddermann:Backlink}.\footnote{The sets of TMs, OBs, and telescopes are referred to as movable optical subassemblies (MOSAs). Each SC has two of them. Their angles are adjustable to compensate for corner angle variations.}
On the OB, the laser beam is used as a local oscillator in three heterodyne interferometers: The interspacecraft interferometer (ISI) measures distance variations between the local and distant OBs by interfering the received distant beam with the local one.
GW signals accumulate in the ISI measurements.
The TM interferometer (TMI) measures distance variations between the OB and the TM by interfering the adjacent beam with the local one after it is reflected off the free-falling TM along the interspacecraft axis.
The reference interferometer (RFI) interferes the adjacent and the local beam without involving the TM.
The \si{megahertz} beatnotes in these interferometers are detected with quadrant photoreceivers (QPRs)~\cite{Colcombet2024}.
Each MOSA is paired with a multichannel phasemeter that simultaneously extracts all beatnote phases~\cite{Shaddock2006,Gerberding2013}.
A single master clock per SC drives both onboard phasemeters.

\textit{A priori} LISA's interferometric measurements are dominated by laser frequency noise, which surpasses typical GW signals expected in the millihertz regime by 8 orders of magnitude.
LISA relies on a sophisticated postprocessing scheme called time-delay interferometry (TDI) to synthesize equal-arm interferometers, in which laser frequency noise naturally cancels~\cite{Tinto1999,Armstrong1999}.~\footnote{
The underlying concept of TDI is similar to displacement-noise free Interferometers~\cite{Kawamura2004,Chen2006,Gefen2024} in the context of ground-based GW detection, which utilize the difference between the light field coupling to GW signals and mirror displacements.}
In the first step, TDI combines the ISI, TMI, and RFI beatnotes to form single-link interspacecraft TM-to-TM measurements, thus suppressing the OB jitter along the interspacecraft axes.
It then combines appropriately delayed versions of these TM-to-TM measurements to form virtual multilink interferometers, which feature equal optical pathlengths in both arms, despite the unequal and nonconstant armlengths of the LISA constellation.
These virtual equal-arm interferometers are insensitive to the laser frequency noise, thus facilitating the search for GWs.
The required delays are the ILTTs, i.e., the absolute SC separations, measured with an accuracy of \SI{1}{\meter} equivalent to \SI{3.3}{\nano\second}~\cite{LisaRed}.
These are obtained from an onboard absolute ranging scheme, which relies on phase modulating the carrier beams with pseudorandom noise (PRN) codes generated according to the onboard clocks.
This technology is called PRN ranging (PRNR)~\cite{Esteban2009,Sutton2010,Heinzel:Ranging}.

In reality, PRNR does not deliver actual ranges but pseudoranges, which entangle the ILTTs with the desynchronizations between the SC clocks.
Originally it was assumed that TDI required clock-synchronized measurements and the ILTTs as delays, i.e., a disentanglement of the pseudoranges~\cite{Reinhardt2024C}.
In contrast, the study~\cite{Hartwig2022} showed that TDI works well with unsynchronized interferometric measurements when the pseudoranges are used as delays.
Hence, we can use the PRNR pseudorange measurements as delays in TDI without the prior disentanglement.
However, that study treated the SC as point masses, neglecting onboard delays in the LISA metrology chain.

Recent studies revealed that onboard delays in the optical, electronic, and digital domains represent non-negligible parts of the above-described TDI combinations.
Prelaunch calibrations are important to accurately correct PRNR-unique onboard delays, thus facilitating the extraction of the TDI-required interspacecraft delays from PRNR data~\cite{Reinhardt2024A}; TDI ranging~\cite{Tinto:TDIR,Staab:Thesis} can serve as a cross-check for these prelaunch calibrations.
Additionally, PRNR-based in-flight calibrations can estimate the electronic onboard delays in the phase-modulation signal chains, QPRs, and phasemeter frontends~\cite{Euringer2023}.
While measuring the ILTTs with interspacecraft PRNR in the ISI, the proposed calibration relies on additional PRNR measurements: the tracking of the local PRN code in the ISI, and the local and received PRN codes in the TMI and RFI.
This in-flight calibration of the electronic onboard delays enhanced confidence in extracting ILTTs from the received PRNR measurement in ISI, as requested by TDI.
Apart from electronic delays, also onboard optical delays cause non-negligible laser noise residuals in TDI if not treated properly.
The study~\cite{Reinhardt2024B} derived a compensation scheme that includes the design values (prelaunch calibration) of the onboard optical pathlengths~\cite{Brzozowski2022} into the TDI algorithm.

This article elaborates on a paradigm shift in TDI, which was first outlined in \cite{Reinhardt2024A}: TDI actually does not require the ILTTs themselves, but the time differences with which laser frequency noise instances appear in the phase measurements on the local and remote SC.
We extend this by also reformulating the intermediate TDI steps from the perspective of the actual TDI needs.
We show that this new perspective enables us to express all delays in the TDI algorithm in terms of onboard measurements, mainly PRNR.
For this purpose, similarly to \cite{Euringer2023}, we make use of both local and received PRNR in all interferometers.
In our scheme, PRNR measurements are naturally integrated into the TDI algorithm, i.e., PRNR does not exist as a separate calibration stage anymore.
Our scheme efficiently integrates all aspects discussed in the previous paragraph into a single algorithm.
The only onboard delays to correct are PRNR delays in the digital domain, which are determined by commandable digital-signal-processing parameters and can be easily managed in postprocessing.
Our scheme can be considered an extension of \cite{Hartwig2022} to the real metrology chain, as it directly works on pseudoranges and does not rely on a prior clock synchronization.
We develop an abstract diagram for the metrology-chain signal paths (see \cref{fig:diagram_notation}) to illustrate that our scheme successfully corrects for onboard optical, electronic, and digital delays.
The reliability of the proposed scheme is, to some extent, guaranteed by the fact that it developed from a ranging scheme already applied to a tabletop experiment (it is called \emph{Scheme 2} in~\cite{Yamamoto2024}).
We expect the proposed framework to make TDI conceptually complete, resulting in various practical advantages.

The paper is structured as follows: \cref{sec:review} provides a brief review of TDI and summarizes the metrology chain of spaceborne GW detectors, focusing on the LISA case.
\cref{sec:framework} formulates the LISA measurements in a realistic metrology chain with onboard delays, first in a very general manner and then in a simpler manner with reasonable assumptions.
\cref{sec:tdi} presents our PRNR-based reformulation of TDI using the framework prepared in \cref{sec:framework}.
Here, after providing the analytical model of the entire algorithm, we conduct a Monte Carlo simulation to analyze the effects of PRNR imperfections on the second-generation TDI Michelson combinations.

\section{Review}\label{sec:review}
\begin{figure}
    \centering
\includegraphics[width=8.6cm]{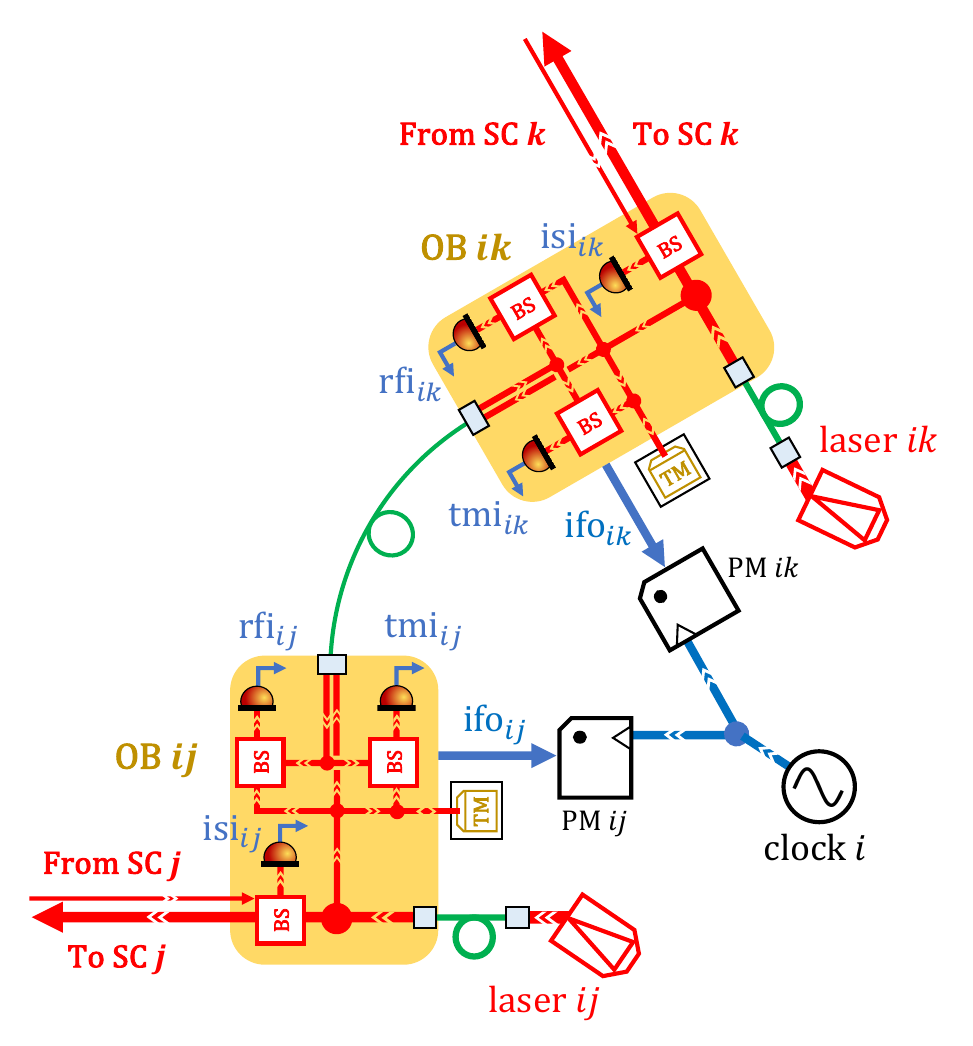}
    \caption{
    Conceptual diagram of the payload on SC $i$.
    Each SC hosts two movable optical subassemblies (MOSAs), including a laser, an optical bench (OB), a test mass (TM), and a phasemeter (PM).
    Red: optical paths; Blue: electronic paths; Green: optical fibers.
    Beam splitters (BSs) combining two beams are denoted by boxes labeled BS, while the other beam-splitting points are denoted by dots.
    }
    \label{fig:sc_notation}
\end{figure}

TDI-based spaceborne GW detectors generally adopt an onboard payload as sketched in \cref{fig:sc_notation} for SC $i \in \{1,2,3\}$; the other SC are labeled clockwise.
Each SC houses two identical MOSAs oriented towards the two distant SC.
Each MOSA is labeled by two indices indicating the host SC (first index) and the respective distant SC (second index).
All components and interferometers are labeled after the host MOSA.

\subsection{Traditional TDI}\label{sub:traditional_tdi}
TDI considers the SC as point masses in its traditional form.
In this section, we first model the heterodyne beatnote phase measurements, neglecting onboard delays.
After that, we review the TDI algorithm in this point-mass framework.
We mostly follow the description in the literature~\cite{Tinto:2020fcc}, using the notation introduced in~\cite{Bayle2023}.

The GW signals are ultimately encoded in the phases of the received distant laser beams.
At the laser source, we represent a beam by its phase in units of cycles:
\begin{align}
    \phi_{ij}(\tau) &= \int_{\tau_0}^{\tau} \left(c/\lambda_0 + O_{ij}(\tau')\right) d\tau' + p_{ij}(\tau),
    \label{eq:phiij}
\end{align}
where $c$ is the speed of light and $\lambda_0$ is the laser central wavelength. $O_{ij}$ denotes the megahertz offset from the central laser frequency, which is controlled via an offset laser lock~\cite{Heinzel2024}. $p_{ij}$ is the stochastic laser phase noise in the millihertz GW observation band.
TDI eventually aims to remove the laser phase noise $p_{ij}$.
In order to model a time delay of a given quantity $x$, a delay operator is defined as
\begin{align}
    \D{}{} x(\tau) \coloneqq x(\tau - d(\tau)).
    \label{eq:D}
\end{align}
Through this article, delays $d$ acquire individual superscripts and subscripts for clear distinguishment; correspondingly, the same superscripts and subscripts will be given to the associated delay operators $\D{}{}$.
In the rest of this article, we mostly omit the time arguments for simplicity.

While we do not focus on macroscopic onboard pathlengths in this section, we do include noise terms to model in-band jitters from distance fluctuations in the beam paths, since these motivate some of the data combinations performed in traditional TDI~\cite{Otto:2015erp}.
Notably, positional jitter of the spacecraft along the line of sight, denoted $N_{ij}^\Delta$, will directly couple as a phase shift of the laser beam.
We call it OB jitter as OBs are rigidly mounted to MOSAs.
OB jitter can be considered stochastic at the nanometer scale in the observation band~\cite{Henri2022}.
These noise terms would deteriorate the LISA target performance and will be compensated for using a dedicated TMI, replacing $N_{ij}^\Delta$ with the TM acceleration noise $N_{ij}^\delta$, as shown below.

Including noise terms to cover any other secondary noises $N_{ij,\mrm{ifo}}$, such as shot noise, the phase measurements of the three types of the interferometer can be written as
\begin{align}
    \isi_{ij} &= \D{}{ij}\phi_{ji}  - \phi_{ij} + \D{}{ij} N_{ji}^\Delta + N_{ij}^\Delta + N_{ij,\isi},
    \label{eq:isi_ij_basic}\\
    \tmi_{ij} &= \phi_{ik} - \phi_{ij} + 2 (N_{ij}^\Delta - N_{ij}^\delta) + N_{ij,\tmi},
    \label{eq:tmi_ij_basic}\\
    \rfi_{ij} &= \phi_{ik} - \phi_{ij} + N_{ij,\rfi},
    \label{eq:rfi_ij_basic}
\end{align}
$\D{}{ij}$ represents the ILTT from SC $j$ to SC $i$, where the GW signals are encoded. The separation between OB and TM $(N_{ij}^\Delta - N_{ij}^\delta)$ enters with a factor of 2 as the laser beam travels back and forth to the TM. We will omit $N_{ij,\mrm{ifo}}$ in the rest of the article to focus on the laser noise and the OB jitter.

TDI configures combinations of the heterodyne beatnotes, which are sensitive to GW signals but mostly insensitive to laser noise. The core TDI algorithm comprises three steps (described below) to suppress laser phase noise and OB jitter, the former of which would otherwise surpass any GW signal by more than 8 orders of magnitude.

The first step composes interspacecraft TM-to-TM measurement, thus removing the OB jitter $N_{ij}^\Delta$ from the interferometric measurements. Most dominant noise terms, in particular laser noise, are common in the RFIs and TMIs, such that their differences can be used to perform this suppression, 
\begin{align}
    \xi_{ij} &= \isi_{ij}+\frac{\rfi_{ij}-\tmi_{ij}}{2} + \D{}{ij}\frac{\rfi_{ji}-\tmi_{ji}}{2}
    \nonumber\\
    &= \D{}{ij}\phi_{ji} - \phi_{ij} + \D{}{ij}N_{ji}^\delta + N_{ij}^\delta .
    \label{eq:xi_basic}
\end{align}
These quantities are referred to as the intermediary TDI $\xi$ variables.
Having shown OB jitter can be suppressed, we will also neglect the remaining $N_{ij}^\delta$ in the following.

The second TDI step reduces six lasers to three lasers by constructing the intermediary $\eta$ variables,
\begin{align}
    \eta_{12} &= \xi_{12} + \D{}{12}\frac{\rfi_{21} - \rfi_{23}}{2} = \D{}{12}\phi_{23}-\phi_{12},
    \label{eq:eta12_basic}\\
    \eta_{13} &= \xi_{13} + \frac{\rfi_{12} - \rfi_{13}}{2} = \D{}{13}\phi_{31}-\phi_{12}.
    \label{eq:eta13_basic}
\end{align}
This combination removes right-handed laser phases $\phi_{21}$ and $\phi_{13}$.
We differentiate two RFIs on the same SC to remove the reciprocal part of the backlink fiber noise.
The other $\eta$ variables can be derived via cyclic permutation.

We can now combine the $\eta$ variables to synthesize single-beam interspacecraft round trips,
\begin{align}
    \eta_{12} + \D{}{12} \eta_{21} &= \left(\D{}{121} - 1\right)\phi_{12},
    \label{eq:eta1221_basic}\\
    \eta_{13} + \D{}{13} \eta_{31} &= \left(\D{}{131} - 1\right)\phi_{12},
    \label{eq:eta1331_basic}
\end{align}
where we apply the short-hand notation $\D{}{ij}\D{}{ji} =\vcentcolon \D{}{iji}$.
These round-trip measurements can be considered arms of a Michelson interferometer, where SC $1$ acts as a central station composed of a laser source, a BS, and a detector, and the respective TMs on SC $2$ and $3$ act as end mirrors.

The last step forms virtual equal-arm interferometers, in which laser phase noise cancels naturally, e.g., the TDI Michelson variables.
Combining \cref{eq:eta1331_basic,eq:eta1221_basic} leads to the zeroth-generation TDI Michelson variable
\begin{align}
    X_0 = \left(\eta_{13} + \D{}{13} \eta_{31}\right) - \left(\eta_{12} + \D{}{12} \eta_{21}\right) &= \left(\D{}{131} - \D{}{121}\right)\phi_{12}.
    \label{eq:X0_basic}
\end{align}
The other two Michelson variables, so-called $Y_0$ and $Z_0$, can be obtained via cyclic permutation of the indices.
The zeroth generation TDI Michelson variables remove the laser noise under the conditions of equal and static armlengths: $\D{}{131} - \D{}{121}=0$.

We must configure more complex linear combinations of delay operators to remove the laser noise beyond such a condition.
The first-generation TDI Michelson variable succeeds in an unequal but constant arm configuration~\cite{Tinto1999}.
A realistic configuration with unequal and drifting armlengths requires the second-generation TDI Michelson variables~\cite{Tinto:2003vj},
\begin{align}
    X_2 &= \left(1 - \D{}{121} - \D{}{12131} + \D{}{1312121}\right)\left(\eta_{13} + \D{}{13}\eta_{31}\right)
    \label{eq:X2_basic}\\
    &\hspace{10mm}- \left(1 - \D{}{131} - \D{}{13121} + \D{}{1213131}\right)\left(\eta_{12} + \D{}{12}\eta_{21}\right)
    \nonumber\\
    &= \left(\D{}{131212131} - \D{}{121313121}\right)\phi_{12}.
    \label{eq:X2_basic_residual}
\end{align}
The delay commutator $\left(\D{}{131212131} - \D{}{121313121}\right)$ remains sufficiently small even in realistic conditions, typically evaluating to a time of arrival mismatch of the two virtual laser beams on the order of picoseconds~\cite{Muratore:2020mdf}. This guarantees the laser phase noise residual to be below \SI{1}{\pico\meter\prtHz} with a laser noise allocation of \SI{30}{Hz\prtHz} and a ranging accuracy of \SI{1}{\meter}.

\subsection{Metrology}\label{sub:metrology}
\begin{figure}[ht]
    \centering
\includegraphics[width=8.6cm]{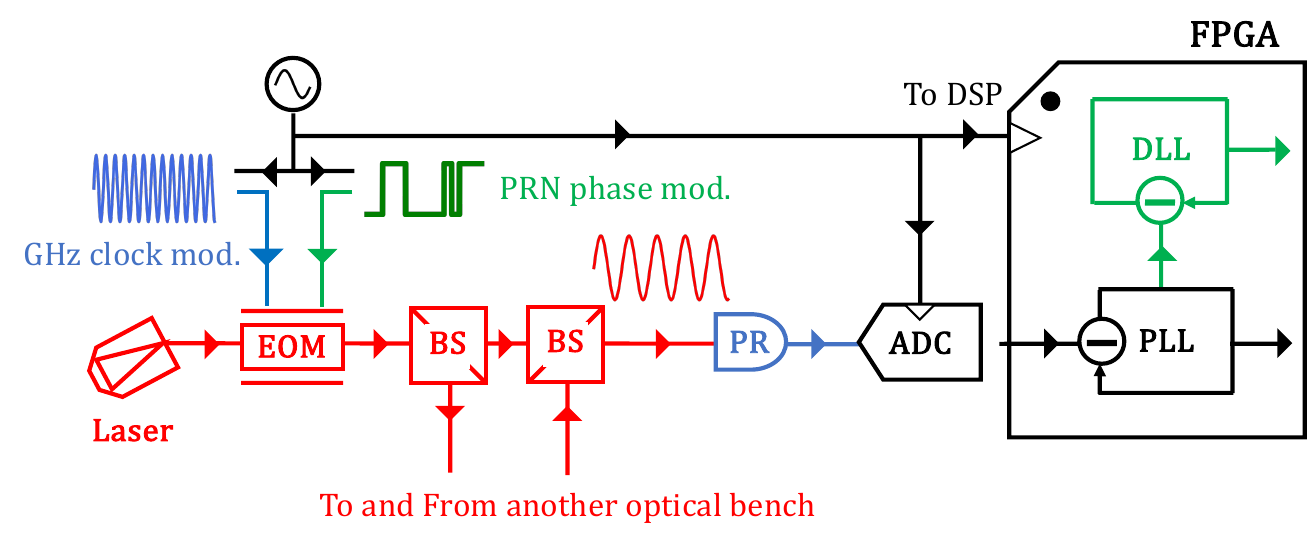}
    \caption{
    Simplified diagram of the metrology chain for heterodyne interferometry, PRNR, and \si{gigahertz} clock noise transfer.
    The extracted heterodyne phases and PRNR delays are delivered to decimation stages to down-sample such high-rate signals to a reasonable data rate for telemetry to the Earth.
    PLL for the clock sideband-sideband beatnote is omitted.
    }
    \label{fig:diagram_metrology}
\end{figure}

The traditional TDI algorithm introduced above does not consider the internal structure of the SC and the measurement systems.
Here, we have a closer look at the metrology chain (see \cref{fig:diagram_metrology}), including not only the laser carrier but also the PRN signals, originally introduced to monitor the ILTTs for TDI.

An electro-optic modulator (EOM) is used to phase modulate the laser beam with a binary PRN signal driven by the onboard clock.~\footnote{Although this article highly focuses on the carrier-carrier beatnote phase and PRNR measurements, the figure also depicts \si{gigahertz} clock modulations for the in-band clock noise transfer~\cite{Hartwig2021,Yamamoto2022} for completeness.}
The phase-modulated beam is then split into a signal path to another OB~\footnote{``Another OB" here can be either of the one on another SC or the adjacent one on the same SC.\label{ft:ob}} and a local oscillator path.
The local oscillator beam interferes with an incoming signal beam from the other OB and generates a heterodyne beatnote between \SIrange[]{5}{25}{\mega\Hz} as both interfering beams feature a \si{megahertz} frequency offset due to Doppler shifts and the offset frequency lock.
A QPR converts the optical beatnote signal to an electronic voltage, which is then digitized in an analog-digital converter (ADC) driven by the onboard clock.
The digitized signal is plugged into a field-programmable gate array (FPGA), in which a phase-locked loop (PLL) extracts the phase of the input sinusoidal signals of the heterodyne beatnote.
In parallel, a delay-locked loop (DLL) correlates the incoming PRN signal, extracted from the PLL error signal, and an identical copy of the PRN sequence generated locally by the FPGA.
Thus, the DLL derives a time difference between the generation of the incoming PRN signal on the transmitter side and the generation of the local copy of the PRN signal on the receiver side.
We can track both the received PRN signal from the other OB (\cref{ft:ob} applies) and the local PRN signal from the local OB~\cite{Sutton2010,Euringer2023,Yamamoto2024}; we call them received PRNR and local PRNR in this article.
Our TDI reformulation in \cref{sec:tdi} highly relies on this dual code tracking.

\section{Framework}\label{sec:framework}
As described in \cref{sub:metrology}, the real metrology system has many components and associated delays.
In \cref{sec:tdi}, we will reformulate TDI in terms of PRNR to efficiently deal with the real system  within the TDI formalism.
This section sets up a framework for this purpose.

In \cref{sub:diagram}, we introduce a diagram that illustrates the metrology chain, focusing on delays. This diagram provides visual assistance to ease the understanding of our TDI algorithm.
Following this, we reformulate the measurements incorporating onboard delays in \cref{sub:measurements}, where we present the most general formulation of the signals. \Cref{sub:assumptions} then clarifies the assumptions regarding the real system, which we apply in this article.
Finally, \cref{sub:setup} presents the simplified formulation of the signals based on those assumptions.
This simplified measurement model will be used in \cref{sec:tdi}.

\subsection{Signal-path diagram}\label{sub:diagram}
\begin{figure*}
    \centering
\includegraphics[width=12.9cm]{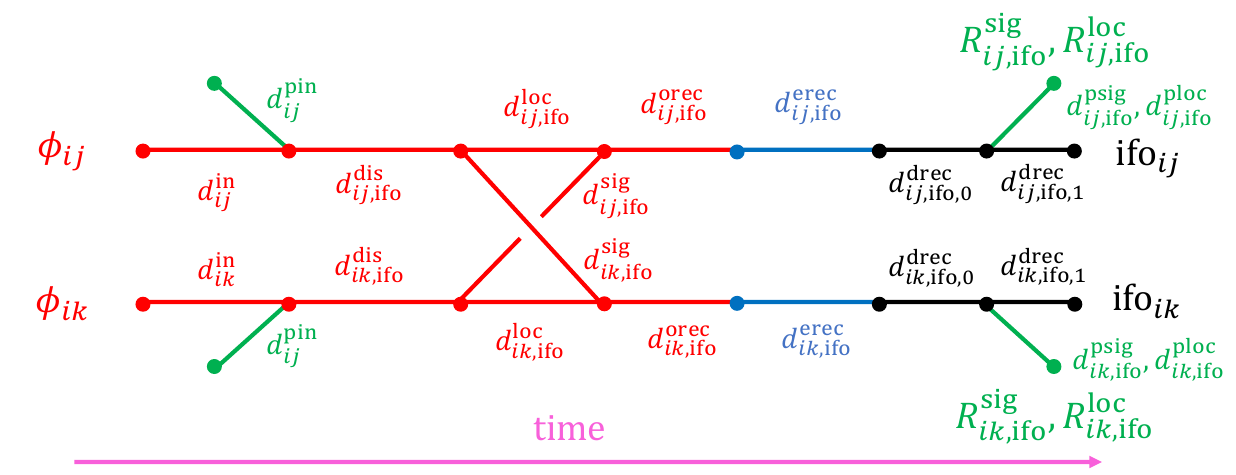}
    \caption{
    The signal-path diagram serves as an abstract version of the metrology chain diagram in \cref{fig:diagram_metrology}.
    Here we depict two heterodyne interferometers formed by a pair of laser beams: in red, optical paths; in blue, electronic paths; in black, digital paths; green lines represent PRN-unique paths: PRN signals generated according to the onboard clock (left green dots) are first delivered to the EOMs (left green paths).
    After following the same paths as the original beam phases, the PLL error signals are routed to DLLs and subsequently decimated (shown by the right green paths), and ultimately recorded in data files as PRNR measurements (indicated by the right green dots).
    See \cref{tab:delays,sub:measurements} for the definition of the delays and the measurements, respectively.
    }
    \label{fig:diagram_notation}
\end{figure*}

We introduce the diagram shown in \cref{fig:diagram_notation}, which we call \emph{signal-path diagram} through the rest of this article.
The diagram illustrates the flow of beams and signals through the metrology system, depicting the delays in delivering beams from laser sources to the measurements to which they couple.
Red, blue, and black represent optical, electronic, and digital delays.
Paths unique to PRN signals are depicted in green to draw special attention.
Note that this diagram focuses on the beam carrier and the PRN signal and their measurements, neglecting the gigahertz clock modulations denoted in \cref{fig:diagram_metrology} and the resulting sideband-sideband beatnotes.

The signal-path diagram in \cref{fig:diagram_notation} has, as inputs, beam phases at the laser sources $\phi_{ij}$ and $\phi_{ik}$, formulated in \cref{eq:phiij}, and PRN signals originating at the green dots on the left.

Outputs of the diagram are measurements: $\ifo{ij}$ and $\ifo{ik}$ are the heterodyne beatnote phase measurements between two lasers, and $R^\mrm{sig}_{ij,\mrm{ifo}}$ and $R^\mrm{loc}_{ij,\mrm{ifo}}$ are received- and local-PRNR measurements associated with the beatnote measurement $\ifo{ij}$.
$\ifo{ij}$ is a placeholder, to which a concrete interferometer (isi, rfi, or tmi) will be later substituted.
These measurements will be formulated in the next section.

Finally, we have delays in delivering the inputs to the outputs.
The definition, the notation, and examples of contributors to each delay are listed in \cref{tab:delays}.
See \cref{sub:assumptions} for the fifth column.
Superscripts tell us delay categories:
They denote the delay’s association with the input beam phase $\phi$ or with the beatnote phase measurement $\mrm{ifo}$: any delay before splitting a beam is labeled after the beam phase, while any delay after the splitting BS is labeled according to the beatnote measurement the phase couples to.
The PRNR-unique delays acquire ``p" in the superscripts.
Also, the subscript ``ifo" expresses a certain interferometer with which the paths are associated (``isi", ``rfi", or ``tmi" will be substituted later).
This means that a delay without ``ifo" is expected to be common for all interferometers.
Note that we push the time frame conversion between two system clocks to the ISI signal interferometric delay $d^\mrm{sig}_{ij,\isi}$ in analogy to \cite{Hartwig2021,Reinhardt2024A}.

The signal-path diagram is applicable to all types of interferometers (ISI, TMI, and RFI), since they are commonly based on heterodyne interferometry with the same measurement principle, and they all come with PRNR.

\begin{table*}[t]
    \caption{\label{tab:delays} Description of delays in a signal-path diagram: first column: name of each delay; second column: most arbitrary notation in \cref{fig:diagram_notation}; third column: domain where a delay occurs (o=optical, e=electronic, d=digital, $\tau$=clock difference); fourth column: example contributors; fifth column: notation based on the assumptions in \cref{sub:assumptions}.
    Many of those delays are already defined in the LISA case concretely, especially the onboard optical delays.
    For example, $d^\mrm{loc}_{ij,\mrm{\isi}}$ is the delay between BS2 and BS12 on OB $ij$~\cite{Brzozowski2022}.
    DSP and DPLL stand for digital signal processing and digital phase-locked loop, respectively.
    }
    \begin{ruledtabular}
    \begin{tabular}{lllll}
    Name & Notation & Domain & Example & Approximation
    \\
    \hline 
    Beam generation delay & $d^\mrm{in}_{ij}$ & o & Optical fibers / EOM &
    \\ \vspace{1.0mm}
    Local distribution delay & $d^\mrm{dis}_{ij,\mrm{ifo}}$ & o & Optical fibers / OB & $d^\mrm{dis}_\mrm{ifo}$
    \\ \vspace{1.0mm}
    Local interferometric delay & $d^\mrm{loc}_{ij,\mrm{ifo}}$ & o & OB & $d^\mrm{loc}_\mrm{ifo}$
    \\ \vspace{1.0mm}
    Signal interferometric delay & $d^\mrm{sig}_{ij,\mrm{ifo}}$ & o, $\tau$ & ILTT / backlink fibers / OB / clocks & $d^\mrm{osig}_{\mrm{ifo}} + \delta\tau^{{ij}}_{ik}$ (for RFIs and TMIs)
    \\ \vspace{1.0mm}
    Optical reception delay & $d^\mrm{orec}_{ij,\mrm{ifo}}$ & o & OB & 
    \\ \vspace{1.0mm}
    Electronic reception delay & $d^\mrm{erec}_{ij,\mrm{ifo}}$ & e & PRs / phasemeter analog electronics &
    \\ \vspace{1.0mm}
    Digital reception delay 0 & $d^\mrm{drec}_{ij,\mrm{ifo},0}$ & d & ADC / any DSP delay before DPLL & 
    \\ \vspace{1.0mm}
    Digital reception delay 1 & $d^\mrm{drec}_{ij,\mrm{ifo},1}$ & d & DPLL / signal decimation stages & 
    \\ \vspace{1.0mm}
    PRN generation delay & $d^\mrm{pin}_{ij}$ & e, d & DSP / analog electronics &
    \\ \vspace{1.0mm}
    PRN reception delays & $d^\mrm{psig}_{ij,\mrm{ifo}}$, $d^\mrm{ploc}_{ij,\mrm{ifo}}$ & d & DPLL / DLL / decimation stages &
    \\
    \end{tabular}
    \end{ruledtabular}
\end{table*}

\subsection{Measurements}\label{sub:measurements}
\cref{sub:traditional_tdi} formulated beatnote phase measurements in \cref{eq:isi_ij_basic,eq:rfi_ij_basic,eq:tmi_ij_basic}.
This section reformulates the beatnote phase measurements, including the onboard delays in \cref{tab:delays}.
In addition, we also model PRNR measurements.
The beatnote phase measurement $\mrm{ifo}_{ij}$ between the local beam phase $\phi_{ij}$ and the signal beam phase $\phi_{ik}$ without secondary noises can be written as,
\begin{widetext}
    \begin{align}
        \ifo{ij} &= \phi_{ik}\left(\tau - d^\mrm{in}_{ik} - d^\mrm{dis}_{ik,\mrm{ifo}} - d^\mrm{sig}_{ij,\mrm{ifo}} -  d^\mrm{orec}_{ij,\mrm{ifo}} - d^\mrm{erec}_{ij,\mrm{ifo}} - d^\mrm{drec}_{ij,\mrm{ifo},0} - d^\mrm{drec}_{ij,\mrm{ifo},1}\right)
        - \phi_{ij}\left(\tau - d^\mrm{in}_{ij} - d^\mrm{dis}_{ij,\mrm{ifo}} - d^\mrm{loc}_{ij,\mrm{ifo}} -  d^\mrm{orec}_{ij,\mrm{ifo}} - d^\mrm{erec}_{ij,\mrm{ifo}} - d^\mrm{drec}_{ij,\mrm{ifo},0} - d^\mrm{drec}_{ij,\mrm{ifo},1}\right)
        \nonumber\\
        &= \D{\mrm{drec}}{ij,\mrm{ifo},1}\D{\mrm{drec}}{ij,\mrm{ifo},0}\D{\mrm{erec}}{ij,\mrm{ifo}}\D{\mrm{orec}}{ij,\mrm{ifo}}\left(\D{\mrm{sig}}{ij,\mrm{ifo}}\D{\mrm{dis}}{ik,\mrm{ifo}}\D{\mrm{in}}{ik}\phi_{ik} - \D{\mrm{loc}}{ij,\mrm{ifo}}\D{\mrm{dis}}{ij,\mrm{ifo}}\D{\mrm{in}}{ij}\phi_{ij}\right).
        \label{eq:ifoij}
    \end{align}
\end{widetext}
Strictly speaking, \cref{eq:ifoij} can represent only RFI or TMI because of the pair of $\phi_{ij}$ and $\phi_{ik}$, see \cref{eq:tmi_ij_basic,eq:rfi_ij_basic}.
However, if we replace $ik$ with $ji$, \cref{eq:ifoij} can also represent ISI in our notation.
For the rest of this paper, we redefine laser phases at an EOM by absorbing a beam generation delay as,
\begin{align}
    \D{\mrm{in}}{ij}\phi_{ij} \rightarrow \phi_{ij},
    \label{eq:Doinphi}
\end{align}
which does not influence any discussion in this paper.

The beatnote phase measurements in \cref{eq:isi_ij_basic,eq:tmi_ij_basic,eq:rfi_ij_basic} can be reformulated based on \cref{eq:ifoij} with onboard delays as
\begin{align}
    \isi_{ij} &= \D{\mrm{rec}}{ij,\isi}\left(\D{\mrm{sig}}{ij,\isi}\D{\mrm{dis}}{ji,\isi}\phi_{ji} - \D{\mrm{loc}}{ij,\isi}\D{\mrm{dis}}{ij,\isi}\phi_{ij}\right.
    \nonumber\\
    &\hspace{35mm}\left.+ \D{\mrm{sig}}{ij,\isi}N_{ji}^\Delta + N_{ij}^\Delta\right),
    \label{eq:isi_ij_general}\\
    \tmi_{ij} &= \D{\mrm{rec}}{ij,\tmi}\left(\D{\mrm{sig}}{ij,\tmi}\D{\mrm{dis}}{ik,\tmi}\phi_{ik} - \D{\mrm{loc}}{ij,\tmi}\D{\mrm{dis}}{ij,\tmi}\phi_{ij}\right.
    \nonumber\\
    &\hspace{35mm}\left. + 2\D{\mrm{loc}}{ij,\tmi}(N_{ij}^\Delta - N_{ij}^\delta)\right),
    \label{eq:tmi_ij_general}\\
    \rfi_{ij} &= \D{\mrm{rec}}{ij,\rfi}\left(\D{\mrm{sig}}{ij,\rfi}\D{\mrm{dis}}{ik,\rfi}\phi_{ik} - \D{\mrm{loc}}{ij,\rfi}\D{\mrm{dis}}{ij,\rfi}\phi_{ij}\right),
    \label{eq:rfi_ij_general}
\end{align}
where the reception delays are bound up by a single operator for simplicity,
\begin{align}
    \D{\mrm{rec}}{ij,\mrm{ifo}} &\coloneqq \D{\mrm{drec}}{ij,\mrm{ifo},1}\D{\mrm{drec}}{ij,\mrm{ifo},0}\D{\mrm{erec}}{ij,\mrm{ifo}}\D{\mrm{orec}}{ij, \mrm{ifo}},
    \label{eq:Ddeorec}\\
    d^{\mrm{rec}}_{ij,\mrm{ifo}} &\coloneqq d^{\mrm{orec}}_{ij, \mrm{ifo}} + d^{\mrm{erec}}_{ij,\mrm{ifo}} + d^{\mrm{drec}}_{ij,\mrm{ifo},0} + d^{\mrm{drec}}_{ij,\mrm{ifo},1}.
    \label{eq:ddeorec}
\end{align}
GW signals are encoded in the ISI signal interferometric delay operator $\D{\mrm{sig}}{ij,\isi}$.
Note that delay operators acting on OB jitter $N_{ij}^\Delta$ and TM acceleration noise $N_{ij}^\delta$ are not strictly correct as those noises can be, in fact, considered stochastic constituents of interferometric delays $\D{\mrm{sig}}{ij,\isi}$ and $\D{\mrm{loc}}{ij,\tmi}$.
However, they are both significantly smaller than laser noise contributions, i.e., $N_{ij}^\Delta$ and $N_{ij}^\delta$ are, respectively, nanometer and picometer scales.
Therefore, the time shift of around $1$-\SI{10}{\nano\second}, expected for onboard optical delays, is not critical when formulating those noises.

In addition to the beatnote phase signals, the PRNR measurements can be modeled as the sum of all of the delays from the PRN generation to PRNR delay readout,
\begin{align}
    R^\mrm{sig}_{ij,\mrm{ifo}} &= d^\mrm{pin}_{ik} + d^\mrm{dis}_{ik,\mrm{ifo}} + d^\mrm{sig}_{ij,\mrm{ifo}}
    \nonumber\\
    &\hspace{7mm} +  d^\mrm{orec}_{ij,\mrm{ifo}} + d^\mrm{erec}_{ij,\mrm{ifo}} + d^\mrm{drec}_{ij,\mrm{ifo},0} + d^\mrm{psig}_{ij,\mrm{ifo}} + N^\mrm{sig}_{ij,\mrm{ifo}},
    \label{eq:Rsig}\\
    R^\mrm{loc}_{ij,\mrm{ifo}} &= d^\mrm{pin}_{ij} + d^\mrm{dis}_{ij,\mrm{ifo}} + d^\mrm{loc}_{ij,\mrm{ifo}}
    \nonumber\\
    &\hspace{7mm} +  d^\mrm{orec}_{ij,\mrm{ifo}} + d^\mrm{erec}_{ij,\mrm{ifo}} + d^\mrm{drec}_{ij,\mrm{ifo},0} + d^\mrm{ploc}_{ij,\mrm{ifo}} + N^\mrm{loc}_{ij,\mrm{ifo}}.
    \label{eq:Rloc}
\end{align}
The $N$ terms represent the stochastic noise of PRNR measurements in the observation band: mainly shot noise and PRN code interference~\cite{Esteban2009,Sutton2010}.
Note that the formulation of PRNR measurements above is assumed to already resolve PRNR ambiguity caused by a finite PRN code length, usually $\sim$ \OSI{100}{\kilo\meter}~\cite{Yamamoto2024}.
This is important only for the ISI received PRNR measurement $R^\mrm{sig}_{ij,\isi}$ because of $d^\mrm{sig}_{ij,\isi}$ that includes large offsets and their rapid time evolution due to the ILTT and the clock difference between local and distant SC.
The ambiguity resolution can be done by either on-ground observations or TDI ranging~\cite{Reinhardt2024A}.
Other PRNR measurements are usually way smaller than the ambiguity; therefore, there is no need to resolve their ambiguity.

Let us elaborate on the explanation of clocks and their differences absorbed into $d^\mrm{sig}_{ij,\mrm{ifo}}$.
In this article, even though the two local phasemeters are driven by a common clock $i$ as denoted in \cref{fig:sc_notation}, we use double indices to distinguish clocks at the phasemeters, i.e., clocks $ij$ and $ik$ originating from the master clock $i$.
This lets us study how the potential relative difference between the two local clocks can be treated in our scheme in \cref{sec:tdi}.
To provide a general formulation of the relation between clocks, we express clock $xy$ in terms of clock $xz$ with its deviation from clock $xz$ denoted by $\delta\tau^{{xz}}_{xy}$,
\begin{align}
    \tau^{{xz}}_{xy}(\tau) &= \tau + \delta\tau^{{xz}}_{xy}(\tau),
    \label{eq:tauxy_tauxz}\\
    \delta\tau^{{xz}}_{xy}(\tau) &= q^{{xz},\epsilon}_{xy}(\tau) + q^{xz,o}_{xy}(\tau) + \delta\tau^{{xz}}_{xy,0}.
    \label{eq:deltauxy_tauxz}
\end{align}
We call $\delta\tau^{{xz}}_{xy}(\tau)$ \emph{clock time deviation}.
For qualitative discussions, \cref{eq:deltauxy_tauxz} decomposes the clock time deviation into three parts: a constant large initial offset $\delta\tau^{{xz}}_{xy,0}$, a slow out-of-band drift due to a clock frequency offset usually around the order of \si{ppm} $q^{xz,o}_{xy}$, and a stochastic jitter in the observation band $q^{{xz},\epsilon}_{xy}$.
By definition, $\delta\tau^{{xz}}_{xy}(0)=\delta\tau^{{xz}}_{xy,0}$.

We assume that the local clock distribution, namely blue lines from clock $i$ in \cref{fig:sc_notation}, is designed to keep the in-band relative jitter between the two local clocks $q^{{ik},\epsilon}_{ij}$ below the target noise level~\cite{Bayle2023}.
Therefore, the clock time deviations read
\begin{align}
    \delta\tau^{{ik}}_{ij}(\tau) &= q^{ik,o}_{ij}(\tau) + \delta\tau^{{ik}}_{ij,0},
    \label{eq:deltauij_tauik}\\
    \delta\tau^{{ji}}_{ij}(\tau) &= q^{{j},\epsilon}_{i}(\tau) + q^{ji,o}_{ij}(\tau) + \delta\tau^{{ji}}_{ij,0},
    \label{eq:deltauij_tauji}
\end{align}
where $\delta\tau^{{ik}}_{ij}$ is between the two local clocks and $\delta\tau^{{ji}}_{ij}$ is between the local and remote clocks.
Note that the stochastic in-band jitter $q^{{j},\epsilon}_{i}$ has a single index notation now because we assumed above that the two local clocks are coherent in the observation band.
Accordingly, the stochastic term is dropped from $\delta\tau^{{ik}}_{ij}$ in \cref{eq:deltauij_tauik}.

To wrap up this section, let us define advancement operators, which will be used in the following sections.
In general, for a given delay $\D{}{}$ in \cref{eq:D}, we can define its inverse advancement operator via the condition of
\begin{align}
    \A{}{} \D{}{} x(t) = \D{}{} \A{}{} x(t) = x(t).
    \label{eq:ADDA1}
\end{align}
This implies the time series of advancements $a(t)$ depends on the corresponding delays $d(t)$ via an implicit equation,
\begin{align}
     \A{}{} \D{}{} x(t) &= \A{}{}x(t - d(t)) = x(t - d(t + a(t)) + a(t)) \stackrel{!}{=} x(t)
     \label{eq:AD1}\\
     & \implies a(t) =  d(t + a(t)).
     \label{eq:adv_del}
\end{align}
In this manuscript, we mostly treat approximately constant on-board delays, in which case this condition simplifies to $a(t) \approx d(t)$.

\subsection{Assumptions}\label{sub:assumptions}
This section clarifies key assumptions that will play a pivotal role in our scheme presented in \cref{sec:tdi}.
\\\\
\emph{Assumption 1, a heterodyne beatnote and a PRN signal sense the same delay along an identical path.}
The formulation in \cref{sub:measurements} and the signal-path diagram in \cref{fig:diagram_notation} already imply that a heterodyne beatnote and PRN signals sense the same delays along identical paths, e.g., from $d^\mrm{dis}_{ij,\mrm{ifo}}$ to $d^\mrm{drec}_{ij,\mrm{ifo},0}$ in \cref{eq:Rloc}.
This assumption neglects the dispersion due to the frequency dependency of delays, especially of the electronic reception path $d^\mrm{erec}_{ij,\mrm{ifo}}$.
Although the resulting difference in delay between the beatnote and the PRN signals highly depends on the QPR specification, the experiment in \cite{Yamamoto2024} observed around \SI{10}{\centi\meter} with the bandwidth of about \SI{35}{\mega\Hz}.
This particular ranging error will be highlighted in noise analysis \cref{sub:noise}.
The interspacecraft light travel also causes such dispersion due to the solar wind; however, it was estimated to be around \SI{10}{\pico\meter}, which is totally negligible~\cite{Reinhardt2024A}.
\\\\
\emph{Assumption 2, signal interferometric delays $d^\mrm{sig}_{ij,\mrm{ifo}}$ in RFIs and TMIs can be explicitly decomposed into optical and clock contributions.}
The optical delay parts (i.e., a backlink fiber and free-space optical paths) are expected to be constant from the ranging perspective.
Hence, for the rest of the article, we explicitly decompose $d^\mrm{sig}_{ij,\mrm{ifo}}$ only for RFIs and TMIs to its optical component $d^\mrm{osig}_{ij,\mrm{ifo}}$ and clock,
\begin{align}
    d^\mrm{sig}_{ij,\mrm{ifo}} &= d^\mrm{osig}_{ij,\mrm{ifo}} + \delta\tau^{{ij}}_{ik},
    \label{eq:dotausig_ifo}
    \\
    \D{\mrm{sig}}{ij,\mrm{ifo}}x(\tau) &= \D{\mrm{osig}}{ij,\mrm{ifo}}\T{ik}{ij} x(\tau)
    \nonumber\\
    &= x(\tau - d^\mrm{osig}_{ij,\mrm{ifo}} - \delta\tau^{{ij}}_{ik}),
    \label{eq:Dotausig_ifo}
\end{align}
To draw attention, we assigned a dedicated operator $\T{ik}{ij}$ for the clock conversion from clock $ik$ to clock $ij$.
Also, $\D{\mrm{osig}}{ij,\mrm{ifo}}$ and $\T{ik}{ij}$ are commutative because they are considered constant or, even if any, drift very slowly.
In addition, we write the advancement operator as
\begin{align}
    \A{\mrm{sig}}{ij,\mrm{ifo}}x(\tau) &= x(\tau + d^\mrm{sig}_{ij,\mrm{ifo}} + \delta\tau^{{ij}}_{ik})
    \nonumber\\
    &= x(\tau + d^\mrm{sig}_{ij,\mrm{ifo}} - \delta\tau^{{ik}}_{ij})
    \nonumber\\
    &= \A{\mrm{osig}}{\mrm{ifo}}\T{ij}{ik}x(\tau),
    \label{eq:Aotausig_ifo}
\end{align}
where $\delta\tau^{{ij}}_{ik}=-\delta\tau^{{ik}}_{ij}$ to get the second line is valid because of the small and slow change of clock time deviations.
\\\\
\emph{Assumption 3, $d^\mrm{psig}_{ij,\mrm{ifo}}$ and $d^\mrm{ploc}_{ij,\mrm{ifo}}$ can be subtracted via simulation.}
These delays are unique to PRN signals; hence, they are considered extra offsets in ranging and need to be removed before use in TDI.
However, importantly, any contributors to them are in the digital domain.
Hence, we can estimate it via numerical simulation based on commandable user-known FPGA parameters to good accuracy (likely a few centimeters or better~\cite{Yamamoto2024}).
This means that those terms can be subtracted from the PRNR measurements in \cref{eq:Rsig,eq:Rloc}.
In the rest of this article, we redefine the PRNR measurements in the form after that subtraction,
\begin{align}
    R^\mrm{sig}_{ij,\mrm{ifo}}-d^\mrm{psig}_{ij,\mrm{ifo}} &\rightarrow R^\mrm{sig}_{ij,\mrm{ifo}},
    \label{eq:Rsig_corr}\\
    R^\mrm{loc}_{ij,\mrm{ifo}}-d^\mrm{ploc}_{ij,\mrm{ifo}} &\rightarrow R^\mrm{loc}_{ij,\mrm{ifo}}.
    \label{eq:Rloc_corr}
\end{align}
\\\\
\noindent \emph{Assumption 4, all OBs in the constellation will be constructed based on the identical design, and their manufacturing tolerance is negligible.}
Similarly to \cite{Reinhardt2024B}, the OBs are assumed to be built based on the same design. This enables the degeneration of many delays arbitrarily indexed in the second column in \cref{tab:delays}.
Each delay is relabeled based on the assumption below,
\begin{align}
    d^\mrm{dis}_{ij,\mrm{ifo}} &\rightarrow d^\mrm{dis}_\mrm{ifo},
    \label{eq:ddist_ifo}\\
    d^\mrm{loc}_{ij,\mrm{ifo}} &\rightarrow d^\mrm{loc}_\mrm{ifo},
    \label{eq:doloc_ifo}\\
    d^\mrm{osig}_{ij,\mrm{ifo}} &\rightarrow d^\mrm{osig}_\mrm{ifo} \hspace{5mm} \text{(only for RFIs and TMIs).}
    \label{eq:dosig_ifo}
\end{align}
$d^\mrm{osig}_{ij,\mrm{ifo}}$ in \cref{eq:dosig_ifo} is the pure optical contribution in the signal interferometric delay for RFIs and TMIs (see \cref{eq:dotausig_ifo}).
ISIs include the ILTTs and the clock differences in $d^\mrm{sig}_{ij,\isi}$; therefore, the relabeling is not applicable.
Concerning the local distribution delay, we actually can allow the length of delivery fibers from the lasers to the OBs to be different between MOSAs, though \cref{eq:ddist_ifo} implies they are assumed the same.
We elaborate on this detail in \cref{sec:dist}, while the main text sticks to \cref{eq:ddist_ifo}.
\\\\
\emph{Assumption 5, most of the nested terms of delays are negligible; therefore, most of delay operators are commutative:} Delays in \cref{tab:delays} are expected to vary, even if any, slowly enough to neglect their nested terms against the scale PRNR looks at: above \SI{1}{\centi\meter}.
The fastest time evolution is expected to be caused by clock differences \OSI{100}{\meter\per\second} with the \si{ppm} clock frequency offset, while ILTT is up to \OSI{10}{\meter\per\second} considering realistic LISA orbit files provided by European Space Agency (ESA)~\cite{LisaOrbits}.
They are both contributors to the ISI signal interferometric delay $d^\mrm{sig}_{ij,\isi}$.
Therefore, to amount to \SI{1}{\centi\meter}, $d^\mrm{sig}_{ij,\isi}$ must be shifted by \SI{0.1}{\milli\second}.
Any delays in \cref{tab:delays} are expected to be much smaller than \SI{0.1}{\milli\second}, except for $d^\mrm{sig}_{ij,\isi}$ itself and the digital reception delay 1 $d^\mrm{drec}_{ij,\mrm{ifo},1}$.
\\\\
\emph{Assumption 6, the digital reception delay 1 $d^\mrm{drec}_{ij,\mrm{ifo},1}$ can be subtracted.}
Most contributions stem from data decimation stages because of antialiasing filtering at lower sampling rates.
However, it is a fully deterministic pure delay: $N$ taps at $f$ \si{Sps} result in $(N-1)/2/f$ \si{seconds}.
Hence, it is assumed that we can easily remove $d^\mrm{drec}_{ij,\mrm{ifo},1}$ from phase measurements.~\footnote{Note that, if we use the identical decimation design for all beatnote phase and PRNR measurements, the decimation delay would not have to be corrected from any measurements.}
This means we can rewrite the beatnote signal in \cref{eq:ifoij} as,
\begin{align}
    \A{d,\mrm{rec}}{ij,\mrm{ifo},1}\ifo{ij} &\rightarrow \ifo{ij}.
    \label{eq:ifoij_rewrite}
\end{align}
Combined with Assumption 5, this assumption enables commutating any operators except for a pair of two signal interferometric delays in ISIs.

\subsection{Setup}\label{sub:setup}
Departing from the general expression of the measurements in \cref{sub:measurements}, we made six assumptions and some relabeling and redefinition in \cref{sub:assumptions}.
This section summarizes the resulting formulation of the measurements based on those assumptions.
In addition, we will provide useful expressions, which will be often used in \cref{sec:tdi}.

Assumptions 4, 5, and 6 simplify the most general measurement equations in \cref{eq:isi_ij_general,eq:tmi_ij_general,eq:rfi_ij_general} as,
\begin{align}
    \isi_{ij} &= \D{\mrm{rec}}{ij,\isi}\D{\mrm{dis}}{\isi}\left[\D{\mrm{sig}}{ij,\isi}\phi_{ji} - \D{\mrm{loc}}{\isi}\phi_{ij} + \D{\mrm{sig}}{ij,\isi}N_{ji}^\Delta + N_{ij}^\Delta \right],
    \label{eq:isi_ij}\\
    \tmi_{ij} &= \D{\mrm{rec}}{ij,\tmi}\D{\mrm{dis}}{\tmi}\left[\D{\mrm{sig}}{ij,\tmi}\phi_{ik} - \D{\mrm{loc}}{\tmi}\phi_{ij} + 2\D{\mrm{loc}}{\tmi}(N_{ij}^\Delta - N_{ij}^\delta)\right],
    \label{eq:tmi_ij}\\
    \rfi_{ij} &= \D{\mrm{rec}}{ij,\rfi}\D{\mrm{dis}}{\rfi}\left[\D{\mrm{sig}}{ij,\rfi}\phi_{ik} - \D{\mrm{loc}}{\rfi}\phi_{ij}\right].
    \label{eq:rfi_ij}
\end{align}
The local distribution delay $\D{\mrm{dis}}{\mrm{ifo}}$ is commuted with $\D{\mrm{loc}}{\mrm{ifo}}$ and $\D{\mrm{sig}}{ij,\mrm{ifo}}$ based on Assumption 5 and taken out of the parentheses for readability.
In addition, Assumptions 3 and 4 simplify the PRNR measurements in \cref{eq:Rsig,eq:Rloc} as,
\begin{align}
    R^\mrm{sig}_{ij,\mrm{ifo}}&= d^\mrm{pin}_{ik} + d^\mrm{dis}_\mrm{ifo} + d^\mrm{sig}_{ij,\mrm{ifo}} +  d^\mrm{rec}_{ij,\mrm{ifo}} + N^\mrm{sig}_{ij,\mrm{ifo}},
    \label{eq:Rsig_simple}\\
    R^\mrm{loc}_{ij,\mrm{ifo}}&= d^\mrm{pin}_{ij} + d^\mrm{dis}_\mrm{ifo} + d^\mrm{loc}_\mrm{ifo} +  d^\mrm{rec}_{ij,\mrm{ifo}} + N^\mrm{loc}_{ij,\mrm{ifo}}.
    \label{eq:Rloc_simple}
\end{align}

Based on the PRNR measurements, we define special operators, which will be used in TDI,
\begin{align}
    \Dc{a}{ij,\mrm{ifo}}x(\tau) \coloneqq x(\tau - \F R^a_{ij,\mrm{ifo}}),
    \label{eq:Dcab}\\
    \Ac{a}{ij,\mrm{ifo}}x(\tau) \coloneqq x(\tau + \F R^a_{ij,\mrm{ifo}}),
    \label{eq:Acab}
\end{align}
where $a=$ sig or loc.
The operator $\F$ represents the filtering of the PRNR measurement $R^a_{ij,\mrm{ifo}}$ to reduce its in-band jitter $N^\mrm{a}_{ij,\mrm{ifo}}$.
We call $\Dc{a}{ij,b}$ ($\Ac{a}{ij,b}$) \emph{PRNR delay (advancement) operator} for the rest of this article in order to highlight that the time shifts applied by these operators directly correspond to the actual PRNR measurements.

\cref{eq:Rsig_simple,eq:Rloc_simple} tell us that the PRNR delay operators can be decomposed into the standard delay operators as,
\begin{align}
    \Dc{\mrm{sig}}{ij,\mrm{ifo}} &= \D{\mrm{rec}}{ij,\mrm{ifo}}\D{\mrm{sig}}{ij,\mrm{ifo}}\D{\mrm{dis}}{\mrm{ifo}}\D{\mrm{pin}}{ik},
    \label{eq:Dc_sig_ijifo}\\
    \Dc{\mrm{loc}}{ij,\mrm{ifo}} &= \D{\mrm{rec}}{ij,\mrm{ifo}}\D{\mrm{loc}}{\mrm{ifo}}\D{\mrm{dis}}{\mrm{ifo}}\D{\mrm{pin}}{ij},
    \label{eq:Dc_loc_ijifo}
\end{align}
where the filtering $\F$ is assumed to properly remove the in-band jitter.
Note that, unlike the standard advancement operator $\A{}{}$ in \cref{eq:ADDA1}, the PRNR advancement operator $\Ac{a}{ij,\mrm{ifo}}$ is defined not as the inverse operator of the PRNR delay operator $\Dc{a}{ij,\mrm{ifo}}$ but as a simple advancement with the PRNR measurement.
The same as the discussions around \cref{eq:adv_del}, $\Ac{a}{ij,\mrm{ifo}}$ becomes inverse to $\Dc{a}{ij,\mrm{ifo}}$ only when the time shift is considered constant over a measurement time window to be analyzed.
Assumptions 5 and 6 suggest that the received PRNR advancement operator in ISI, therefore, cannot be written as the inverse of its delay operator: $\Ac{\mrm{sig}}{ij,\isi}\Dc{\mrm{sig}}{ij,\isi}\neq\Dc{\mrm{sig}}{ij,\isi}\Ac{\mrm{sig}}{ij,\isi}\neq \mathds{1}$.
However, our scheme, presented in \cref{sec:tdi}, never uses $\Ac{\mrm{sig}}{ij,\isi}$.
Hence, we write the PRNR advancement operators as the inverse to the delay operators in \cref{eq:Dc_sig_ijifo,eq:Dc_loc_ijifo} within our framework:
\begin{align}
    \Ac{\mrm{sig}}{ij,\mrm{ifo}} &= \A{\mrm{pin}}{ik}\A{\mrm{dis}}{\mrm{ifo}}\A{\mrm{sig}}{ij,\mrm{ifo}}\A{\mrm{rec}}{ij,\mrm{ifo}},
    \label{eq:Ac_sig_ijifo}\\
    \Ac{\mrm{loc}}{ij,\mrm{ifo}} &= \A{\mrm{pin}}{ij}\A{\mrm{dis}}{\mrm{ifo}}\A{\mrm{loc}}{\mrm{ifo}}\A{\mrm{rec}}{ij,\mrm{ifo}}.
    \label{eq:Ac_loc_ijifo}
\end{align}

Regarding the filtering process $\F$ in \cref{eq:Dcab,eq:Acab}, a simple low-pass filter or polynomial fit must be enough for most PRNR measurements that evolve slowly and slightly.
Only when it comes to the ISI received-PRNR measurement $R^\mrm{sig}_{ij,\isi}$, we also need to make use of the phase measurement of a clock sideband-sideband beatnote to remove relative in-band clock jitter~\cite{Hartwig2021}.
We can use either of the averaging method~\cite{Hartwig2022} or a more sophisticated Kalman filter approach~\cite{Reinhardt2024A}.
The clock modulation signal is delayed over different paths from the PRN signal paths, e.g., the blue modulation path in \cref{fig:diagram_metrology}.
The process to use the clock sideband signals is absorbed into $\F$ with no discussion on such relative delays because they are negligible up to \OSI{1}{\milli\second} if we aim to suppress the in-band relative clock jitter of \OSI{10}{\pico\second\prtHz} down to the target level \OSI{10}{\femto\second\prtHz}.
We expect the relative delay cannot be as large as \OSI{1}{\milli\second}.

In \cref{sec:tdi}, the PRNR operators will always be used as a pair of the delay and advancement operators in the TDI algorithm.
Hence, it is useful to leave the general formulas of a beatnote phase measurement advanced by each of the associated received and local PRNR operators,
\begin{align}
    \Ac{\mrm{sig}}{ij,\mrm{ifo}}\ifo{ij}(\tau) &= \A{\mrm{pin}}{ik}\left(\phi_{ik} - \A{\mrm{sig}}{ij,\mrm{ifo}}\D{\mrm{loc}}{\mrm{ifo}}\phi_{ij}\right),
    \label{eq:Asig_ifoij}\\
    \Ac{\mrm{loc}}{ij,\mrm{ifo}}\ifo{ij}(\tau) &= \A{\mrm{pin}}{ij}\left(\A{\mrm{loc}}{\mrm{ifo}}\D{\mrm{sig}}{ij,\mrm{ifo}}\phi_{ik} - \phi_{ij}\right),
    \label{eq:Aloc_ifoij}
\end{align}
where OB jitter and TM acceleration noise in ISIs and TMIs are omitted.

\section{Application to TDI}\label{sec:tdi}
Using the framework prepared in \cref{sec:framework}, we reformulate TDI with the PRNR operators in \cref{eq:Dcab,eq:Acab}.
The characteristic usage of PRNR operators in our scheme is described as follows: we first \emph{pull} beatnote phase measurements with the associated PRNR advancement operators $\Ac{a}{ij,\mrm{ifo}}$ (see \cref{eq:Asig_ifoij,eq:Aloc_ifoij}) and then \emph{push} them to another beatnote phase measurement with a PRNR delay operator $\Dc{a}{ij,\mrm{ifo}}$ associated with the target phase measurement.
This way, we directly commit to the actual TDI needs, described in \cite{Reinhardt2024A}: TDI requires the time differences with which laser frequency noise instances appear in the phase measurements on the local and remote SC.
Not limited to the interspacecraft measurements, we generalize this to all interferometers by reformulating also the intermediate TDI steps based on the pull-and-push action with PRNR.
We will also show that such a PRNR action cancels the PRN generation delay $d^\mrm{pin}_{ij}$ by itself, which otherwise requires separate calibration.
This characteristic action with PRNR measurements has already been demonstrated experimentally at the Albert Einstein Institute Hannover~\cite{Yamamoto2024} with a hexagonal optical test bed.
This fact advances our confidence in the proposed scheme.

\subsection{TDI $\xi$ variables}\label{sub:xi}
Going straight to the point, we propose to reformulate the $\xi$ variables in \cref{eq:xi_basic} based on the PRNR operators (\cref{eq:Dcab,eq:Acab}) as
\begin{align}
    \xi_{ij}&=\isi_{ij} + \Dc{\mrm{loc}}{ij,\isi}\frac{\Ac{\mrm{loc}}{ij,\rfi}\rfi_{ij}-\Ac{\mrm{loc}}{ij,\tmi}\tmi_{ij}}{2}
    \nonumber\\
    &\hspace{10mm} + \Dc{\mrm{sig}}{ij,\isi}\frac{\Ac{\mrm{loc}}{ji,\rfi}\rfi_{ji}-\Ac{\mrm{loc}}{ji,\tmi}\tmi_{ji}}{2}.
    \label{eq:xi_prnr}
\end{align}
Note that the PRNR operators act on the phase measurements as a pair of advancement and delay, as mentioned above.
The representation of operator actions by the signal-path diagram is shown in \cref{fig:diagram_rfi_isi_xi}.
For visibility, \cref{fig:diagram_rfi_isi_xi} only illustrates the action on RFIs, neglecting the one on TMIs.

\begin{figure}[ht]
    \centering
\includegraphics[width=8.6cm]{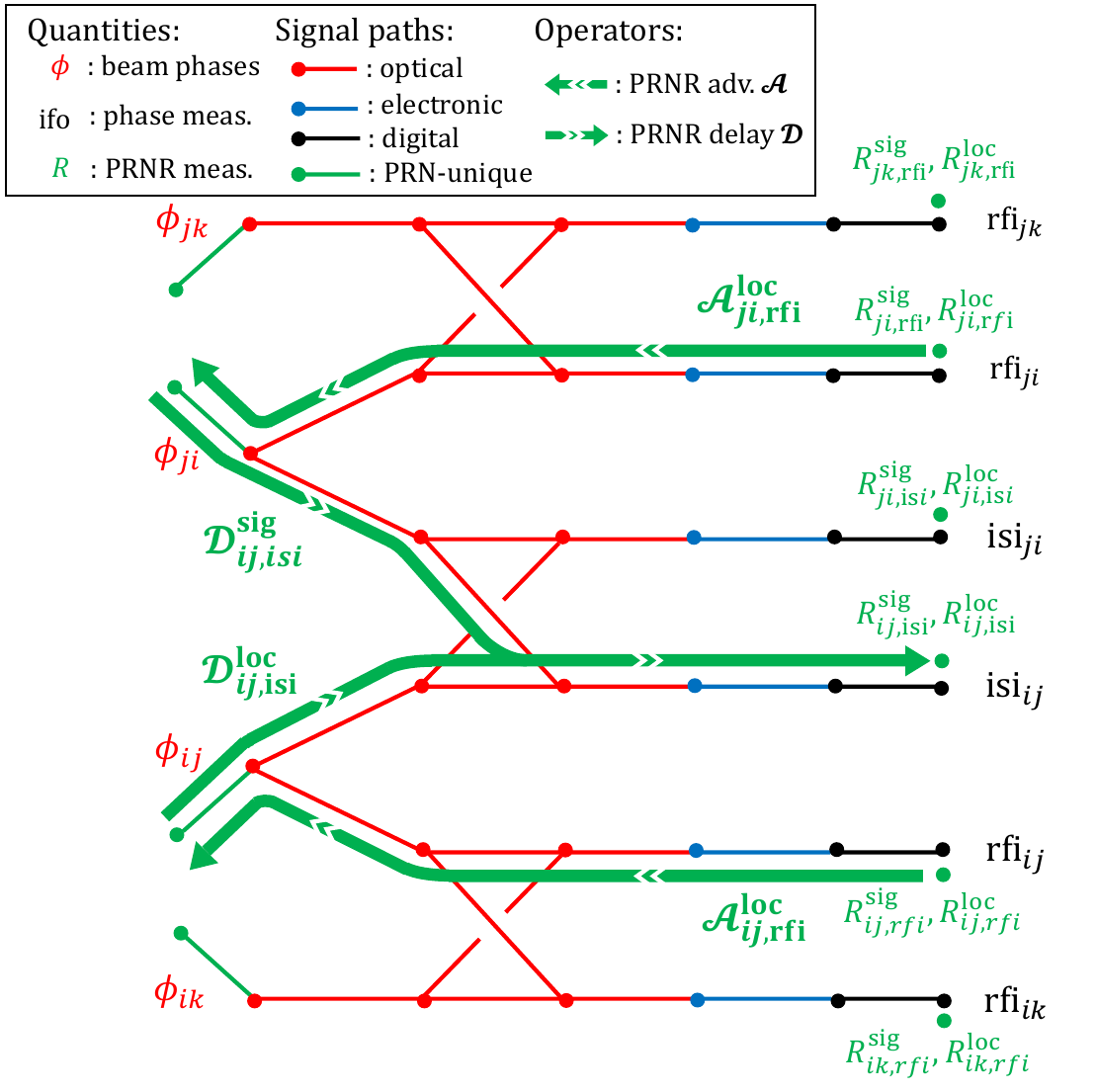}
    \caption{
    Signal-path diagram showing the actions of PRNR operators on the RFI measurements ($\rfi_{ij}$ and $\rfi_{ji}$) in $\xi_{ij}$ (see \cref{eq:xi_prnr}).
    The RFI measurements are time adjusted to $\isi_{ij}$.
    Contrary to the general diagram in \cref{fig:diagram_notation}, this diagram (and the ones below) considers the assumptions made in \cref{sub:assumptions}.
    A PRNR operator effectively shifts a beatnote phase measurement in time from one green dot to another.
    Green dots either correspond to PRNR measurement times (right green dots) or to PRN signal generation times (left green dots).
    Note that the PRN generation delays (green signal paths on the left) cancel between PRNR advancement and delay operators, such that this processing step is inherently unaffected by these delays.
    }
    \label{fig:diagram_rfi_isi_xi}
\end{figure}

The general formulas, \cref{eq:Asig_ifoij,eq:Aloc_ifoij}, let us easily calculate the numerators composed of one RFI and one TMI,
\begin{align}
    &\Ac{\mrm{loc}}{ij,\rfi}\rfi_{ij}-\Ac{\mrm{loc}}{ij,\tmi}\tmi_{ij}
    \nonumber\\
    &= \A{\mrm{pin}}{ij}\left[\left(\A{\mrm{loc}}{\rfi}\D{\mrm{sig}}{ij,\rfi} - \A{\mrm{loc}}{\tmi}\D{\mrm{sig}}{ij,\tmi}\right)\phi_{ik} - 2(N_{ij}^\Delta - N_{ij}^\delta)\right]
    \nonumber\\
    &= -2\A{\mrm{pin}}{ij}(N_{ij}^\Delta - N_{ij}^\delta),
    \label{eq:xi_prnr_cond2}
\end{align}
since
\begin{align}
    &\left(\A{\mrm{loc}}{\rfi}\D{\mrm{sig}}{ij, \rfi} - \A{\mrm{loc}}{\tmi}\D{\mrm{sig}}{ij, \tmi}\right)\phi_{ik}
    \nonumber\\
    &\hspace{5mm}= \left(\A{\mrm{loc}}{\rfi}\D{\mrm{osig}}{\rfi} - \A{\mrm{loc}}{\tmi}\D{\mrm{osig}}{ \tmi}\right)\T{ik}{ij}\phi_{ik}
    \nonumber\\
    &\hspace{5mm}= 0,\hspace{2mm}\text{if $d^\mrm{sig}_\rfi - d^\mrm{loc}_\rfi=d^\mrm{sig}_\tmi - d^\mrm{loc}_\tmi$}.
\end{align}
In the second line, we used the decomposition of the signal interferometric delays into their optical and clock parts as formulated in Eq. (31).
The laser phase $\phi_{ik}$ cancels to the first-order time derivative by minimizing onboard optical delay effects if $d^\mrm{sig}_\rfi - d^\mrm{loc}_\rfi=d^\mrm{sig}_\tmi - d^\mrm{loc}_\tmi$.
Such a relative delay difference between RFI and TMI is around \SI{2}{\centi\meter} in the LISA case~\cite{Brzozowski2022}, and its effect is small enough against the target final performance~\cite{Reinhardt2024B}.
This is an important condition when designing OBs.
At the end, \cref{eq:xi_prnr} results in
\begin{widetext}
    \begin{align}
        \xi_{ij} 
        &= \isi_{ij} - \D{\mrm{rec}}{ij,\isi}\D{\mrm{dis}}{\isi}\left[\D{\mrm{loc}}{\isi}(N_{ij}^\Delta - N_{ij}^\delta) + \D{\mrm{sig}}{ij,\isi}(N_{ji}^\Delta - N_{ji}^\delta)\right]
        \nonumber\\
        &= \D{\mrm{rec}}{ij,\isi}\D{\mrm{dis}}{\isi}\left[\D{\mrm{sig}}{ij,\isi}\phi_{ji} - \D{\mrm{loc}}{\isi}\phi_{ij} + \D{\mrm{sig}}{ij,\isi}N_{ji}^\Delta + N_{ij}^\Delta - \D{\mrm{loc}}{\isi}(N_{ij}^\Delta - N_{ij}^\delta) - \D{\mrm{sig}}{ij,\isi}(N_{ji}^\Delta - N_{ji}^\delta)\right]
        \nonumber\\
        &\approx \D{\mrm{rec}}{ij,\isi}\D{\mrm{dis}}{\isi}\left[\D{\mrm{sig}}{ij,\isi}\phi_{ji} - \D{\mrm{loc}}{\isi}\phi_{ij} + \D{\mrm{loc}}{\isi}N_{ij}^\delta + \D{\mrm{sig}}{ij,\isi}N_{ji}^\delta\right] \hspace{5mm} \because (1-\D{\mrm{loc}}{\isi})N_{ij}^\Delta \approx 0.
        \label{eq:xi_prnr_result}
    \end{align}
\end{widetext}
This successfully ends up with $\isi_{ij}$ in \cref{eq:isi_ij}, but without the OB jitter $N_{ij}^\Delta$ and $N_{ji}^\Delta$.
The same as \cref{sub:traditional_tdi}, we will neglect the TM acceleration noises $N_{ij}^\delta$ in the following.

\cref{eq:xi_prnr_result} shows one benefit of using PRNR operators as a pair of advancement and delay operators: the cancellation of the PRN generation delays $\D{\mrm{pin}}{ij}$.
Unlike PRN reception delays in the digital domain, the PRN generation delays include analog electronics and an EOM, whose delays are not direct measures of any signal available from the metrology system.
This means we would need dedicated calibration schemes like the PRNR-based calibration algorithm~\cite{Euringer2023} and/or the combination of prelaunch component characterization and in-flight environmental monitors like temperature sensors~\cite{Dollase2023}.
Hence, constructing a signal combination insensitive to such delays enhances the robustness of the detector performance.

\subsection{TDI $\eta$ variables}\label{sub:eta}
We reformulate $\eta$ variables in \cref{eq:eta12_basic,eq:eta13_basic} as,
\begin{align}
    \eta_{12} &= \xi_{12} + \Dc{\mrm{sig}}{12,\isi}\frac{\Ac{\mrm{loc}}{21,\rfi}\rfi_{21} - \Ac{\mrm{sig}}{23,\rfi}\rfi_{23}}{2},
    \label{eq:eta12_prnr}\\
    \eta_{13} &= \xi_{13} + \Dc{\mrm{loc}}{13,\isi}\frac{\Ac{\mrm{sig}}{12,\rfi}\rfi_{12} - \Ac{\mrm{loc}}{13,\rfi}\rfi_{13}}{2}.
    \label{eq:eta13_prnr}
\end{align}
The action of the PRNR operators in $\eta_{12}$ is visualized in the signal-path diagram in \cref{fig:diagram_rfi_isi_eta}.

\begin{figure}
    \centering
\includegraphics[width=8.6cm]{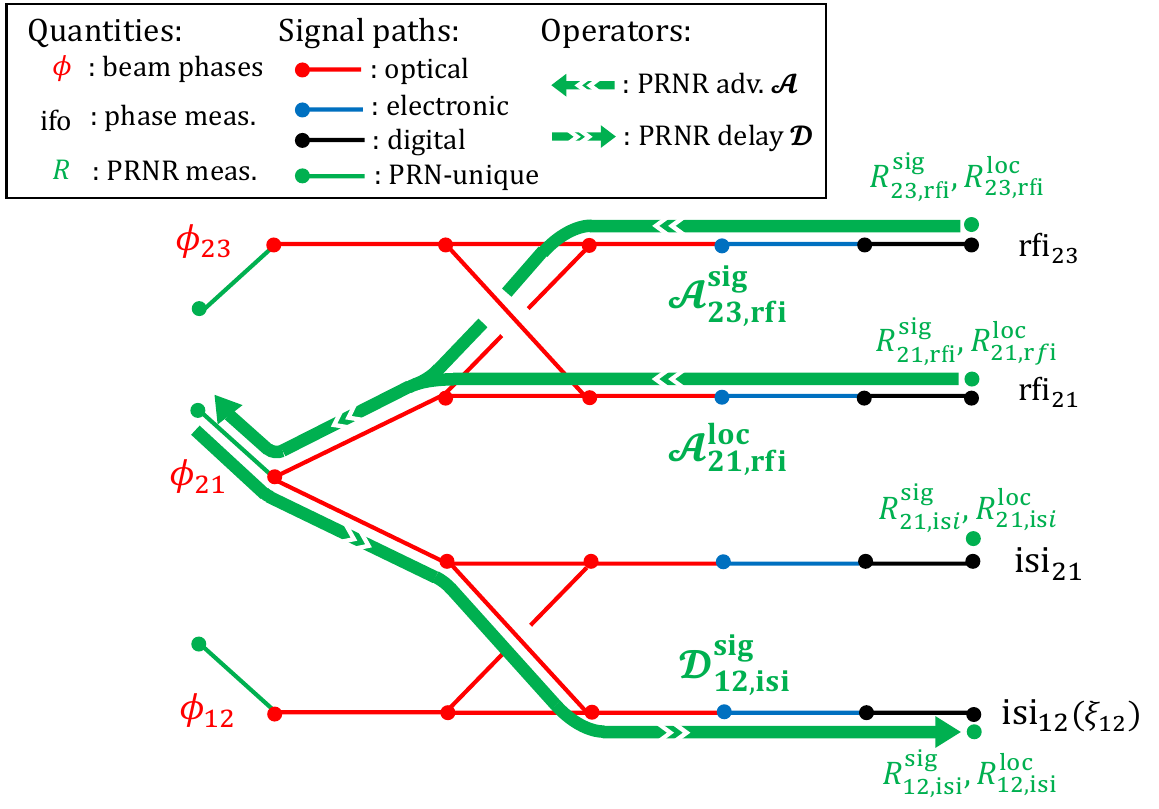}
    \caption{
    Signal-path diagram showing the actions of PRNR operators on the RFI measurements ($\rfi_{21}$ and $\rfi_{23}$) in $\eta_{12}$ (see \cref{eq:eta12_prnr}).
    They are time adjusted to $\xi_{12}$, i.e., to $\isi_{12}$.
    The potential clock drift between the two local RFI measurements is automatically corrected by combining $\Ac{\mrm{loc}}{21,\rfi}$ and $\Ac{\mrm{sig}}{23,\rfi}$.
    }
    \label{fig:diagram_rfi_isi_eta}
\end{figure}

Starting with \cref{eq:eta12_prnr}, the pair of RFIs can be calculated via \cref{eq:Asig_ifoij,eq:Aloc_ifoij},
\begin{align}
    \Ac{\mrm{loc}}{21,\rfi}\rfi_{21} &- \Ac{\mrm{sig}}{23,\rfi}\rfi_{23}
    \nonumber\\
    &= \A{\mrm{pin}}{21}\left(\left(\A{\mrm{sig}}{23,\rfi}\D{\mrm{loc}}{\rfi} + \A{\mrm{loc}}{\rfi}\D{\mrm{sig}}{21,\rfi}\right)\phi_{23} - 2\phi_{21}\right)
    \nonumber\\
    &= \A{\mrm{pin}}{21}\left(\left(\A{\mrm{osig}}{\rfi}\D{\mrm{loc}}{\rfi} + \A{\mrm{loc}}{\rfi}\D{\mrm{osig}}{\rfi}\right)\T{23}{21}\phi_{23} - 2\phi_{21}\right)
    \nonumber\\
    &= 2\A{\mrm{pin}}{21}\left(\T{23}{21}\phi_{23} - \phi_{21}\right),
    \label{eq:eta_rfi21_rfi23}
\end{align}
where the first-order time derivative of laser phases cancels out, and the second and higher orders are neglected.
The same as the $\xi$ variables, this means that the onboard optical delay effect is adequately compensated for.
Substituting \cref{eq:xi_prnr_result,eq:eta_rfi21_rfi23} into \cref{eq:eta12_prnr}, we get
\begin{align}
    \eta_{12} &= \xi_{12} + \Dc{\mrm{sig}}{12,\isi}\A{\mrm{pin}}{21}\left(\T{23}{21}\phi_{23} - \phi_{21}\right)
    \nonumber\\
    &= \D{\mrm{rec}}{12,\isi}\D{\mrm{dis}}{\isi}\left(\D{\mrm{sig}}{12,\isi}\T{23}{21}\phi_{23} - \D{\mrm{loc}}{\isi}\phi_{12}\right),
    \label{eq:eta12_prnr_result}
\end{align}
where the right-handed laser noise $\phi_{21}$ is replaced with $\phi_{12}$, as traditionally targeted by the $\eta$ variable.

Similarly, \cref{eq:eta13_prnr} results in,
\begin{align}
    \eta_{13} &= \xi_{13} + \Dc{\mrm{loc}}{13,\isi}\A{\mrm{pin}}{13}\left(\phi_{13} - \T{12}{13}\phi_{12}\right)
    \nonumber\\
    &= \D{\mrm{rec}}{13,\isi}\D{\mrm{dis}}{\isi}\left(\D{\mrm{sig}}{13,\isi}\phi_{31} - \D{\mrm{loc}}{\isi}\T{12}{13}\phi_{12}\right),
    \label{eq:eta13_prnr_result}
\end{align}
where the right-handed laser noise $\phi_{13}$ is successfully removed.
However, importantly, $\eta_{13}$ is still evaluated at the time of the $\isi_{13}$ measurement, to which $\phi_{12}$ couples only virtually and not physically.

Note that the $\eta$ variables compensate for the potential clock time deviation between the two local phasemeters $\T{ik}{ij}$, to the nanosecond PRNR accuracy.
For example, $\Ac{\mrm{sig}}{23,\rfi}$ in \cref{eq:eta12_prnr} remove such a clock time deviation between phasemeter $23$ and phasemeter $21$ from $\rfi_{23}$, as represented by $\T{23}{21}$ in \cref{eq:eta12_prnr_result}.
Compared with estimating such potential drifts with assistance from separate environmental monitors, we believe it is more robust to use PRN signals, generated by the clocks themselves, as direct monitors in this way.

\subsection{Second-generation TDI Michelson variable $X_2$}\label{sub:x2}
We can finally formulate the second-generation TDI Michelson variable $X_2$ to form the laser-noise-free TDI combination.
For the sake of simplicity, similarly to \cite{Reinhardt2024A,Reinhardt2024B}, we define a single delay operator as the combination of PRNR delay and advancement operators for ISIs,
\begin{align}
    \Dc{}{ij} &\coloneqq \Dc{\mrm{sig}}{ij,\isi}\Ac{\mrm{loc}}{ji,\isi}
    \nonumber\\
    &= \D{\mrm{rec}}{ij,\isi}\D{\mrm{sig}}{ij,\isi}\cdot\A{\mrm{loc}}{\isi}\A{\mrm{rec}}{ji,\isi},
    \label{eq:dotDcij}
\end{align}
where $\D{\mrm{pin}}{ji}$ and $\D{\mrm{dis}}{\isi}$ are canceled.
A round-trip operator $\Dc{}{iji}$ gives
\begin{align}
    \Dc{}{iji} &= \Dc{}{ij}\Dc{}{ji}
    \nonumber\\
    &= \D{\mrm{rec}}{ij,\isi}\D{\mrm{sig}}{ij,\isi}\A{\mrm{loc}}{\isi}\A{\mrm{rec}}{ji,\isi}\cdot\D{\mrm{rec}}{ji,\isi}\D{\mrm{sig}}{ji,\isi}\A{\mrm{loc}}{\isi}\A{\mrm{rec}}{ij,\isi}
    \nonumber\\
    &= \D{\mrm{sig}}{ij,\isi}\A{\mrm{loc}}{\isi}\cdot\D{\mrm{sig}}{ji,\isi}\A{\mrm{loc}}{\isi},
    \label{eq:dotDciji}
\end{align}
where $\A{\mrm{rec}}{ij,\isi}$ commutes with other operators based on Assumption 5 in \cref{sub:assumptions} and cancel $\D{\mrm{rec}}{ij,\isi}$.

Our scheme formulates the combination for a round trip over the interspacecraft links as,
\begin{align}
    \eta_{12} &+ \Dc{}{12} \eta_{21}
    \nonumber\\
    &= \D{\mrm{rec}}{12,\isi}\D{\mrm{dis}}{\isi}\left(\D{\mrm{sig}}{12,\isi}\A{\mrm{loc}}{\isi}\D{\mrm{sig}}{21,\isi} - \D{\mrm{loc}}{\isi}\right)\phi_{12}
    \nonumber\\
    &= \D{\mrm{rec}}{12,\isi}\D{\mrm{loc}}{\isi}\D{\mrm{dis}}{\isi}\left(\Dc{}{121} - 1\right)\phi_{12},
    \label{eq:eta12_eta21_prnr}\\
    \eta_{13} &+ \Dc{}{13} \eta_{31}
    \nonumber\\
    &= \D{\mrm{rec}}{13,\isi}\D{\mrm{dis}}{\isi}\left(\D{\mrm{sig}}{13,\isi}\A{\mrm{loc}}{\isi}\D{\mrm{sig}}{31,\isi} - \D{\mrm{loc}}{\isi}\right)\T{12}{13}\phi_{12}
    \nonumber\\
    &= \D{\mrm{rec}}{13,\isi}\D{\mrm{loc}}{\isi}\D{\mrm{dis}}{\isi}\left(\Dc{}{131} - 1\right)\T{12}{13}\phi_{12},
    \label{eq:eta13_eta31_prnr}
\end{align}
where only $\phi_{12}$ remains as $\phi_{23}$ and $\phi_{31}$ are canceled.
The last lines are to write these equations in analogy to the traditional forms in \cref{eq:eta1331_basic,eq:eta1221_basic}, but with the local delays in the local oscillator path.
The signal-path diagram is shown in \cref{fig:diagram_x2}.

\begin{figure}[ht]
    \centering
\includegraphics[width=8.6cm]{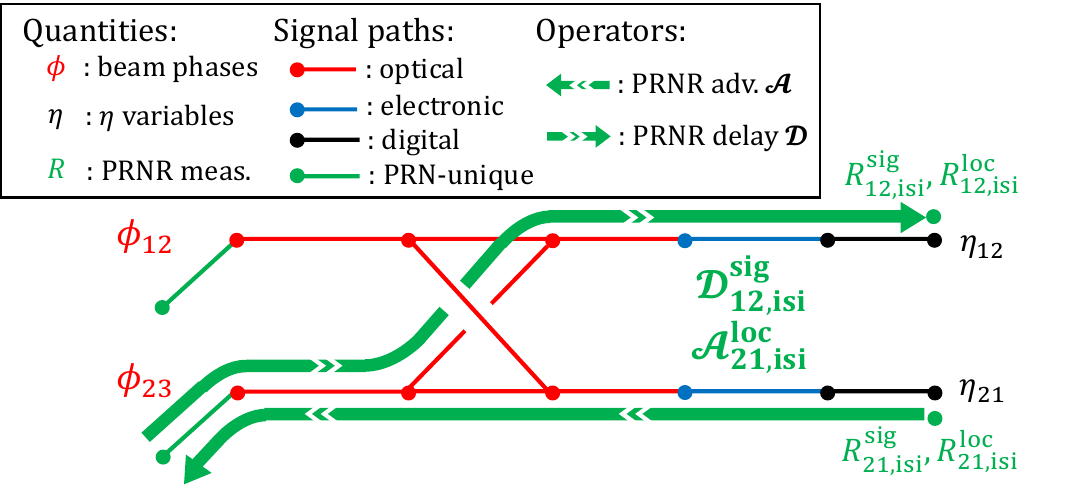}
    \caption{
    Signal-path diagram showing the action of the PRNR operators on $\eta_{21}$ to construct an interspacecraft round-trip measurement between SC 1 and 2 (see \cref{eq:eta12_eta21_prnr}).
    $\eta_{ij}$ is timestamped according to $\isi_{ij}$.
    Hence, we directly write the $\eta$ variables in the diagram as if they were real measurements. 
    }
    \label{fig:diagram_x2}
\end{figure}

To synthesize a Michelson-like equal-arm interferometer, the $X_2$ combination combines the left and right interspacecraft round trips in \cref{eq:eta12_eta21_prnr,eq:eta13_eta31_prnr}, evaluated at the measurement times of $\isi_{12}$ and $\isi_{13}$, respectively.
This means we need to shift the measurement time of the one round-trip combination to that of the other with PRNR, as done so far.
Naively speaking, such time synchronization seems to be achieved by an operator combination like $\Dc{\mrm{loc}}{ij,\isi}\Ac{\mrm{loc}}{ik,\isi}$.
However, it leaves the difference in the PRNR generation delays $d^\mrm{pin}_{ij}-d^\mrm{pin}_{ik}$ because the two operators do not share the PRN generation chains.
This discrepancy stems from the fact that, as discussed below \cref{eq:eta13_prnr_result}, $\phi_{12}$ does not physically couple to $\isi_{13}$, which means there is no direct optical link between \cref{eq:eta12_eta21_prnr,eq:eta13_eta31_prnr} that a PRN signal can trace.

Therefore, we need to acquire $d^\mrm{pin}_{ij}-d^\mrm{pin}_{ik}$ as is, which is fortunately possible with the following combination,
\begin{align}
    R^\mrm{sig}_{ik,\rfi} &- R^\mrm{loc}_{ik,\rfi} - \left(R^\mrm{sig}_{ij,\rfi} - R^\mrm{loc}_{ij,\rfi}\right)
    \nonumber\\
    &\hspace{5mm}=  2(d^\mrm{pin}_{ij} - d^\mrm{pin}_{ik}) + 2\delta\tau^{{ik}}_{ij} + N^\mrm{sig}_{ik,\rfi} - N^\mrm{loc}_{ik,\rfi} - N^\mrm{sig}_{ij,\rfi} + N^\mrm{loc}_{ij,\rfi}.
    \label{eq:Rsig_Rloc_2}
\end{align}
This leaves $2(d^\mrm{pin}_{ij} - d^\mrm{pin}_{ik})$ and $2\delta\tau^{{ik}}_{ij}$ after the same filtering process as the PRNR operators in \cref{eq:Dcab,eq:Acab}.

Using the PRNR combination, we define a new operator that represents a \emph{jump} between two PRN generation points,
\begin{align}
    \Tc{\mrm{p},ik}{\mrm{p},ij}x(\tau) &\coloneqq x\left(\tau - \F\left[\frac{R^\mrm{sig}_{ik,\rfi} - R^\mrm{loc}_{ik,\rfi} - R^\mrm{sig}_{ij,\rfi} + R^\mrm{loc}_{ij,\rfi}}{2}\right]\right)
    \label{eq:Tcbijbik}\\
    &= x(\tau + d^\mrm{pin}_{ik} - d^\mrm{pin}_{ij} - \delta\tau^{{ik}}_{ij})
    \nonumber\\
    &= \A{\mrm{pin}}{ik}\D{\mrm{pin}}{ij}\T{ij}{ik}x(\tau).
    \label{eq:Tcbijbik_DA}
\end{align}
Its inverse operator is written as $\left(\Tc{\mrm{p},ik}{\mrm{p},ij}\right)^{-1}\coloneqq\A{\mrm{pin}}{ij}\D{\mrm{pin}}{ik}\T{ik}{ij} = \Tc{\mrm{p},ij}{\mrm{p},ik}$.

We can combine the operators to virtually link a $\eta$ variable to the adjacent one on the same SC,
\begin{align}
    \Tc{ik}{ij} &\coloneqq \Dc{\mrm{loc}}{ij,\isi}\left(\Tc{\mrm{p},ik}{\mrm{p},ij}\right)^{-1}\Ac{\mrm{loc}}{ik,\isi}
    \nonumber\\
    &= \D{\mrm{rec}}{ij,\isi}\T{ik}{ij}\A{\mrm{rec}}{ik,\isi}.
    \label{eq:Tcijik}
\end{align}
For example, the zeroth-generation TDI $X$ variable, traditionally in \cref{eq:X0_basic}, needs $\Tc{13}{12}$ as,
\begin{align}
    X_0 &= \Tc{13}{12}\left(\eta_{13} + \Dc{}{13}\eta_{31}\right) -\left(\eta_{12} + \Dc{}{12}\eta_{21}\right)
    \nonumber\\
    &= \D{\mrm{rec}}{12,\isi}\D{\mrm{dis}}{\isi}\left(\T{13}{12}\D{\mrm{sig}}{13,\isi}\A{\mrm{loc}}{\isi}\D{\mrm{sig}}{31,\isi}\T{12}{13} - \D{\mrm{sig}}{12,\isi}\A{\mrm{loc}}{\isi}\D{\mrm{sig}}{21,\isi}\right)\phi_{12}
    \nonumber\\
    &= \D{\mrm{rec}}{12,\isi}\D{\mrm{loc}}{\isi}\D{\mrm{dis}}{\isi}\left(\Dc{}{131} - \Dc{}{121}\right)\phi_{12}.
    \label{eq:eta1331_eta1221}
\end{align}
$\T{12}{13}$ and $\T{13}{12}$ cancel in the last line, which is analogous to the traditional formula in \cref{eq:X0_basic}.
\cref{fig:diagram_T1312_x0} visualizes \cref{eq:eta1331_eta1221}, including the interspacecraft-round-trip signal paths synthesized in \cref{eq:eta12_eta21_prnr,eq:eta13_eta31_prnr} in gray.
The chain composed of the orange, pink, and purple arrows depicts the action of $\Tc{13}{12}$.

\begin{figure}
    \centering
\includegraphics[width=8.6cm]{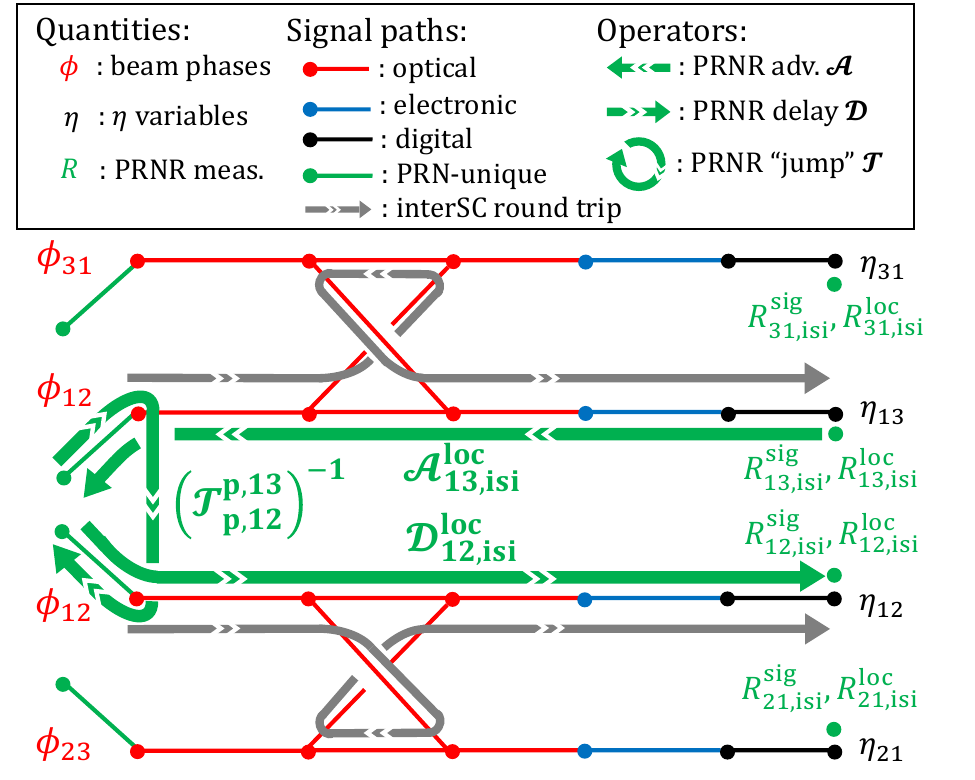}
    \caption{
    Signal-path diagram showing the action of $\Tc{13}{12}\left(\coloneqq\Dc{\mrm{loc}}{12,\isi}\left(\Tc{\mrm{p},13}{\mrm{p},12}\right)^{-1}\Ac{\mrm{loc}}{13,\isi}\right)$ on the virtual interspacecraft round trip $\eta_{13} + \Dc{}{13} \eta_{31}$ evaluated at the time of $\eta_{13}$.
    The interspacecraft round trip between SC 1 and 2, as synthesized in \cref{fig:diagram_x2}, is indicated by the lower gray arrow.
    The round trip between SC 1 and 3 is shown by the upper gray arrow.
    The PRNR operators combine both round-trip measurements to configure a Michelson interferometer corresponding to the zeroth-generation TDI $X$ variable (see \cref{eq:eta1331_eta1221}) evaluated at the time of $\eta_{12}$.
    }
    \label{fig:diagram_T1312_x0}
\end{figure}

Finally, with $\Tc{ik}{ij}$ inserted every time the $\eta$ variables are linked from the left to the right, or vice versa, the second-generation TDI Michelson $X_2$ variable, traditionally in \cref{eq:X2_basic}, can be reformulated as,
\begin{widetext}
    \begin{align}
        X_2 &= \left(\Tc{13}{12} - \Dc{}{121}\Tc{13}{12} - \Dc{}{121}\Tc{13}{12}\Dc{}{131} + \Tc{13}{12}\Dc{}{131}\Tc{12}{13}\Dc{}{12121}\Tc{13}{12}\right)\left(\eta_{13} + \Dc{}{13}\eta_{31}\right)
        \nonumber\\
        &\hspace{10mm}- \left(1 - \Tc{13}{12}\Dc{}{131}\Tc{12}{13} - \Tc{13}{12}\Dc{}{131}\Tc{12}{13}\Dc{}{121} + \Dc{}{121}\Tc{13}{12}\Dc{}{13131}\Tc{12}{13}\right)\left(\eta_{12} + \Dc{}{12}\eta_{21}\right)
        \nonumber\\
        &= \left(1 - \Dc{}{121} - \Dc{}{12131} + \Dc{}{1312121}\right)\Tc{13}{12}\left(\eta_{13} + \Dc{}{13}\eta_{31}\right) - \left(1 - \Dc{}{131} - \Dc{}{13121} + \Dc{}{1213131}\right)\left(\eta_{12} + \Dc{}{12}\eta_{21}\right)
        \label{eq:X2_prnr_result}\\
        &= \D{\mrm{rec}}{12,\isi}\left(\Dc{}{131212131} - \Dc{}{121313121}\right)\D{\mrm{loc}}{\isi}\D{\mrm{dis}}{\isi}\phi_{12}.
        \label{eq:X2_prnr_result_residual}
    \end{align}
\end{widetext}
\cref{eq:X2_prnr_result} commuted the operators based on Assumption 5 in \cref{sub:assumptions} and used $\Tc{ij}{ik}\Tc{ik}{ij}=\mathds{1}$.
At the end, we can apply the operator $\Tc{13}{12}$ only once.
\cref{eq:X2_prnr_result_residual} gives the laser phase noise residual.
The algebraic structure $\Dc{}{131212131} - \Dc{}{121313121}$, which determines the residual, is the same as $\D{}{131212131} - \D{}{121313121}$ in \cref{eq:X2_basic_residual}.
Also, departing from the pure mathematical insight, we can relate these two cases in the physical interpretation: the traditional TDI assumes the SC as point masses, and the interspacecraft delay operators $\D{}{131212131} - \D{}{121313121}$ are correspondingly defined between the points.
Our case (i.e., $\Dc{}{131212131} - \Dc{}{121313121}$) converges on the exact same interpretation if we consider the combining BSs in the ISIs as the points.
It is because, as formulated in \cref{eq:dotDciji}, our round-trip PRNR operator $\Dc{}{iji}$ first pulls a phase from a combining BS to a splitting BS by $\A{\mrm{loc}}{\isi}$ and then pushes it from the splitting BS to a combining BS on the remote SC by $\D{\mrm{sig}}{ij, \isi}$.
Then, it subsequently repeats this operation to bring the phase back to the original combining BS on the local SC.
Therefore, the physical interpretation of $\Dc{}{131212131} - \Dc{}{121313121}$ is identical to that of $\D{}{131212131} - \D{}{121313121}$: this represents the difference in the arrival time in the two interfering virtual photons.
Hence, we conclude that the residual noise is at the same level as the traditional expression.
Finally, as represented by $\D{\mrm{rec}}{12,\isi}$, the virtual beam interference at the combining BS is brought to the measurement time of $\isi_{12}$ according to its driving clock.
For the purpose of multimessenger astronomy, $X_2$ has to be time shifted from the onboard time of SC 1 to the Barycentric Celestial Reference System (BCRS) frame~\cite{LisaRed}.
This can be achieved via one-way clock synchronization similar to Gaia with submillisecond accuracy~\cite{Klioner2017}.

\subsection{Ranging error coupling}\label{sub:noise}
We wrap up this study by analytically investigating the coupling of ranging errors to the $X_2$ variable in \cref{eq:X2_prnr_result}.
Qualitatively, the PRNR error can be categorized into two types: in-band stochastic noises (i.e., $N$ terms in \cref{eq:Rsig,eq:Rloc}) and constant or slowly drifting offsets.
Concerning the in-band noises, the PRNR operators explicitly treat them via filtering; see \cref{eq:Dcab,eq:Acab}.
Therefore, we focus on the impact of the ranging offset in this section.

The most intuitive and straightforward way of this analysis would be to introduce an additional delay operator $\D{\delta,a}{b}$ representing a ranging error of $\delta^a_b$ to the PRNR operators as
\begin{align}
    \Dc{a}{b} \rightarrow \D{\delta,a}{b}\Dc{a}{b}.
    \label{eq:Dcabdel}
\end{align}
Then, when applying the PRNR operators to beatnote signals in TDI, the signals can be written down to the first-order Taylor expansion like
\begin{align}
    \D{\delta,a}{b}\Dc{a}{b}x &= \Dc{a}{b}x - \delta^a_b \Dc{a}{b}\dot{x}.
    \label{eq:Dcabdel_x}
\end{align}
However, because we have a lot of PRNR operators in the algorithm, this way of analysis could be very complex and hard to follow.

Therefore, we tackle noise analysis the other way around: we push the ranging error to physical delays, shown in the measurement equations in \cref{eq:isi_ij,eq:tmi_ij,eq:rfi_ij}, which the beam phases actually experience over the metrology chains.
This simplifies the calculation and makes the noise propagation much easier to track, as described below.

Let us first rewrite the beatnote phase measurements in \cref{eq:isi_ij,eq:tmi_ij,eq:rfi_ij}, focusing on laser noises $p_{ij}$ without OB jitter and TM acceleration noise, in a simpler manner as
\begin{align}
    \isi_{ij} &= \D{\mrm{rec}}{ij,\isi}\D{\mrm{dis}}{\isi}\left(\D{\mrm{sig}}{ij,\isi}p_{ji} - \D{\mrm{loc}}{\isi}p_{ij}\right)
    \nonumber\\
    &= \D{\mrm{tsig}}{ij,\isi}p_{ji} - \D{\mrm{tloc}}{ij,\isi}p_{ij},
    \label{eq:isi_ij_phy}\\
    \tmi_{ij} &= \D{\mrm{rec}}{ij,\tmi}\D{\mrm{dis}}{\tmi}\left(\D{\mrm{sig}}{\tmi}p_{ik} - \D{\mrm{loc}}{\tmi}p_{ij}\right)
    \nonumber\\
    &= \D{\mrm{tsig}}{ij,\tmi}p_{ik} - \D{\mrm{tloc}}{ij,\tmi}p_{ij},
    \label{eq:tmi_ij_phy}\\
    \rfi_{ij} &= \D{\mrm{rec}}{ij,\rfi}\D{\mrm{dis}}{\rfi}\left(\D{\mrm{sig}}{\rfi}p_{ik} - \D{\mrm{loc}}{\rfi}p_{ij}\right)
    \nonumber\\
    &= \D{\mrm{tsig}}{ij,\rfi}p_{ik} - \D{\mrm{tloc}}{ij,\rfi}p_{ij},
    \label{eq:rfi_ij_phy}
\end{align}
where we condense individual delay operators for a laser noise to a single delay operator $\D{\mrm{tsig}}{ij,\mrm{ifo}}$ for a signal phase and $\D{\mrm{tloc}}{ij,\mrm{ifo}}$ for a local phase.
Let us call such an operator a \emph{total physical delay operator} to distinguish it from individual physical delays.

We then consider the PRNR error.
We neglect the PRN generation delays $d^\mrm{pin}_{ij}$ because our PRNR scheme is, by definition, insensitive to this delay.
Under this condition, the PRNR operator $\Dc{a}{b}$ relates to the total physical delay operator $\D{\mrm{t}a}{b}$ via a ranging error placeholder $\A{\delta,a}{b}$ as
\begin{align}
    \Dc{a}{b} &= \A{\delta,a}{b}\D{\mrm{t}a}{b}.
    \label{eq:Dcab_del_tot}
\end{align}

Then, importantly, we write \cref{eq:Dcab_del_tot} in terms of $\D{\mrm{t}a}{b}$
\begin{align}
    \D{\mrm{t}a}{b} &= \D{\delta,a}{b}\Dc{a}{b}.
    \label{eq:Dab_del_tot}
\end{align}
We substitute \cref{eq:Dab_del_tot} into \cref{eq:isi_ij_phy,eq:tmi_ij_phy,eq:rfi_ij_phy} and apply the expansion of \cref{eq:Dcabdel_x},
\begin{align}
    \isi_{ij} &= \D{\delta,\mrm{sig}}{ij,\isi}\Dc{\mrm{sig}}{\isi}p_{ji} - \D{\delta,\mrm{loc}}{ij,\isi}\Dc{\mrm{loc}}{\isi}p_{ij}
    \nonumber\\
    &= \left(\Dc{\mrm{sig}}{\isi}p_{ji}-\Dc{\mrm{loc}}{\isi}p_{ij}\right) - \left(\delta^{\mrm{sig}}_{ij,\isi}\Dc{\mrm{sig}}{\isi}\dot{p}_{ji} - \delta^{\mrm{loc}}_{ij,\isi}\Dc{\mrm{loc}}{\isi}\dot{p}_{ij}\right),
    \label{eq:isi_ij_bias}\\
    \tmi_{ij} &= \D{\delta,\mrm{sig}}{ij,\tmi}\Dc{\mrm{sig}}{\tmi}p_{ik} - \D{\delta,\mrm{loc}}
    {ij,\tmi}\Dc{\mrm{loc}}{\tmi}p_{ij}
    \nonumber\\
    &= \left(\Dc{\mrm{sig}}{\tmi}p_{ik}-\Dc{\mrm{loc}}{\tmi}p_{ij}\right) - \left(\delta^{\mrm{sig}}_{ij,\tmi}\Dc{\mrm{sig}}{\tmi}\dot{p}_{ik} - \delta^{\mrm{loc}}_{ij,\tmi}\Dc{\mrm{loc}}{\tmi}\dot{p}_{ij}\right),
    \label{eq:tmi_ij_bias}\\
    \rfi_{ij} &= \D{\delta,\mrm{sig}}{ij,\rfi}\Dc{\mrm{sig}}{\rfi}p_{ik} - \D{\delta,\mrm{loc}}{ij,\rfi}\Dc{\mrm{loc}}{\rfi}p_{ij}
    \nonumber\\
    &= \left(\Dc{\mrm{sig}}{\rfi}p_{ik}-\Dc{\mrm{loc}}{\rfi}p_{ij}\right) - \left(\delta^{\mrm{sig}}_{ij,\rfi}\Dc{\mrm{sig}}{\rfi}\dot{p}_{ik} - \delta^{\mrm{loc}}_{ij,\rfi}\Dc{\mrm{loc}}{\rfi}\dot{p}_{ij}\right),
    \label{eq:rfi_ij_bias}
\end{align}
where we also omit the $ijk$ subscriptions in the PRNR delay operators.
This means that we assume every reception delay to be the same ($\D{\mrm{rec}}{ij,\mrm{ifo}}\rightarrow\D{\mrm{rec}}{\mrm{ifo}}$) and all interspacecraft arms to be equal and constant.
The former is applied because we do not expect a big impact of keeping the different reception delays on noise analysis as long as we keep the $ijk$ subscriptions for error terms.
The latter is the usual assumption for this type of noise analysis in TDI for conciseness~\cite{Hartwig2021}.

\cref{eq:isi_ij_bias,eq:tmi_ij_bias,eq:rfi_ij_bias} tell us the advantage of expanding the total physical delay operators instead of expanding the PRNR operators through the TDI algorithm: the first terms $\left(\Dc{\mrm{sig}}{\isi}p_{ji}-\Dc{\mrm{loc}}{\isi}p_{ij}\right)$ can be dropped in the following noise analysis.
This is because they have the identical algebraic structure as \cref{eq:isi_ij_phy,eq:tmi_ij_phy,eq:rfi_ij_phy}; therefore, they should cancel through our TDI procedure as studied in the previous sections.
As a result, we can only analyze how the error terms scaling with $\delta$ propagate through our TDI algorithm.

Based on this formulation, a ranging error of the $\xi_{ij}$ variable, formulated in \cref{eq:xi_prnr}, reads
\begin{align}
    C^\xi_{ij} &= \frac{1}{2}\Dc{\mrm{loc}}{\isi}\left[(2\delta^\mrm{loc}_{ij,\isi} + \delta^\mrm{loc}_{ij,\rfi}-\delta^\mrm{loc}_{ij,\tmi})\dot{p}_{ij} - (\delta^\mrm{sig}_{ij,\rfi}-\delta^\mrm{sig}_{ij,\tmi})\Dc{\Delta}{\rfi}\dot{p}_{ik}\right]
    \nonumber\\
    &\hspace{3mm} + \frac{1}{2}\Dc{\mrm{sig}}{\isi}\left[(\delta^\mrm{loc}_{ji,\rfi}-\delta^\mrm{loc}_{ji,\tmi} - 2\delta^\mrm{sig}_{ij,\isi})\dot{p}_{ji} - (\delta^\mrm{sig}_{ji,\rfi}-\delta^\mrm{sig}_{ji,\tmi})\Dc{\Delta}{\rfi}\dot{p}_{jk}\right],
    \label{eq:C_xi_ij_n}
\end{align}
where, for the sake of simplicity, we define
\begin{align}
    \Dc{\Delta}{\rfi} &\coloneqq \Dc{\mrm{sig}}{\rfi}\Ac{\mrm{loc}}{\rfi},
    \label{eq:Dc_Delta_rfi}\\
    \Ac{\Delta}{\rfi} &\coloneqq \Dc{\mrm{loc}}{\rfi}\Ac{\mrm{sig}}{\rfi}.
    \label{eq:Ac_Delta_rfi}
\end{align}
As discussed in \cref{sub:xi}, \cref{eq:C_xi_ij_n} used the condition of $d^\mrm{sig}_\rfi - d^\mrm{loc}_\rfi=d^\mrm{sig}_\tmi - d^\mrm{loc}_\tmi$, which lets us use only the RFI operator $\Dc{\Delta}{\rfi}$ for $\dot{p}_{ik}$ and $\dot{p}_{jk}$.
The effect of the break of this condition has been estimated in~\cite{Reinhardt2024A}.

Next, ranging error terms $C^{\eta}_{13}$ and $C^{\eta}_{12}$ of the $\eta$ variables in \cref{eq:eta13_prnr,eq:eta12_prnr}, in addition to $C^{\xi}_{ij}$, become
\begin{align}
    C^\eta_{12} &= \frac{1}{2}\Dc{\mrm{sig}}{\isi}\left[\left(\delta^\mrm{loc}_{21,\rfi} + \delta^\mrm{sig}_{23,\rfi}\right)\dot{p}_{21} - \left(\delta^\mrm{loc}_{23,\rfi}\Ac{\Delta}{\rfi} + \delta^\mrm{sig}_{21,\rfi}\Dc{\Delta}{\rfi}\right)\dot{p}_{23}\right],
    \label{eq:C_eta_12_n}\\
    C^\eta_{13} &= \frac{1}{2}\Dc{\mrm{loc}}{\isi}\left[-\left(\delta^\mrm{loc}_{13,\rfi} + \delta^\mrm{sig}_{12,\rfi}\right)\dot{p}_{13} + \left(\delta^\mrm{loc}_{12,\rfi}\Ac{\Delta}{\rfi} + \delta^\mrm{sig}_{13,\rfi}\Dc{\Delta}{\rfi}\right)\dot{p}_{12}\right].
    \label{eq:C_eta_13_n}
\end{align}
Hence, the total error of $\eta_{ij}$ becomes $C^{\xi}_{ij} + C^{\eta}_{ij}$.

We then express the error terms of the interspacecraft round trips ($C^{\eta}_{121}$ and $C^{\eta}_{131}$) in \cref{eq:eta12_eta21_prnr,eq:eta13_eta31_prnr},
\begin{widetext}
    \begin{align}
        C^{\eta}_{121} &= C^{\xi}_{12} + C^{\eta}_{12} + \Dc{\Delta}{\isi}\left(C^{\xi}_{21} + C^{\eta}_{21}\right)
        \nonumber\\
        &= \left(\delta^\mrm{loc}_{21,\rfi} - \delta^\mrm{loc}_{21,\tmi} + \delta^\mrm{loc}_{21,\isi} - \delta^\mrm{sig}_{12,\isi}\right)\Dc{\mrm{sig}}{\isi}\dot{p}_{21}
        - \left(\delta^\mrm{sig}_{21,\rfi} - \delta^\mrm{sig}_{21,\tmi}\right)\Dc{\Delta}{\rfi}\Dc{\mrm{sig}}{\isi}\dot{p}_{23}
        - \frac{1}{2}\left(\delta^\mrm{sig}_{12,\rfi} - \delta^\mrm{sig}_{12,\tmi}\right)\left(1 + (\Dc{\Delta}{\isi})^2\right)\Dc{\mrm{loc}}{\isi}\Dc{\Delta}{\rfi}\dot{p}_{13}
        \nonumber\\
        &\hspace{5mm} + \frac{1}{2}\left[\left(2\delta^\mrm{loc}_{12,\isi} + \delta^\mrm{loc}_{12,\rfi} - \delta^\mrm{loc}_{12,\tmi}\right) + \left(\delta^\mrm{loc}_{12,\rfi} - \delta^\mrm{loc}_{12,\tmi} - 2\delta^\mrm{sig}_{21,\isi}\right)(\Dc{\Delta}{\isi})^2\right]\Dc{\mrm{loc}}{\isi}\dot{p}_{12},
        \label{eq:C_eta_121_n}\\
        C^{\eta}_{131} &= C^{\xi}_{13} + C^{\eta}_{13} + \Dc{\Delta}{\isi}\left(C^{\xi}_{31} + C^{\eta}_{31}\right)
        \nonumber\\
        &=\left(\delta^\mrm{loc}_{31,\rfi} - \delta^\mrm{loc}_{31,\tmi} + \delta^\mrm{loc}_{31,\isi} - \delta^\mrm{sig}_{13,\isi}\right)\Dc{\mrm{sig}}{\isi}\dot{p}_{31}
        - \left(\delta^\mrm{sig}_{31,\rfi} - \delta^\mrm{sig}_{31,\tmi}\right)\Dc{\Delta}{\rfi}\Dc{\mrm{sig}}{\isi}\dot{p}_{32}
        \nonumber\\
        &\hspace{5mm} + \frac{1}{2}\left[\left(\delta^\mrm{sig}_{13,\tmi} + \delta^\mrm{loc}_{12,\rfi}(\Ac{\Delta}{\rfi})^2\right) + \left((\delta^\mrm{sig}_{13,\tmi}-2\delta^\mrm{sig}_{13,\rfi}) - \delta^\mrm{loc}_{12,\rfi}(\Ac{\Delta}{\rfi})^2\right)(\Dc{\Delta}{\isi})^2\right]\Dc{\Delta}{\rfi}\Dc{\mrm{loc}}{\isi}\dot{p}_{12}
        \nonumber\\
        &\hspace{5mm} + \frac{1}{2}\left[\left(2\delta^\mrm{loc}_{13,\isi} - \delta^\mrm{sig}_{12,\rfi} - \delta^\mrm{loc}_{13,\tmi}\right) + \left(2\delta^\mrm{loc}_{13,\rfi} + \delta^\mrm{sig}_{12,\rfi} - \delta^\mrm{loc}_{13,\tmi} - 2\delta^\mrm{sig}_{31,\isi}\right)(\Dc{\Delta}{\isi})^2\right]\Dc{\mrm{loc}}{\isi}\dot{p}_{13},
        \label{eq:C_eta_131_n}
    \end{align}
\end{widetext}
where, for conciseness, we define
\begin{align}
    \Dc{\Delta}{\isi} &\coloneqq \Dc{\mrm{sig}}{\isi}\Ac{\mrm{loc}}{\isi}.
    \label{eq:Dc_Delta_isi}
\end{align}
This operator is basically just a reformulation of \cref{eq:dotDcij} under the notation in this noise analysis.

Finally, we plug those error terms into the Michelson $X_2$ combination in \cref{eq:X2_prnr_result}.
As \cref{eq:Tcijik} shows, $\Tc{ik}{ij}$ can be understood as the translation between the reception delays.
Because our noise analysis assumes the same nominal reception delays for every path, as discussed below \cref{eq:rfi_ij_bias}, $\Tc{ik}{ij}$ becomes unity; therefore, we omit it.

As a result, \cref{eq:X2_prnr_result} gives us the $X_2$ noise term $C^{X_2}$ in the time domain as
\begin{align}
    C^{X_2} &= C^{X_2}_{\isi}C^{X_2}_L,
    \label{eq:C_x2}\\
    C^{X_2}_{\isi} &\coloneqq \left(1 - (\Dc{\Delta}{\isi})^2 - (\Dc{\Delta}{\isi})^4 + (\Dc{\Delta}{\isi})^6\right),
    \label{eq:C_x2_isi}\\
    C^{X_2}_L &\coloneqq C^{\eta}_{131} - C^{\eta}_{121}.
    \label{eq:C_x2_L}
\end{align}

We are ready to calculate the noise term in the $X_2$ variable in the frequency domain, namely the power spectral density (PSD) of $C^{X_2}$.
The product of the Fourier transform of $C^{X_2}_{\isi}$ and its conjugate reads
\begin{align}
    F^{X_2}_{\isi} &= 16 \sin^2(2\omega d^{\Delta}_{\isi}) \sin^2(\omega d^{\Delta}_{\isi}).
    \label{eq:S_x2_isi}
\end{align}

To calculate the PSD of $C^{X_2}_L$, all of the laser frequency noises are assumed to have the same PSD $S_{\dot{p}}$ and are incoherent.
This assumption neglects the fact that the lasers in the constellation are actually coherent to some extent because of laser offset lock.
The effect of this assumption was discussed in \cite{Reinhardt2024A,Wang2024}.
The assumption leads us to
\begin{widetext}
    \begin{align}
        S^{X_2}_L &= \left(F^{X_2}_{12} + F^{X_2}_{23} + F^{X_2}_{31} + F^{X_2}_{21} + F^{X_2}_{13} + F^{X_2}_{32}\right) S_{\dot{p}},
        \label{eq:S_x2_L}\\
        F^{X_2}_{12} &= \frac{1}{4}\Bigg|\Bigg.\left[\left(\delta^\mrm{sig}_{13,\tmi} + \delta^\mrm{loc}_{12,\rfi}(\Ac{\Delta}{\rfi})^2\right) + \left((\delta^\mrm{sig}_{13,\tmi}-2\delta^\mrm{sig}_{13,\rfi}) - \delta^\mrm{loc}_{12,\rfi}(\Ac{\Delta}{\rfi})^2\right)(\Dc{\Delta}{\isi})^2\right]\Dc{\Delta}{\rfi}
        \nonumber\\
        &\hspace{60mm} - \left(2\delta^\mrm{loc}_{12,\isi} + \delta^\mrm{loc}_{12,\rfi} - \delta^\mrm{loc}_{12,\tmi}\right) + \left(\delta^\mrm{loc}_{12,\rfi} - \delta^\mrm{loc}_{12,\tmi} - 2\delta^\mrm{sig}_{21,\isi}\right)(\Dc{\Delta}{\isi})^2\Bigg.\Bigg|^2.
        \label{eq:F_x2_12}\\
        F^{X_2}_{23} &= \left(\delta^\mrm{sig}_{21,\rfi} - \delta^\mrm{sig}_{21,\tmi}\right)^2,
        \label{eq:F_x2_23}\\
        F^{X_2}_{31} &= \left(\delta^\mrm{loc}_{31,\rfi} - \delta^\mrm{loc}_{31,\tmi} + \delta^\mrm{loc}_{31,\isi} - \delta^\mrm{sig}_{13,\isi}\right)^2,
        \label{eq:F_x2_31}\\
        F^{X_2}_{21} &= \left(\delta^\mrm{loc}_{21,\rfi} - \delta^\mrm{loc}_{21,\tmi} + \delta^\mrm{loc}_{21,\isi} - \delta^\mrm{sig}_{12,\isi}\right)^2,
        \label{eq:F_x2_21}\\
        F^{X_2}_{13} &= \frac{1}{4}\left|\left(2\delta^\mrm{loc}_{13,\isi} - \delta^\mrm{sig}_{12,\rfi} - \delta^\mrm{loc}_{13,\tmi}\right) + \left(2\delta^\mrm{loc}_{13,\rfi} + \delta^\mrm{sig}_{12,\rfi} - \delta^\mrm{loc}_{13,\tmi} - 2\delta^\mrm{sig}_{31,\isi}\right)(\Dc{\Delta}{\isi})^2 + \left(\delta^\mrm{sig}_{12,\rfi} - \delta^\mrm{sig}_{12,\tmi}\right)\left(1 + (\Dc{\Delta}{\isi})^2\right)\Dc{\Delta}{\rfi}\right|^2,
        \label{eq:F_x2_13}\\
        F^{X_2}_{32} &= \left(\delta^\mrm{sig}_{31,\rfi} - \delta^\mrm{sig}_{31,\tmi}\right)^2.
        \label{eq:F_x2_32}
    \end{align}
\end{widetext}
Combining \cref{eq:S_x2_isi,eq:S_x2_L}, we obtain the ranging error contribution in the $X_2$ variable in the unit of \si{cycle^2\per\Hz} as,
\begin{align}
    S^{X_2}_{p} &= F^{X_2}_{\isi}S^{X_2}_L
    \nonumber\\
    &= 16 \sin^2(2\omega d^{\Delta}_{\isi}) \sin^2(\omega d^{\Delta}_{\isi})
    \nonumber\\
    &\hspace{5mm}\cdot \left(F^{X_2}_{12} + F^{X_2}_{23} + F^{X_2}_{31} + F^{X_2}_{21} + F^{X_2}_{13} + F^{X_2}_{32}\right) S_{\dot{p}}.
    \label{eq:S_x2_dotp}
\end{align}

We performed a Monte Carlo simulation in terms of ranging errors $\delta^a_b$.
We used the following values with a speed of light $c$,
\begin{itemize}
    \item $\delta^a_b$: random number generated in the uniform distribution with a standard deviation of 1.0/$c$ (second),
    \item $d^\Delta_{\isi} = 2.5\times10^9/c$ (second); dominated by ILTT,
    \item $d^\Delta_{\rfi} = 10/c$ (second); dominated by a backlink fiber,
    \item $\sqrt{S_{\dot{p}}} = 30\cdot\sqrt{1+(2\si{\milli\Hz}/f)^4}$ (\si{\Hz\prtHz}); the LISA laser noise allocation.
\end{itemize}
Although the previous studies suggested PRNR performs at submeter accuracy~\cite{Esteban2009,Sutton2010,Yamamoto2024}, we use the classical \SI{1}{\meter} mark to simulate a relatively bad accuracy.
The number of Monte Carlo samples was 500000.
Two cases were considered in this simulation: (1) all PRNR errors are incoherent, (2) the received- and local-PRNR errors associated with the same beatnote are the same, i.e., $\delta^\mrm{sig}_b=\delta^\mrm{loc}_b$.
The latter is to simulate the effect of specific ranging errors in the reception delays.
Especially, as discussed in Assumption 1 in \cref{sub:assumptions}, the electronic reception delay $d^\mrm{erec}_{ij,\mrm{ifo}}$ is expected to differ between a carrier-carrier beatnote and PRN signals in reality.

The resulting performance of the $X_2$ combination under the ranging imperfection in our TDI scheme is shown in \cref{fig:Sx2_mc} with the unit converted to \si{\meter\prtHz}.
The top panel is for the all-incoherent case, while the bottom is for the case where the two PRNR errors in a beatnote are identical.
Gray is the maximum-to-minimum area, blue and red are the $\pm 1\sigma$ areas, and the dashed gold and green are the mean curves.
The highest noise floor of each curve is summarized in \cref{tab:noise_x2}.
There is no significant difference in noise floors between the two cases.
Both cases successfully suppress the laser noise down to the \SI{1}{\pico\meter\prtHz} level or lower with the relatively high ranging errors of $\pm$ 1.0/$c$ second.
This result suggests that our observable-based TDI can perform at the usual \SI{1}{\pico\meter\prtHz} mark for spaceborne GW detectors aiming for the millihertz detection regime.
Finally, qualitatively speaking, we would not expect a significant difference in the design performance between our scheme and traditional TDI with separate calibration.
This is because they share the same number of interspacecraft ranging measurements put into TDI, while such interspacecraft measurements would be dominant sources of ranging errors~\cite{Euringer2023,YamamotoPhD}.

\begin{figure*}
    \centering
\includegraphics[width=17.2cm]{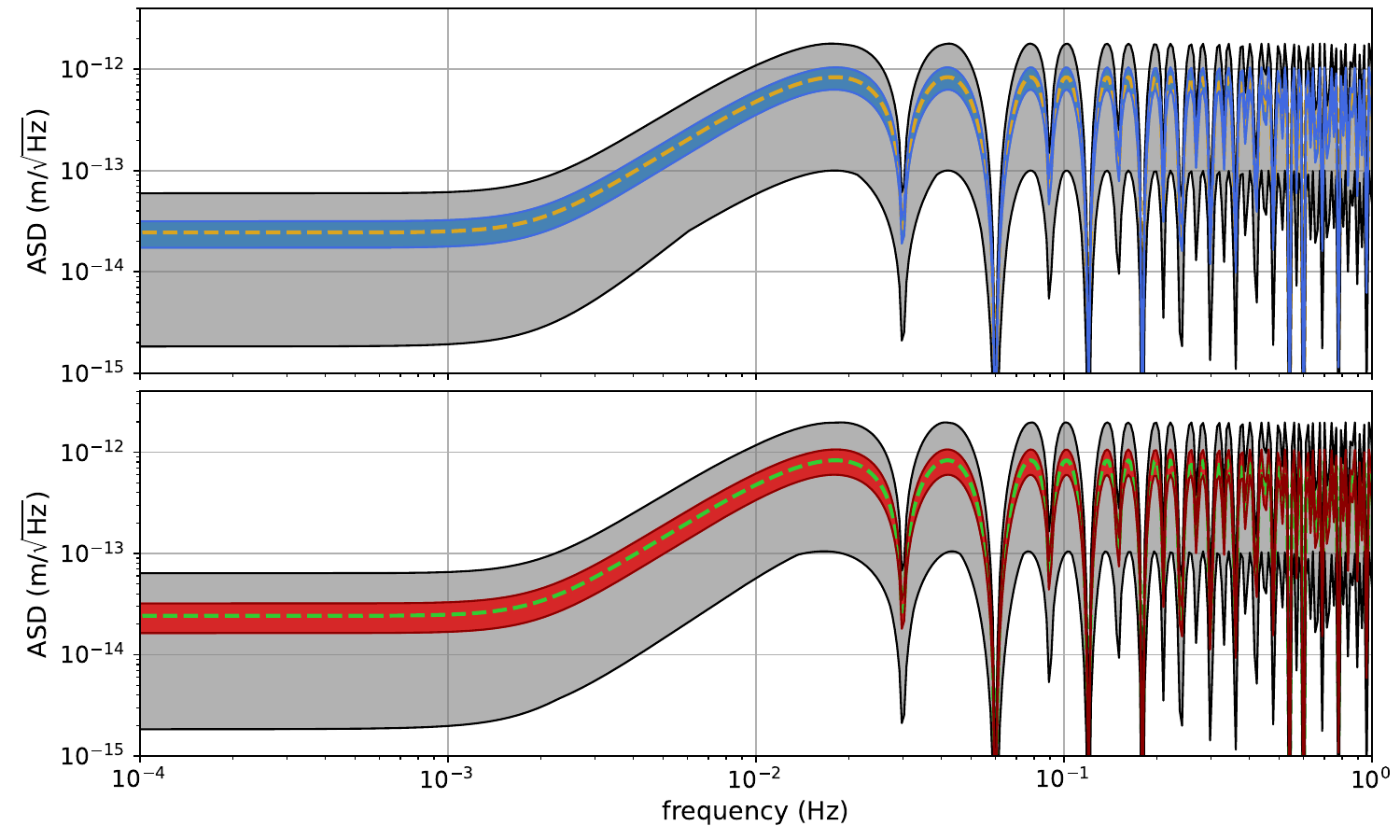}
    \caption{
    PRNR error coupling to the $X_2$ variable in our TDI scheme, as formulated in \cref{eq:S_x2_dotp}, computed by Monte Carlo simulation.
    The number of Monte Carlo samples is 500000, and the uniformly distributed random numbers with a standard deviation of 1.0/$c$ second are applied to different PRNR errors $\delta^a_b$.
    The unit of $S^{X_2}_{p}$ in \cref{eq:S_x2_dotp} has been converted via $\sqrt{S^{X_2}_{p}}\cdot\lambda_0$.
    Top: all PRNR errors are incoherent; Bottom: the local and received PRNR errors associated with the same optical beatnote are identical.
    Gray: the maximum-to-minimum area; blue and red: the $\pm 1\sigma$ areas; dashed gold and green: the mean curves.
    }
    \label{fig:Sx2_mc}
\end{figure*}

\begin{table}[b]
    \caption{\label{tab:noise_x2}
    Highest reading of the $X_2$ noise floor for each statistical curve in Monte Carlo simulation from \cref{fig:Sx2_mc}.
    The left is for the case where all PRNR errors are incoherent, corresponding to the top of \cref{fig:Sx2_mc}; the right is where the two PRNR errors in the same beatnote are identical, i.e., the bottom of \cref{fig:Sx2_mc}.
    The unit is \si{\pico\meter\prtHz}.
    }
    \begin{ruledtabular}
    \begin{tabular}{ccc}
     & All incoherent  & Identical in a beatnote \\
    \hline \\\vspace{1.5mm}
    Maximum & 1.80 & 1.97 \\\vspace{1.5mm}
    $+1\sigma$ & 1.05 & 1.07 \\\vspace{1.5mm}
    Mean & 0.84 & 0.83 \\\vspace{1.5mm}
    $-1\sigma$ & 0.63 & 0.60 \\\vspace{1.5mm}
    Minimum & 0.10 & 0.11 \\
    \end{tabular}
    \end{ruledtabular}
\end{table}


\section{Conclusion}\label{sec:conclusion}
TDI is a postprocessing technique for spaceborne GW detectors to synthesize equal-arm interferometers from their interferometric measurements.
These virtual interferometers are insensitive to laser frequency noise but sensitive to GW signals.
Apart from the interspacecraft signal propagation delays, onboard delays constitute non-negligible parts of the TDI combinations.
Previous investigations updated the TDI algorithm to include onboard delays obtained from prelaunch and in-flight calibrations~\cite{Reinhardt2024B,Euringer2023}.
These studies do not address onboard delays and interspacecraft delays equally: while interspacecraft delays are incorporated in the TDI algorithm, onboard delays are compensated for separately.
This approach would not provide the most natural treatment of the onboard delays, as they are, in fact, constituents of the virtual equal-armlength interferometers.

In this article, we applied two recent advancements in the interpretation of TDI: (1) TDI can work on pseudoranges~\cite{Hartwig2022}; (2) the required TDI delays are the differences between the recording times of two phase measurements capturing an instantaneous feature of the laser frequency noise, and these delays differ from the ILTTs, when onboard delays are considered~\cite{Reinhardt2024A}.
We merged these points by directly expressing all delays in the TDI algorithm through received and local PRNR observables in all interferometers.
In this formulation, the PRNR measurements are not used as calibration ingredients but as direct constituents of the TDI combinations.
We introduced the signal-path diagram to visualize onboard delays (optical, electronic, and digital) in the TDI combinations and to show that the onboard signal paths can be synthesized from PRNR measurements.
While the proposed TDI framework mainly advances the conceptual consistency and completeness, it also comes with unique practical advantages:
\begin{itemize}
    \item It enhances the performance robustness as it only requires correcting the onboard delays in the digital domain, which are determined by commandable or fixed parameters in the phasemeter design.
    \item It prevents potential failure cases due to relative drifts between the clocks driving local phase measurement systems.
    \item It reduces the risk of calibration errors as the algorithm is inherently unaffected by delays in the generation of PRN signals.
\end{itemize}
A Monte Carlo simulation suggests that it performs at the target picometer level with the ranging standard deviation of \SI{1}{\meter} (see \cref{fig:Sx2_mc}).
The simulation also featured a specific source of ranging error in the bottom panel in \cref{fig:Sx2_mc}: the difference in the electronic reception delay between a beatnote signal and a PRN sequence, as outlined in Assumption 1 in \cref{sub:assumptions}. 
To overcome this error, which our scheme cannot treat as is, and reach better laser noise suppression, we could perform fine calibration with the assistance of TDI ranging, the QPR and phasemeter specifications, and environmental monitors, as needed.

While this article introduces the concept of a fully PRNR-based TDI processing scheme, a follow-up investigation should numerically implement it.
This first requires the implementation of the various onboard delays into the LISA simulation tools~\cite{Bayle2023} and a simulation of local and received PRNR measurements in all interferometers, including these onboard delays.
This enables us to numerically test our processing scheme and compare its results with those of the Monte Carlo simulation conducted in this paper.
Including onboard delays in the LISA simulation tools opens up further perspectives: TDI ranging can be investigated in the presence of onboard delays.
This approach would necessitate expanding the TDI ranging parameter space to include all interferometers.
If TDI ranging proves successful in this context, it could serve as a valuable validation method for our TDI scheme.
The update of the simulation tools also makes it possible to consider onboard delays in other TDI processing schemes, such as principal component interferometry~\cite{Baghi2021A} and TDI infinity~\cite{Vallisneri2021}, which would advance their practicality.
Furthermore, we plan to generalize our processing scheme so that it can be applied to other TDI combinations, such as the Sagnac variables or the orthogonal TDI variables $A$, $E$, and $T$~\cite{Vallisneri:2005}.

The proposed observable-based formulation closely links TDI to the actual metrology system, and it explicitly shows how to handle onboard measurements in postprocessing.
In addition, we showed in~\cite{Yamamoto2024} that the essence of the scheme is testable in a lab experiment.
Because the nearly self-contained algorithm does not rely on separate measurements, we expect that such a lab verification of our scheme gives us significant confidence in both the algorithm and the instruments.
As a side note, our scheme can easily be extended to data in the units of frequency~\cite{Bayle2021} by defining PRNR Doppler operators as $\dotDc{a}{ij,\mrm{ifo}}x(\tau)\coloneqq \left(1 - d\left(\mathbf{F} R^a_{ij,\mrm{ifo}}\right)/dt\right)\cdot x(\tau - \mathbf{F} R^a_{ij,\mrm{ifo}})$, and so is the PRNR advancement operator.
Finally, as the proposed TDI algorithm is directly written in terms of onboard measurements, we believe this paper enhances collaborations or eases communication between instrumentalists and data scientists in the field.

\begin{acknowledgments}
The authors thank Peter Wolf and Gerhard Heinzel for useful discussions.
K. Y.'s work is supported by NASA under Award No. 80GSFC24M0006.
The authors gratefully acknowledge financial support by the German Aerospace Center (DLR) with funds from the German Federal Ministry for Economic Affairs and Climate Action (BMWK) under Grants No. 50OQ1801, and No. 50OQ2301.
Furthermore, this work was supported by the LEGACY cooperation on low-frequency gravitational-wave astronomy (M.IF.A.QOP18098) and by the Bundesministerium für Wirtschaft und Klimaschutz based on a resolution of the German Bundestag (Project Ref. No. 50 OQ 1801).
Kohei Yamamoto acknowledges support from the Cluster of Excellence “QuantumFrontiers: Light and Matter at the Quantum Frontier: Foundations and Applications in Metrology” (EXC-2123, Project No. 390837967).
Jan Niklas Reinhardt acknowledges the funding by the Deutsche Forschungsgemeinschaft (DFG, German Research Foundation) under Germany's Excellence Strategy within the Cluster of Excellence PhoenixD (EXC 2122, Project ID No. 390833453).
He also acknowledges the support of the IMPRS on Gravitational Wave Astronomy at the Max-Planck-Institut für Gravitationsphysik in Hannover, Germany.
\end{acknowledgments}

\appendix
\section{Local distribution delay}\label{sec:dist}
Assumption 4 in \cref{sub:assumptions} drops the $ijk$ subscript from the local distribution delay $d^\mrm{dis}_{ij,\mrm{ifo}}$, as shown in \cref{eq:ddist_ifo}.
This section provides a supplementary explanation of this assumption and clarifies what the assumption actually constrains.

This type of delay internally includes an important optical component, i.e., a first BS: the first red dot right after the optical fiber delivering a laser beam to an OB in \cref{fig:sc_notation}.
Let us divide the local distribution delay into delays before and after the first BS ($d^\mrm{\rightarrow bs}_{ij}$ and $d^\mrm{bs\rightarrow}_{ij,\mrm{ifo}}$) as,
\begin{align}
    d^\mrm{dis}_{ij,\mrm{ifo}} &= d^\mrm{\rightarrow bs}_{ij} + d^\mrm{bs\rightarrow}_{ij,\mrm{ifo}},
    \label{eq:ddis_split}
\end{align}
where the delay before the BS $d^\mrm{\rightarrow bs}_{ij}$ drops the subscript $\mrm{ifo}$ since it is common for any beatnote phase measurements.

Contrary to Assumption 4 in \cref{sub:assumptions}, we can actually keep the $ijk$ subscript for $d^\mrm{\rightarrow bs}_{ij}$.
This is because the local distribution delay is common for any PRNR operators associated with the same laser source, and the PRNR action as a pair of the PRNR delay and advancement operators in our scheme cancels $d^\mrm{\rightarrow bs}_{ij}$, as they do for the PRN generation delay.
This means that our PRNR scheme does nothing to the local distribution delay; hence, $d^\mrm{\rightarrow bs}_{ij}$ always appears as $d^\mrm{\rightarrow bs}_{ij}\phi_{ij}$ in the algorithm.
As a result, we can rewrite the original beam phase so that it absorbs $d^\mrm{\rightarrow bs}_{ij}$ exactly as we did for the beam generation delay in \cref{eq:Doinphi}.
The necessary assumption for the local distribution delay is
\begin{align}
    d^\mrm{dis}_{ij,\mrm{ifo}} &\rightarrow d^\mrm{\rightarrow bs}_{ij} + d^\mrm{bs\rightarrow}_\mrm{ifo},
    \label{eq:ddis_split_simple}
\end{align}
which drops the $ijk$ subscript only from the delay after the BS.

In summary, there are two conclusions: (1) the difference in a local distribution delay before a BS between laser sources is unimportant in our PRNR scheme; and (2) a local distribution delay after a BS has to be the ``same" between OBs for a certain interferometer type. Practically speaking, conclusion 1, for example, suggests that we can tolerate any difference in the length of delivery fibers between laser sources and OBs. Conclusion 2 requires the manufacturing tolerance of the OBs to be small enough to maintain the final performance, which is the original intention of introducing Assumption 4.

\bibliography{references}

\begin{thebibliography}{52}%
\makeatletter
\providecommand \@ifxundefined [1]{%
 \@ifx{#1\undefined}
}%
\providecommand \@ifnum [1]{%
 \ifnum #1\expandafter \@firstoftwo
 \else \expandafter \@secondoftwo
 \fi
}%
\providecommand \@ifx [1]{%
 \ifx #1\expandafter \@firstoftwo
 \else \expandafter \@secondoftwo
 \fi
}%
\providecommand \natexlab [1]{#1}%
\providecommand \enquote  [1]{``#1''}%
\providecommand \bibnamefont  [1]{#1}%
\providecommand \bibfnamefont [1]{#1}%
\providecommand \citenamefont [1]{#1}%
\providecommand \href@noop [0]{\@secondoftwo}%
\providecommand \href [0]{\begingroup \@sanitize@url \@href}%
\providecommand \@href[1]{\@@startlink{#1}\@@href}%
\providecommand \@@href[1]{\endgroup#1\@@endlink}%
\providecommand \@sanitize@url [0]{\catcode `\\12\catcode `\$12\catcode
  `\&12\catcode `\#12\catcode `\^12\catcode `\_12\catcode `\%12\relax}%
\providecommand \@@startlink[1]{}%
\providecommand \@@endlink[0]{}%
\providecommand \url  [0]{\begingroup\@sanitize@url \@url }%
\providecommand \@url [1]{\endgroup\@href {#1}{\urlprefix }}%
\providecommand \urlprefix  [0]{URL }%
\providecommand \Eprint [0]{\href }%
\providecommand \doibase [0]{https://doi.org/}%
\providecommand \selectlanguage [0]{\@gobble}%
\providecommand \bibinfo  [0]{\@secondoftwo}%
\providecommand \bibfield  [0]{\@secondoftwo}%
\providecommand \translation [1]{[#1]}%
\providecommand \BibitemOpen [0]{}%
\providecommand \bibitemStop [0]{}%
\providecommand \bibitemNoStop [0]{.\EOS\space}%
\providecommand \EOS [0]{\spacefactor3000\relax}%
\providecommand \BibitemShut  [1]{\csname bibitem#1\endcsname}%
\let\auto@bib@innerbib\@empty
\bibitem [{\citenamefont {Abbott}\ \emph {et~al.}(2016)\citenamefont {Abbott}
  \emph {et~al.}}]{GW150914}%
  \BibitemOpen
  \bibfield  {author} {\bibinfo {author} {\bibfnamefont {B.~P.}\ \bibnamefont
  {Abbott}} \emph {et~al.} (\bibinfo {collaboration} {LIGO and Virgo
  Collaboration}),\ }\bibfield  {title} {\bibinfo {title} {Observation of
  gravitational waves from a binary black hole merger},\ }\bibfield  {journal}
  {\bibinfo  {journal} {Physical Review Letters}\ }\textbf {\bibinfo {volume}
  {116}},\ \href {https://doi.org/10.1103/physrevlett.116.061102}
  {10.1103/physrevlett.116.061102} (\bibinfo {year} {2016})\BibitemShut
  {NoStop}%
\bibitem [{LIG(2019)}]{LIGO:Catalog1}%
  \BibitemOpen
  \bibfield  {title} {\bibinfo {title} {{GWTC-1: A Gravitational-Wave Transient
  Catalog of Compact Binary Mergers Observed by LIGO and Virgo during the First
  and Second Observing Runs}},\ }\href
  {https://doi.org/10.1103/PhysRevX.9.031040} {\bibfield  {journal} {\bibinfo
  {journal} {Phys. Rev. X}\ }\textbf {\bibinfo {volume} {9}},\ \bibinfo {pages}
  {031040} (\bibinfo {year} {2019})}\BibitemShut {NoStop}%
\bibitem [{LIG(2021)}]{LIGO:Catalog2}%
  \BibitemOpen
  \bibfield  {title} {\bibinfo {title} {{GWTC-2: Compact Binary Coalescences
  Observed by LIGO and Virgo during the First Half of the Third Observing
  Run}},\ }\href {https://doi.org/10.1103/PhysRevX.11.021053} {\bibfield
  {journal} {\bibinfo  {journal} {Phys. Rev. X}\ }\textbf {\bibinfo {volume}
  {11}},\ \bibinfo {pages} {021053} (\bibinfo {year} {2021})}\BibitemShut
  {NoStop}%
\bibitem [{LIG(2023)}]{LIGO:Catalog3}%
  \BibitemOpen
  \bibfield  {title} {\bibinfo {title} {{GWTC-3: Compact Binary Coalescences
  Observed by LIGO and Virgo during the Second Part of the Third Observing
  Run}},\ }\href {https://doi.org/10.1103/PhysRevX.13.041039} {\bibfield
  {journal} {\bibinfo  {journal} {Phys. Rev. X}\ }\textbf {\bibinfo {volume}
  {13}},\ \bibinfo {pages} {041039} (\bibinfo {year} {2023})}\BibitemShut
  {NoStop}%
\bibitem [{\citenamefont {Amaro-Seoane}\ \emph {et~al.}(2023)\citenamefont
  {Amaro-Seoane}, \citenamefont {Andrews}, \citenamefont {Sedda} \emph
  {et~al.}}]{Seoane2023}%
  \BibitemOpen
  \bibfield  {author} {\bibinfo {author} {\bibfnamefont {P.}~\bibnamefont
  {Amaro-Seoane}}, \bibinfo {author} {\bibfnamefont {J.}~\bibnamefont
  {Andrews}}, \bibinfo {author} {\bibfnamefont {M.~A.}\ \bibnamefont {Sedda}},
  \emph {et~al.},\ }\bibfield  {title} {\bibinfo {title} {Astrophysics with the
  laser interferometer space antenna},\ }\bibfield  {journal} {\bibinfo
  {journal} {Living Reviews in Relativity}\ }\textbf {\bibinfo {volume} {26}},\
  \href {https://doi.org/10.1007/s41114-022-00041-y}
  {10.1007/s41114-022-00041-y} (\bibinfo {year} {2023})\BibitemShut {NoStop}%
\bibitem [{\citenamefont {Luo}\ \emph {et~al.}(2016)\citenamefont {Luo},
  \citenamefont {Chen}, \citenamefont {Duan}, \citenamefont {Gong},
  \citenamefont {Hu}, \citenamefont {Ji}, \citenamefont {Liu}, \citenamefont
  {Mei}, \citenamefont {Milyukov}, \citenamefont {Sazhin}, \citenamefont
  {Shao}, \citenamefont {Toth}, \citenamefont {Tu}, \citenamefont {Wang},
  \citenamefont {Wang}, \citenamefont {Yeh}, \citenamefont {Zhan},
  \citenamefont {Zhang}, \citenamefont {Zharov},\ and\ \citenamefont
  {Zhou}}]{TianQin}%
  \BibitemOpen
  \bibfield  {author} {\bibinfo {author} {\bibfnamefont {J.}~\bibnamefont
  {Luo}}, \bibinfo {author} {\bibfnamefont {L.-S.}\ \bibnamefont {Chen}},
  \bibinfo {author} {\bibfnamefont {H.-Z.}\ \bibnamefont {Duan}}, \bibinfo
  {author} {\bibfnamefont {Y.-G.}\ \bibnamefont {Gong}}, \bibinfo {author}
  {\bibfnamefont {S.}~\bibnamefont {Hu}}, \bibinfo {author} {\bibfnamefont
  {J.}~\bibnamefont {Ji}}, \bibinfo {author} {\bibfnamefont {Q.}~\bibnamefont
  {Liu}}, \bibinfo {author} {\bibfnamefont {J.}~\bibnamefont {Mei}}, \bibinfo
  {author} {\bibfnamefont {V.}~\bibnamefont {Milyukov}}, \bibinfo {author}
  {\bibfnamefont {M.}~\bibnamefont {Sazhin}}, \bibinfo {author} {\bibfnamefont
  {C.-G.}\ \bibnamefont {Shao}}, \bibinfo {author} {\bibfnamefont {V.~T.}\
  \bibnamefont {Toth}}, \bibinfo {author} {\bibfnamefont {H.-B.}\ \bibnamefont
  {Tu}}, \bibinfo {author} {\bibfnamefont {Y.}~\bibnamefont {Wang}}, \bibinfo
  {author} {\bibfnamefont {Y.}~\bibnamefont {Wang}}, \bibinfo {author}
  {\bibfnamefont {H.-C.}\ \bibnamefont {Yeh}}, \bibinfo {author} {\bibfnamefont
  {M.-S.}\ \bibnamefont {Zhan}}, \bibinfo {author} {\bibfnamefont
  {Y.}~\bibnamefont {Zhang}}, \bibinfo {author} {\bibfnamefont
  {V.}~\bibnamefont {Zharov}},\ and\ \bibinfo {author} {\bibfnamefont {Z.-B.}\
  \bibnamefont {Zhou}},\ }\bibfield  {title} {\bibinfo {title} {{TianQin: a
  space-borne gravitational wave detector}},\ }\href
  {https://doi.org/10.1088/0264-9381/33/3/035010} {\bibfield  {journal}
  {\bibinfo  {journal} {Classical and Quantum Gravity}\ }\textbf {\bibinfo
  {volume} {33}},\ \bibinfo {pages} {035010} (\bibinfo {year}
  {2016})}\BibitemShut {NoStop}%
\bibitem [{\citenamefont {Luo}\ \emph {et~al.}(2020)\citenamefont {Luo},
  \citenamefont {Wang}, \citenamefont {Wu}, \citenamefont {Hu},\ and\
  \citenamefont {Jin}}]{Taiji}%
  \BibitemOpen
  \bibfield  {author} {\bibinfo {author} {\bibfnamefont {Z.}~\bibnamefont
  {Luo}}, \bibinfo {author} {\bibfnamefont {Y.}~\bibnamefont {Wang}}, \bibinfo
  {author} {\bibfnamefont {Y.}~\bibnamefont {Wu}}, \bibinfo {author}
  {\bibfnamefont {W.}~\bibnamefont {Hu}},\ and\ \bibinfo {author}
  {\bibfnamefont {G.}~\bibnamefont {Jin}},\ }\bibfield  {title} {\bibinfo
  {title} {{The Taiji program: A concise overview}},\ }\href
  {https://doi.org/10.1093/ptep/ptaa083} {\bibfield  {journal} {\bibinfo
  {journal} {Progress of Theoretical and Experimental Physics}\ }\textbf
  {\bibinfo {volume} {2021}},\ \bibinfo {pages} {05A108} (\bibinfo {year}
  {2020})}\BibitemShut {NoStop}%
\bibitem [{\citenamefont {Seto}\ \emph {et~al.}(2001)\citenamefont {Seto},
  \citenamefont {Kawamura},\ and\ \citenamefont {Nakamura}}]{DECIGO}%
  \BibitemOpen
  \bibfield  {author} {\bibinfo {author} {\bibfnamefont {N.}~\bibnamefont
  {Seto}}, \bibinfo {author} {\bibfnamefont {S.}~\bibnamefont {Kawamura}},\
  and\ \bibinfo {author} {\bibfnamefont {T.}~\bibnamefont {Nakamura}},\
  }\bibfield  {title} {\bibinfo {title} {Possibility of direct measurement of
  the acceleration of the universe using 0.1 hz band laser interferometer
  gravitational wave antenna in space},\ }\href
  {https://doi.org/10.1103/PhysRevLett.87.221103} {\bibfield  {journal}
  {\bibinfo  {journal} {Phys. Rev. Lett.}\ }\textbf {\bibinfo {volume} {87}},\
  \bibinfo {pages} {221103} (\bibinfo {year} {2001})}\BibitemShut {NoStop}%
\bibitem [{\citenamefont {Colpi}\ \emph {et~al.}(2024)\citenamefont {Colpi},
  \citenamefont {Danzmann}, \citenamefont {Hewitson}, \citenamefont
  {Holley-Bockelmann}, \citenamefont {Jetzer}, \citenamefont {Nelemans},
  \citenamefont {Petiteau}, \citenamefont {Shoemaker}, \citenamefont
  {Sopuerta}, \citenamefont {Stebbins} \emph {et~al.}}]{LisaRed}%
  \BibitemOpen
  \bibfield  {author} {\bibinfo {author} {\bibfnamefont {M.}~\bibnamefont
  {Colpi}}, \bibinfo {author} {\bibfnamefont {K.}~\bibnamefont {Danzmann}},
  \bibinfo {author} {\bibfnamefont {M.}~\bibnamefont {Hewitson}}, \bibinfo
  {author} {\bibfnamefont {K.}~\bibnamefont {Holley-Bockelmann}}, \bibinfo
  {author} {\bibfnamefont {P.}~\bibnamefont {Jetzer}}, \bibinfo {author}
  {\bibfnamefont {G.}~\bibnamefont {Nelemans}}, \bibinfo {author}
  {\bibfnamefont {A.}~\bibnamefont {Petiteau}}, \bibinfo {author}
  {\bibfnamefont {D.}~\bibnamefont {Shoemaker}}, \bibinfo {author}
  {\bibfnamefont {C.}~\bibnamefont {Sopuerta}}, \bibinfo {author}
  {\bibfnamefont {R.}~\bibnamefont {Stebbins}}, \emph {et~al.},\ }\href
  {https://arxiv.org/abs/2402.07571} {\bibinfo {title} {{LISA Definition Study
  Report}}} (\bibinfo {year} {2024}),\ \Eprint
  {https://arxiv.org/abs/2402.07571} {arXiv:2402.07571 [astro-ph.CO]}
  \BibitemShut {NoStop}%
\bibitem [{\citenamefont {Otto}(2015)}]{Otto:2015erp}%
  \BibitemOpen
  \bibfield  {author} {\bibinfo {author} {\bibfnamefont {M.}~\bibnamefont
  {Otto}},\ }\emph {\bibinfo {title} {{{Time-Delay Interferometry Simulations
  for the Laser Interferometer Space Antenna}}}},\ \href
  {https://doi.org/10.15488/8545} {Ph.D. thesis},\ \bibinfo  {school}
  {Gottfried Wilhelm Leibniz Universit{\"a}t Hannover} (\bibinfo {year}
  {2015})\BibitemShut {NoStop}%
\bibitem [{\citenamefont {Armano}\ \emph {et~al.}(2016)\citenamefont {Armano},
  \citenamefont {Audley}, \citenamefont {Auger}, \citenamefont {Baird},
  \citenamefont {Bassan}, \citenamefont {Binetruy}, \citenamefont {Born},
  \citenamefont {Bortoluzzi}, \citenamefont {Brandt}, \citenamefont {Caleno}
  \emph {et~al.}}]{Armano:LPF-I}%
  \BibitemOpen
  \bibfield  {author} {\bibinfo {author} {\bibfnamefont {M.}~\bibnamefont
  {Armano}}, \bibinfo {author} {\bibfnamefont {H.}~\bibnamefont {Audley}},
  \bibinfo {author} {\bibfnamefont {G.}~\bibnamefont {Auger}}, \bibinfo
  {author} {\bibfnamefont {J.~T.}\ \bibnamefont {Baird}}, \bibinfo {author}
  {\bibfnamefont {M.}~\bibnamefont {Bassan}}, \bibinfo {author} {\bibfnamefont
  {P.}~\bibnamefont {Binetruy}}, \bibinfo {author} {\bibfnamefont
  {M.}~\bibnamefont {Born}}, \bibinfo {author} {\bibfnamefont {D.}~\bibnamefont
  {Bortoluzzi}}, \bibinfo {author} {\bibfnamefont {N.}~\bibnamefont {Brandt}},
  \bibinfo {author} {\bibfnamefont {M.}~\bibnamefont {Caleno}}, \emph
  {et~al.},\ }\bibfield  {title} {\bibinfo {title} {{Sub-Femto-$g$ Free Fall
  for Space-Based Gravitational Wave Observatories: LISA Pathfinder Results}},\
  }\href {https://doi.org/10.1103/PhysRevLett.116.231101} {\bibfield  {journal}
  {\bibinfo  {journal} {Phys. Rev. Lett.}\ }\textbf {\bibinfo {volume} {116}},\
  \bibinfo {pages} {231101} (\bibinfo {year} {2016})}\BibitemShut {NoStop}%
\bibitem [{\citenamefont {Armano}\ \emph {et~al.}(2018)\citenamefont {Armano},
  \citenamefont {Audley}, \citenamefont {Baird}, \citenamefont {Binetruy},
  \citenamefont {Born}, \citenamefont {Bortoluzzi}, \citenamefont {Castelli},
  \citenamefont {Cavalleri}, \citenamefont {Cesarini}, \citenamefont {Cruise}
  \emph {et~al.}}]{Armano:LPF-II}%
  \BibitemOpen
  \bibfield  {author} {\bibinfo {author} {\bibfnamefont {M.}~\bibnamefont
  {Armano}}, \bibinfo {author} {\bibfnamefont {H.}~\bibnamefont {Audley}},
  \bibinfo {author} {\bibfnamefont {J.}~\bibnamefont {Baird}}, \bibinfo
  {author} {\bibfnamefont {P.}~\bibnamefont {Binetruy}}, \bibinfo {author}
  {\bibfnamefont {M.}~\bibnamefont {Born}}, \bibinfo {author} {\bibfnamefont
  {D.}~\bibnamefont {Bortoluzzi}}, \bibinfo {author} {\bibfnamefont
  {E.}~\bibnamefont {Castelli}}, \bibinfo {author} {\bibfnamefont
  {A.}~\bibnamefont {Cavalleri}}, \bibinfo {author} {\bibfnamefont
  {A.}~\bibnamefont {Cesarini}}, \bibinfo {author} {\bibfnamefont {A.~M.}\
  \bibnamefont {Cruise}}, \emph {et~al.},\ }\bibfield  {title} {\bibinfo
  {title} {{Beyond the Required LISA Free-Fall Performance: New LISA Pathfinder
  Results down to $20\text{ }\text{ }\ensuremath{\mu}\mathrm{Hz}$}},\ }\href
  {https://doi.org/10.1103/PhysRevLett.120.061101} {\bibfield  {journal}
  {\bibinfo  {journal} {Phys. Rev. Lett.}\ }\textbf {\bibinfo {volume} {120}},\
  \bibinfo {pages} {061101} (\bibinfo {year} {2018})}\BibitemShut {NoStop}%
\bibitem [{\citenamefont {Brzozowski}\ \emph {et~al.}(2022)\citenamefont
  {Brzozowski}, \citenamefont {Robertson}, \citenamefont {Fitzsimons},
  \citenamefont {Ward}, \citenamefont {Keogh}, \citenamefont {Taylor},
  \citenamefont {Milanova}, \citenamefont {Perreur-Lloyd}, \citenamefont {Ali},
  \citenamefont {Earle}, \citenamefont {Clarkson}, \citenamefont {Sharman},
  \citenamefont {Wells},\ and\ \citenamefont {Parr-Burman}}]{Brzozowski2022}%
  \BibitemOpen
  \bibfield  {author} {\bibinfo {author} {\bibfnamefont {W.}~\bibnamefont
  {Brzozowski}}, \bibinfo {author} {\bibfnamefont {D.}~\bibnamefont
  {Robertson}}, \bibinfo {author} {\bibfnamefont {E.}~\bibnamefont
  {Fitzsimons}}, \bibinfo {author} {\bibfnamefont {H.}~\bibnamefont {Ward}},
  \bibinfo {author} {\bibfnamefont {J.}~\bibnamefont {Keogh}}, \bibinfo
  {author} {\bibfnamefont {A.}~\bibnamefont {Taylor}}, \bibinfo {author}
  {\bibfnamefont {M.}~\bibnamefont {Milanova}}, \bibinfo {author}
  {\bibfnamefont {M.}~\bibnamefont {Perreur-Lloyd}}, \bibinfo {author}
  {\bibfnamefont {Z.}~\bibnamefont {Ali}}, \bibinfo {author} {\bibfnamefont
  {A.}~\bibnamefont {Earle}}, \bibinfo {author} {\bibfnamefont
  {D.}~\bibnamefont {Clarkson}}, \bibinfo {author} {\bibfnamefont
  {R.}~\bibnamefont {Sharman}}, \bibinfo {author} {\bibfnamefont
  {M.}~\bibnamefont {Wells}},\ and\ \bibinfo {author} {\bibfnamefont
  {P.}~\bibnamefont {Parr-Burman}},\ }\bibfield  {title} {\bibinfo {title}
  {{The LISA optical bench: an overview and engineering challenges}},\ }in\
  \href {https://doi.org/10.1117/12.2627465} {\emph {\bibinfo {booktitle}
  {Space Telescopes and Instrumentation 2022: Optical, Infrared, and Millimeter
  Wave}}},\ Vol.\ \bibinfo {volume} {12180},\ \bibinfo {editor} {edited by\
  \bibinfo {editor} {\bibfnamefont {L.~E.}\ \bibnamefont {Coyle}}, \bibinfo
  {editor} {\bibfnamefont {S.}~\bibnamefont {Matsuura}},\ and\ \bibinfo
  {editor} {\bibfnamefont {M.~D.}\ \bibnamefont {Perrin}}},\ \bibinfo
  {organization} {International Society for Optics and Photonics}\ (\bibinfo
  {publisher} {SPIE},\ \bibinfo {year} {2022})\ p.\ \bibinfo {pages}
  {121800O}\BibitemShut {NoStop}%
\bibitem [{\citenamefont {Numata}\ \emph {et~al.}(2023)\citenamefont {Numata},
  \citenamefont {Yu}, \citenamefont {Brambora}, \citenamefont {Camp},
  \citenamefont {Fahey}, \citenamefont {Feizi}, \citenamefont {Heesh},
  \citenamefont {Jiao}, \citenamefont {Konoplev}, \citenamefont {Mullin},
  \citenamefont {Merritt}, \citenamefont {Mamakos}, \citenamefont {Marlow},
  \citenamefont {Poulios}, \citenamefont {Pruessner}, \citenamefont
  {Rodriguez}, \citenamefont {Vasilyev}, \citenamefont {Wu}, \citenamefont
  {Xu},\ and\ \citenamefont {Yevick}}]{Numata2023}%
  \BibitemOpen
  \bibfield  {author} {\bibinfo {author} {\bibfnamefont {K.}~\bibnamefont
  {Numata}}, \bibinfo {author} {\bibfnamefont {A.}~\bibnamefont {Yu}}, \bibinfo
  {author} {\bibfnamefont {C.}~\bibnamefont {Brambora}}, \bibinfo {author}
  {\bibfnamefont {J.}~\bibnamefont {Camp}}, \bibinfo {author} {\bibfnamefont
  {M.}~\bibnamefont {Fahey}}, \bibinfo {author} {\bibfnamefont
  {A.}~\bibnamefont {Feizi}}, \bibinfo {author} {\bibfnamefont
  {K.}~\bibnamefont {Heesh}}, \bibinfo {author} {\bibfnamefont
  {H.}~\bibnamefont {Jiao}}, \bibinfo {author} {\bibfnamefont {O.}~\bibnamefont
  {Konoplev}}, \bibinfo {author} {\bibfnamefont {M.}~\bibnamefont {Mullin}},
  \bibinfo {author} {\bibfnamefont {S.}~\bibnamefont {Merritt}}, \bibinfo
  {author} {\bibfnamefont {W.}~\bibnamefont {Mamakos}}, \bibinfo {author}
  {\bibfnamefont {S.}~\bibnamefont {Marlow}}, \bibinfo {author} {\bibfnamefont
  {D.}~\bibnamefont {Poulios}}, \bibinfo {author} {\bibfnamefont
  {P.}~\bibnamefont {Pruessner}}, \bibinfo {author} {\bibfnamefont
  {M.}~\bibnamefont {Rodriguez}}, \bibinfo {author} {\bibfnamefont
  {A.}~\bibnamefont {Vasilyev}}, \bibinfo {author} {\bibfnamefont
  {S.}~\bibnamefont {Wu}}, \bibinfo {author} {\bibfnamefont {X.}~\bibnamefont
  {Xu}},\ and\ \bibinfo {author} {\bibfnamefont {A.}~\bibnamefont {Yevick}},\
  }\bibfield  {title} {\bibinfo {title} {{Development of LISA laser system at
  NASA}},\ }in\ \href {https://doi.org/10.1117/12.2690342} {\emph {\bibinfo
  {booktitle} {International Conference on Space Optics — ICSO 2022}}},\
  Vol.\ \bibinfo {volume} {12777},\ \bibinfo {editor} {edited by\ \bibinfo
  {editor} {\bibfnamefont {K.}~\bibnamefont {Minoglou}}, \bibinfo {editor}
  {\bibfnamefont {N.}~\bibnamefont {Karafolas}},\ and\ \bibinfo {editor}
  {\bibfnamefont {B.}~\bibnamefont {Cugny}}},\ \bibinfo {organization}
  {International Society for Optics and Photonics}\ (\bibinfo  {publisher}
  {SPIE},\ \bibinfo {year} {2023})\ p.\ \bibinfo {pages} {1277738}\BibitemShut
  {NoStop}%
\bibitem [{\citenamefont {Verlaan}\ \emph {et~al.}(2012)\citenamefont
  {Verlaan}, \citenamefont {Hogenhuis}, \citenamefont {Pijnenburg},
  \citenamefont {Lemmen}, \citenamefont {Lucarelli}, \citenamefont {Scheulen},\
  and\ \citenamefont {Ende}}]{Verlaan:Telescope}%
  \BibitemOpen
  \bibfield  {author} {\bibinfo {author} {\bibfnamefont {A.~L.}\ \bibnamefont
  {Verlaan}}, \bibinfo {author} {\bibfnamefont {H.}~\bibnamefont {Hogenhuis}},
  \bibinfo {author} {\bibfnamefont {J.}~\bibnamefont {Pijnenburg}}, \bibinfo
  {author} {\bibfnamefont {M.}~\bibnamefont {Lemmen}}, \bibinfo {author}
  {\bibfnamefont {S.}~\bibnamefont {Lucarelli}}, \bibinfo {author}
  {\bibfnamefont {D.}~\bibnamefont {Scheulen}},\ and\ \bibinfo {author}
  {\bibfnamefont {D.}~\bibnamefont {Ende}},\ }\bibfield  {title} {\bibinfo
  {title} {{LISA telescope assembly optical stability characterization for
  ESA}},\ }in\ \href {https://doi.org/10.1117/12.925112} {\emph {\bibinfo
  {booktitle} {Modern Technologies in Space- and Ground-based Telescopes and
  Instrumentation II}}},\ Vol.\ \bibinfo {volume} {8450},\ \bibinfo {editor}
  {edited by\ \bibinfo {editor} {\bibfnamefont {R.}~\bibnamefont {Navarro}},
  \bibinfo {editor} {\bibfnamefont {C.~R.}\ \bibnamefont {Cunningham}},\ and\
  \bibinfo {editor} {\bibfnamefont {E.}~\bibnamefont {Prieto}}},\ \bibinfo
  {organization} {International Society for Optics and Photonics}\ (\bibinfo
  {publisher} {SPIE},\ \bibinfo {year} {2012})\ p.\ \bibinfo {pages}
  {845003}\BibitemShut {NoStop}%
\bibitem [{\citenamefont {Fleddermann}\ \emph {et~al.}(2018)\citenamefont
  {Fleddermann}, \citenamefont {Diekmann}, \citenamefont {Steier},
  \citenamefont {Tröbs}, \citenamefont {Heinzel},\ and\ \citenamefont
  {Danzmann}}]{Fleddermann:Backlink}%
  \BibitemOpen
  \bibfield  {author} {\bibinfo {author} {\bibfnamefont {R.}~\bibnamefont
  {Fleddermann}}, \bibinfo {author} {\bibfnamefont {C.}~\bibnamefont
  {Diekmann}}, \bibinfo {author} {\bibfnamefont {F.}~\bibnamefont {Steier}},
  \bibinfo {author} {\bibfnamefont {M.}~\bibnamefont {Tröbs}}, \bibinfo
  {author} {\bibfnamefont {G.}~\bibnamefont {Heinzel}},\ and\ \bibinfo {author}
  {\bibfnamefont {K.}~\bibnamefont {Danzmann}},\ }\bibfield  {title} {\bibinfo
  {title} {{Sub-pm{$\sqrt{{\rm Hz}}^{-1}$} non-reciprocal noise in the LISA
  backlink fiber}},\ }\href {https://doi.org/10.1088/1361-6382/aaa276}
  {\bibfield  {journal} {\bibinfo  {journal} {Classical and Quantum Gravity}\
  }\textbf {\bibinfo {volume} {35}},\ \bibinfo {pages} {075007} (\bibinfo
  {year} {2018})}\BibitemShut {NoStop}%
\bibitem [{\citenamefont {Colcombet}\ \emph {et~al.}(2024)\citenamefont
  {Colcombet}, \citenamefont {Dinu-Jaeger}, \citenamefont {Inguimbert},
  \citenamefont {Nuns}, \citenamefont {Bruhier}, \citenamefont {Christensen},
  \citenamefont {Hofverberg}, \citenamefont {van Bakel}, \citenamefont {van
  Beuzekom}, \citenamefont {Mistry}, \citenamefont {Visser}, \citenamefont
  {Pascucci}, \citenamefont {Izumi}, \citenamefont {Komori}, \citenamefont
  {Heinzel}, \citenamefont {Fernández~Barranco}, \citenamefont {in~t Zand},
  \citenamefont {Laubert},\ and\ \citenamefont {Frericks}}]{Colcombet2024}%
  \BibitemOpen
  \bibfield  {author} {\bibinfo {author} {\bibfnamefont {P.}~\bibnamefont
  {Colcombet}}, \bibinfo {author} {\bibfnamefont {N.}~\bibnamefont
  {Dinu-Jaeger}}, \bibinfo {author} {\bibfnamefont {C.}~\bibnamefont
  {Inguimbert}}, \bibinfo {author} {\bibfnamefont {T.}~\bibnamefont {Nuns}},
  \bibinfo {author} {\bibfnamefont {S.}~\bibnamefont {Bruhier}}, \bibinfo
  {author} {\bibfnamefont {N.}~\bibnamefont {Christensen}}, \bibinfo {author}
  {\bibfnamefont {P.}~\bibnamefont {Hofverberg}}, \bibinfo {author}
  {\bibfnamefont {N.}~\bibnamefont {van Bakel}}, \bibinfo {author}
  {\bibfnamefont {M.}~\bibnamefont {van Beuzekom}}, \bibinfo {author}
  {\bibfnamefont {T.}~\bibnamefont {Mistry}}, \bibinfo {author} {\bibfnamefont
  {G.}~\bibnamefont {Visser}}, \bibinfo {author} {\bibfnamefont
  {D.}~\bibnamefont {Pascucci}}, \bibinfo {author} {\bibfnamefont
  {K.}~\bibnamefont {Izumi}}, \bibinfo {author} {\bibfnamefont
  {K.}~\bibnamefont {Komori}}, \bibinfo {author} {\bibfnamefont
  {G.}~\bibnamefont {Heinzel}}, \bibinfo {author} {\bibfnamefont
  {G.}~\bibnamefont {Fernández~Barranco}}, \bibinfo {author} {\bibfnamefont
  {J.~J.~M.}\ \bibnamefont {in~t Zand}}, \bibinfo {author} {\bibfnamefont
  {P.}~\bibnamefont {Laubert}},\ and\ \bibinfo {author} {\bibfnamefont
  {M.}~\bibnamefont {Frericks}},\ }\bibfield  {title} {\bibinfo {title}
  {{Radiation Tolerance of Low-Noise Photoreceivers for the LISA Space
  Mission}},\ }\href {https://doi.org/10.1109/TNS.2024.3401047} {\bibfield
  {journal} {\bibinfo  {journal} {IEEE Transactions on Nuclear Science}\
  }\textbf {\bibinfo {volume} {71}},\ \bibinfo {pages} {1914} (\bibinfo {year}
  {2024})}\BibitemShut {NoStop}%
\bibitem [{\citenamefont {Shaddock}\ \emph {et~al.}(2006)\citenamefont
  {Shaddock}, \citenamefont {Ware}, \citenamefont {Halverson}, \citenamefont
  {Spero},\ and\ \citenamefont {Klipstein}}]{Shaddock2006}%
  \BibitemOpen
  \bibfield  {author} {\bibinfo {author} {\bibfnamefont {D.}~\bibnamefont
  {Shaddock}}, \bibinfo {author} {\bibfnamefont {B.}~\bibnamefont {Ware}},
  \bibinfo {author} {\bibfnamefont {P.~G.}\ \bibnamefont {Halverson}}, \bibinfo
  {author} {\bibfnamefont {R.~E.}\ \bibnamefont {Spero}},\ and\ \bibinfo
  {author} {\bibfnamefont {B.}~\bibnamefont {Klipstein}},\ }\bibfield  {title}
  {\bibinfo {title} {{Overview of the LISA Phasemeter}},\ }\bibfield  {journal}
  {\bibinfo  {journal} {AIP Conference Proceedings}\ }\textbf {\bibinfo
  {volume} {873}},\ \href {https://doi.org/10.1063/1.2405113}
  {10.1063/1.2405113} (\bibinfo {year} {2006})\BibitemShut {NoStop}%
\bibitem [{\citenamefont {Gerberding}\ \emph {et~al.}(2013)\citenamefont
  {Gerberding}, \citenamefont {Sheard}, \citenamefont {Bykov}, \citenamefont
  {Kullmann}, \citenamefont {Delgado}, \citenamefont {Danzmann},\ and\
  \citenamefont {Heinzel}}]{Gerberding2013}%
  \BibitemOpen
  \bibfield  {author} {\bibinfo {author} {\bibfnamefont {O.}~\bibnamefont
  {Gerberding}}, \bibinfo {author} {\bibfnamefont {B.}~\bibnamefont {Sheard}},
  \bibinfo {author} {\bibfnamefont {I.}~\bibnamefont {Bykov}}, \bibinfo
  {author} {\bibfnamefont {J.}~\bibnamefont {Kullmann}}, \bibinfo {author}
  {\bibfnamefont {J.~J.~E.}\ \bibnamefont {Delgado}}, \bibinfo {author}
  {\bibfnamefont {K.}~\bibnamefont {Danzmann}},\ and\ \bibinfo {author}
  {\bibfnamefont {G.}~\bibnamefont {Heinzel}},\ }\bibfield  {title} {\bibinfo
  {title} {Phasemeter core for intersatellite laser heterodyne interferometry:
  modelling, simulations and experiments},\ }\href
  {https://doi.org/10.1088/0264-9381/30/23/235029} {\bibfield  {journal}
  {\bibinfo  {journal} {Classical and Quantum Gravity}\ }\textbf {\bibinfo
  {volume} {30}},\ \bibinfo {pages} {235029} (\bibinfo {year}
  {2013})}\BibitemShut {NoStop}%
\bibitem [{\citenamefont {Tinto}\ and\ \citenamefont
  {Armstrong}(1999)}]{Tinto1999}%
  \BibitemOpen
  \bibfield  {author} {\bibinfo {author} {\bibfnamefont {M.}~\bibnamefont
  {Tinto}}\ and\ \bibinfo {author} {\bibfnamefont {J.~W.}\ \bibnamefont
  {Armstrong}},\ }\bibfield  {title} {\bibinfo {title} {Cancellation of laser
  noise in an unequal-arm interferometer detector of gravitational radiation},\
  }\bibfield  {journal} {\bibinfo  {journal} {Physical Review D}\ }\textbf
  {\bibinfo {volume} {59}},\ \href {https://doi.org/10.1103/physrevd.59.102003}
  {10.1103/physrevd.59.102003} (\bibinfo {year} {1999})\BibitemShut {NoStop}%
\bibitem [{\citenamefont {Armstrong}\ \emph {et~al.}(1999)\citenamefont
  {Armstrong}, \citenamefont {Estabrook},\ and\ \citenamefont
  {Tinto}}]{Armstrong1999}%
  \BibitemOpen
  \bibfield  {author} {\bibinfo {author} {\bibfnamefont {J.~W.}\ \bibnamefont
  {Armstrong}}, \bibinfo {author} {\bibfnamefont {F.~B.}\ \bibnamefont
  {Estabrook}},\ and\ \bibinfo {author} {\bibfnamefont {M.}~\bibnamefont
  {Tinto}},\ }\bibfield  {title} {\bibinfo {title} {{Time-Delay Interferometry
  for Space-based Gravitational Wave Searches}},\ }\href
  {https://doi.org/10.1086/308110} {\bibfield  {journal} {\bibinfo  {journal}
  {The Astrophysical Journal}\ }\textbf {\bibinfo {volume} {527}},\ \bibinfo
  {pages} {814} (\bibinfo {year} {1999})}\BibitemShut {NoStop}%
\bibitem [{\citenamefont {Kawamura}\ and\ \citenamefont
  {Chen}(2004)}]{Kawamura2004}%
  \BibitemOpen
  \bibfield  {author} {\bibinfo {author} {\bibfnamefont {S.}~\bibnamefont
  {Kawamura}}\ and\ \bibinfo {author} {\bibfnamefont {Y.}~\bibnamefont
  {Chen}},\ }\bibfield  {title} {\bibinfo {title} {{Displacement-Noise-Free
  Gravitational-Wave Detection}},\ }\href
  {https://doi.org/10.1103/PhysRevLett.93.211103} {\bibfield  {journal}
  {\bibinfo  {journal} {Phys. Rev. Lett.}\ }\textbf {\bibinfo {volume} {93}},\
  \bibinfo {pages} {211103} (\bibinfo {year} {2004})}\BibitemShut {NoStop}%
\bibitem [{\citenamefont {Chen}\ and\ \citenamefont
  {Kawamura}(2006)}]{Chen2006}%
  \BibitemOpen
  \bibfield  {author} {\bibinfo {author} {\bibfnamefont {Y.}~\bibnamefont
  {Chen}}\ and\ \bibinfo {author} {\bibfnamefont {S.}~\bibnamefont
  {Kawamura}},\ }\bibfield  {title} {\bibinfo {title} {{Displacement- and
  Timing-Noise-Free Gravitational-Wave Detection}},\ }\href
  {https://doi.org/10.1103/PhysRevLett.96.231102} {\bibfield  {journal}
  {\bibinfo  {journal} {Phys. Rev. Lett.}\ }\textbf {\bibinfo {volume} {96}},\
  \bibinfo {pages} {231102} (\bibinfo {year} {2006})}\BibitemShut {NoStop}%
\bibitem [{\citenamefont {Gefen}\ \emph {et~al.}(2024)\citenamefont {Gefen},
  \citenamefont {Tarafder}, \citenamefont {Adhikari},\ and\ \citenamefont
  {Chen}}]{Gefen2024}%
  \BibitemOpen
  \bibfield  {author} {\bibinfo {author} {\bibfnamefont {T.}~\bibnamefont
  {Gefen}}, \bibinfo {author} {\bibfnamefont {R.}~\bibnamefont {Tarafder}},
  \bibinfo {author} {\bibfnamefont {R.~X.}\ \bibnamefont {Adhikari}},\ and\
  \bibinfo {author} {\bibfnamefont {Y.}~\bibnamefont {Chen}},\ }\bibfield
  {title} {\bibinfo {title} {{Quantum Precision Limits of Displacement
  Noise-Free Interferometers}},\ }\href
  {https://doi.org/10.1103/PhysRevLett.132.020801} {\bibfield  {journal}
  {\bibinfo  {journal} {Phys. Rev. Lett.}\ }\textbf {\bibinfo {volume} {132}},\
  \bibinfo {pages} {020801} (\bibinfo {year} {2024})}\BibitemShut {NoStop}%
\bibitem [{\citenamefont {Esteban}\ \emph {et~al.}(2009)\citenamefont
  {Esteban}, \citenamefont {Bykov}, \citenamefont {MarÃ­n}, \citenamefont
  {Heinzel},\ and\ \citenamefont {Danzmann}}]{Esteban2009}%
  \BibitemOpen
  \bibfield  {author} {\bibinfo {author} {\bibfnamefont {J.~J.}\ \bibnamefont
  {Esteban}}, \bibinfo {author} {\bibfnamefont {I.}~\bibnamefont {Bykov}},
  \bibinfo {author} {\bibfnamefont {A.~F.~G.}\ \bibnamefont {MarÃ­n}},
  \bibinfo {author} {\bibfnamefont {G.}~\bibnamefont {Heinzel}},\ and\ \bibinfo
  {author} {\bibfnamefont {K.}~\bibnamefont {Danzmann}},\ }\bibfield  {title}
  {\bibinfo {title} {Optical ranging and data transfer development for lisa},\
  }\href {https://doi.org/10.1088/1742-6596/154/1/012025} {\bibfield  {journal}
  {\bibinfo  {journal} {Journal of Physics: Conference Series}\ }\textbf
  {\bibinfo {volume} {154}},\ \bibinfo {pages} {012025} (\bibinfo {year}
  {2009})}\BibitemShut {NoStop}%
\bibitem [{\citenamefont {Sutton}\ \emph {et~al.}(2010)\citenamefont {Sutton},
  \citenamefont {McKenzie}, \citenamefont {Ware},\ and\ \citenamefont
  {Shaddock}}]{Sutton2010}%
  \BibitemOpen
  \bibfield  {author} {\bibinfo {author} {\bibfnamefont {A.}~\bibnamefont
  {Sutton}}, \bibinfo {author} {\bibfnamefont {K.}~\bibnamefont {McKenzie}},
  \bibinfo {author} {\bibfnamefont {B.}~\bibnamefont {Ware}},\ and\ \bibinfo
  {author} {\bibfnamefont {D.~A.}\ \bibnamefont {Shaddock}},\ }\bibfield
  {title} {\bibinfo {title} {Laser ranging and communications for lisa},\
  }\href {https://doi.org/10.1364/OE.18.020759} {\bibfield  {journal} {\bibinfo
   {journal} {Opt. Express}\ }\textbf {\bibinfo {volume} {18}},\ \bibinfo
  {pages} {20759} (\bibinfo {year} {2010})}\BibitemShut {NoStop}%
\bibitem [{\citenamefont {Heinzel}\ \emph {et~al.}(2011)\citenamefont
  {Heinzel}, \citenamefont {Esteban}, \citenamefont {Barke}, \citenamefont
  {Otto}, \citenamefont {Wang}, \citenamefont {Garcia},\ and\ \citenamefont
  {Danzmann}}]{Heinzel:Ranging}%
  \BibitemOpen
  \bibfield  {author} {\bibinfo {author} {\bibfnamefont {G.}~\bibnamefont
  {Heinzel}}, \bibinfo {author} {\bibfnamefont {J.~J.}\ \bibnamefont
  {Esteban}}, \bibinfo {author} {\bibfnamefont {S.}~\bibnamefont {Barke}},
  \bibinfo {author} {\bibfnamefont {M.}~\bibnamefont {Otto}}, \bibinfo {author}
  {\bibfnamefont {Y.}~\bibnamefont {Wang}}, \bibinfo {author} {\bibfnamefont
  {A.~F.}\ \bibnamefont {Garcia}},\ and\ \bibinfo {author} {\bibfnamefont
  {K.}~\bibnamefont {Danzmann}},\ }\bibfield  {title} {\bibinfo {title}
  {{Auxiliary functions of the LISA laser link: ranging, clock noise transfer
  and data communication}},\ }\href
  {https://doi.org/10.1088/0264-9381/28/9/094008} {\bibfield  {journal}
  {\bibinfo  {journal} {Classical and Quantum Gravity}\ }\textbf {\bibinfo
  {volume} {28}},\ \bibinfo {pages} {094008} (\bibinfo {year}
  {2011})}\BibitemShut {NoStop}%
\bibitem [{\citenamefont {Reinhardt}\ \emph
  {et~al.}(2024{\natexlab{a}})\citenamefont {Reinhardt}, \citenamefont
  {Hartwig},\ and\ \citenamefont {Heinzel}}]{Reinhardt2024C}%
  \BibitemOpen
  \bibfield  {author} {\bibinfo {author} {\bibfnamefont {J.~N.}\ \bibnamefont
  {Reinhardt}}, \bibinfo {author} {\bibfnamefont {O.}~\bibnamefont {Hartwig}},\
  and\ \bibinfo {author} {\bibfnamefont {G.}~\bibnamefont {Heinzel}},\
  }\bibfield  {title} {\bibinfo {title} {{Clock synchronization and
  light-travel-time estimation for space-based gravitational-wave detectors}},\
  }\href {http://iopscience.iop.org/article/10.1088/1361-6382/ada2d3}
  {\bibfield  {journal} {\bibinfo  {journal} {Classical and Quantum Gravity}\ }
  (\bibinfo {year} {2024}{\natexlab{a}})}\BibitemShut {NoStop}%
\bibitem [{\citenamefont {Hartwig}\ \emph {et~al.}(2022)\citenamefont
  {Hartwig}, \citenamefont {Bayle}, \citenamefont {Staab}, \citenamefont
  {Hees}, \citenamefont {Lilley},\ and\ \citenamefont {Wolf}}]{Hartwig2022}%
  \BibitemOpen
  \bibfield  {author} {\bibinfo {author} {\bibfnamefont {O.}~\bibnamefont
  {Hartwig}}, \bibinfo {author} {\bibfnamefont {J.-B.}\ \bibnamefont {Bayle}},
  \bibinfo {author} {\bibfnamefont {M.}~\bibnamefont {Staab}}, \bibinfo
  {author} {\bibfnamefont {A.}~\bibnamefont {Hees}}, \bibinfo {author}
  {\bibfnamefont {M.}~\bibnamefont {Lilley}},\ and\ \bibinfo {author}
  {\bibfnamefont {P.}~\bibnamefont {Wolf}},\ }\bibfield  {title} {\bibinfo
  {title} {Time-delay interferometry without clock synchronization},\ }\href
  {https://doi.org/10.1103/PhysRevD.105.122008} {\bibfield  {journal} {\bibinfo
   {journal} {Phys. Rev. D}\ }\textbf {\bibinfo {volume} {105}},\ \bibinfo
  {pages} {122008} (\bibinfo {year} {2022})}\BibitemShut {NoStop}%
\bibitem [{\citenamefont {Reinhardt}\ \emph
  {et~al.}(2024{\natexlab{b}})\citenamefont {Reinhardt}, \citenamefont {Staab},
  \citenamefont {Yamamoto}, \citenamefont {Bayle}, \citenamefont {Hees},
  \citenamefont {Hartwig}, \citenamefont {Wiesner}, \citenamefont {Shah},\ and\
  \citenamefont {Heinzel}}]{Reinhardt2024A}%
  \BibitemOpen
  \bibfield  {author} {\bibinfo {author} {\bibfnamefont {J.~N.}\ \bibnamefont
  {Reinhardt}}, \bibinfo {author} {\bibfnamefont {M.}~\bibnamefont {Staab}},
  \bibinfo {author} {\bibfnamefont {K.}~\bibnamefont {Yamamoto}}, \bibinfo
  {author} {\bibfnamefont {J.-B.}\ \bibnamefont {Bayle}}, \bibinfo {author}
  {\bibfnamefont {A.}~\bibnamefont {Hees}}, \bibinfo {author} {\bibfnamefont
  {O.}~\bibnamefont {Hartwig}}, \bibinfo {author} {\bibfnamefont
  {K.}~\bibnamefont {Wiesner}}, \bibinfo {author} {\bibfnamefont
  {S.}~\bibnamefont {Shah}},\ and\ \bibinfo {author} {\bibfnamefont
  {G.}~\bibnamefont {Heinzel}},\ }\bibfield  {title} {\bibinfo {title}
  {{Ranging sensor fusion in LISA data processing: Treatment of ambiguities,
  noise, and onboard delays in LISA ranging observables}},\ }\href
  {https://doi.org/10.1103/PhysRevD.109.022004} {\bibfield  {journal} {\bibinfo
   {journal} {Phys. Rev. D}\ }\textbf {\bibinfo {volume} {109}},\ \bibinfo
  {pages} {022004} (\bibinfo {year} {2024}{\natexlab{b}})}\BibitemShut
  {NoStop}%
\bibitem [{\citenamefont {Tinto}\ \emph {et~al.}(2005)\citenamefont {Tinto},
  \citenamefont {Vallisneri},\ and\ \citenamefont {Armstrong}}]{Tinto:TDIR}%
  \BibitemOpen
  \bibfield  {author} {\bibinfo {author} {\bibfnamefont {M.}~\bibnamefont
  {Tinto}}, \bibinfo {author} {\bibfnamefont {M.}~\bibnamefont {Vallisneri}},\
  and\ \bibinfo {author} {\bibfnamefont {J.~W.}\ \bibnamefont {Armstrong}},\
  }\bibfield  {title} {\bibinfo {title} {{Time-delay interferometric ranging
  for space-borne gravitational-wave detectors}},\ }\href
  {https://doi.org/10.1103/PhysRevD.71.041101} {\bibfield  {journal} {\bibinfo
  {journal} {Phys. Rev. D}\ }\textbf {\bibinfo {volume} {71}},\ \bibinfo
  {pages} {041101} (\bibinfo {year} {2005})}\BibitemShut {NoStop}%
\bibitem [{\citenamefont {Staab}(2023)}]{Staab:Thesis}%
  \BibitemOpen
  \bibfield  {author} {\bibinfo {author} {\bibfnamefont {M.~B.}\ \bibnamefont
  {Staab}},\ }\emph {\bibinfo {title} {Time-delay interferometric ranging for
  {LISA}: Statistical analysis of bias-free ranging using laser noise
  minimization}},\ \href {https://doi.org/10.15488/15739} {Ph.D. thesis},\
  \bibinfo  {school} {Gottfried Wilhelm Leibniz Universit{\"a}t Hannover}
  (\bibinfo {year} {2023})\BibitemShut {NoStop}%
\bibitem [{\citenamefont {Euringer}\ \emph {et~al.}(2024)\citenamefont
  {Euringer}, \citenamefont {Houba}, \citenamefont {Hechenblaikner},
  \citenamefont {Mandel}, \citenamefont {Soualle},\ and\ \citenamefont
  {Fichter}}]{Euringer2023}%
  \BibitemOpen
  \bibfield  {author} {\bibinfo {author} {\bibfnamefont {P.}~\bibnamefont
  {Euringer}}, \bibinfo {author} {\bibfnamefont {N.}~\bibnamefont {Houba}},
  \bibinfo {author} {\bibfnamefont {G.}~\bibnamefont {Hechenblaikner}},
  \bibinfo {author} {\bibfnamefont {O.}~\bibnamefont {Mandel}}, \bibinfo
  {author} {\bibfnamefont {F.}~\bibnamefont {Soualle}},\ and\ \bibinfo {author}
  {\bibfnamefont {W.}~\bibnamefont {Fichter}},\ }\bibfield  {title} {\bibinfo
  {title} {{Compensation of front-end and modulation delays in phase and
  ranging measurements for time-delay interferometry}},\ }\href
  {https://doi.org/10.1103/PhysRevD.109.083024} {\bibfield  {journal} {\bibinfo
   {journal} {Phys. Rev. D}\ }\textbf {\bibinfo {volume} {109}},\ \bibinfo
  {pages} {083024} (\bibinfo {year} {2024})}\BibitemShut {NoStop}%
\bibitem [{\citenamefont {Reinhardt}\ \emph
  {et~al.}(2024{\natexlab{c}})\citenamefont {Reinhardt}, \citenamefont
  {Euringer}, \citenamefont {Hartwig}, \citenamefont {Hechenblaikner},
  \citenamefont {Heinzel},\ and\ \citenamefont {Yamamoto}}]{Reinhardt2024B}%
  \BibitemOpen
  \bibfield  {author} {\bibinfo {author} {\bibfnamefont {J.~N.}\ \bibnamefont
  {Reinhardt}}, \bibinfo {author} {\bibfnamefont {P.}~\bibnamefont {Euringer}},
  \bibinfo {author} {\bibfnamefont {O.}~\bibnamefont {Hartwig}}, \bibinfo
  {author} {\bibfnamefont {G.}~\bibnamefont {Hechenblaikner}}, \bibinfo
  {author} {\bibfnamefont {G.}~\bibnamefont {Heinzel}},\ and\ \bibinfo {author}
  {\bibfnamefont {K.}~\bibnamefont {Yamamoto}},\ }\bibfield  {title} {\bibinfo
  {title} {Time-delay interferometry with onboard optical delays},\ }\href
  {https://doi.org/10.1103/PhysRevD.110.082005} {\bibfield  {journal} {\bibinfo
   {journal} {Phys. Rev. D}\ }\textbf {\bibinfo {volume} {110}},\ \bibinfo
  {pages} {082005} (\bibinfo {year} {2024}{\natexlab{c}})}\BibitemShut
  {NoStop}%
\bibitem [{\citenamefont {Yamamoto}\ \emph {et~al.}(2024)\citenamefont
  {Yamamoto}, \citenamefont {Bykov}, \citenamefont {Reinhardt}, \citenamefont
  {Bode}, \citenamefont {Grafe}, \citenamefont {Staab}, \citenamefont
  {Messied}, \citenamefont {Clark}, \citenamefont {Barranco}, \citenamefont
  {Schwarze}, \citenamefont {Hartwig}, \citenamefont {Delgado},\ and\
  \citenamefont {Heinzel}}]{Yamamoto2024}%
  \BibitemOpen
  \bibfield  {author} {\bibinfo {author} {\bibfnamefont {K.}~\bibnamefont
  {Yamamoto}}, \bibinfo {author} {\bibfnamefont {I.}~\bibnamefont {Bykov}},
  \bibinfo {author} {\bibfnamefont {J.~N.}\ \bibnamefont {Reinhardt}}, \bibinfo
  {author} {\bibfnamefont {C.}~\bibnamefont {Bode}}, \bibinfo {author}
  {\bibfnamefont {P.}~\bibnamefont {Grafe}}, \bibinfo {author} {\bibfnamefont
  {M.}~\bibnamefont {Staab}}, \bibinfo {author} {\bibfnamefont
  {N.}~\bibnamefont {Messied}}, \bibinfo {author} {\bibfnamefont
  {M.}~\bibnamefont {Clark}}, \bibinfo {author} {\bibfnamefont {G.~F.}\
  \bibnamefont {Barranco}}, \bibinfo {author} {\bibfnamefont {T.~S.}\
  \bibnamefont {Schwarze}}, \bibinfo {author} {\bibfnamefont {O.}~\bibnamefont
  {Hartwig}}, \bibinfo {author} {\bibfnamefont {J.~J.~E.}\ \bibnamefont
  {Delgado}},\ and\ \bibinfo {author} {\bibfnamefont {G.}~\bibnamefont
  {Heinzel}},\ }\bibfield  {title} {\bibinfo {title} {{Experimental end-to-end
  demonstration of intersatellite absolute ranging for the Laser Interferometer
  Space Antenna}},\ }\href {https://doi.org/10.1103/PhysRevApplied.22.054020}
  {\bibfield  {journal} {\bibinfo  {journal} {Phys. Rev. Appl.}\ }\textbf
  {\bibinfo {volume} {22}},\ \bibinfo {pages} {054020} (\bibinfo {year}
  {2024})}\BibitemShut {NoStop}%
\bibitem [{\citenamefont {Tinto}\ and\ \citenamefont
  {Dhurandhar}(2021)}]{Tinto:2020fcc}%
  \BibitemOpen
  \bibfield  {author} {\bibinfo {author} {\bibfnamefont {M.}~\bibnamefont
  {Tinto}}\ and\ \bibinfo {author} {\bibfnamefont {S.~V.}\ \bibnamefont
  {Dhurandhar}},\ }\bibfield  {title} {\bibinfo {title} {{Time-delay
  interferometry}},\ }\href {https://doi.org/10.1007/s41114-020-00029-6}
  {\bibfield  {journal} {\bibinfo  {journal} {Living Rev. Rel.}\ }\textbf
  {\bibinfo {volume} {24}},\ \bibinfo {pages} {1} (\bibinfo {year}
  {2021})}\BibitemShut {NoStop}%
\bibitem [{\citenamefont {Bayle}\ and\ \citenamefont
  {Hartwig}(2023)}]{Bayle2023}%
  \BibitemOpen
  \bibfield  {author} {\bibinfo {author} {\bibfnamefont {J.-B.}\ \bibnamefont
  {Bayle}}\ and\ \bibinfo {author} {\bibfnamefont {O.}~\bibnamefont
  {Hartwig}},\ }\bibfield  {title} {\bibinfo {title} {{Unified model for the
  LISA measurements and instrument simulations}},\ }\href
  {https://doi.org/10.1103/PhysRevD.107.083019} {\bibfield  {journal} {\bibinfo
   {journal} {Phys. Rev. D}\ }\textbf {\bibinfo {volume} {107}},\ \bibinfo
  {pages} {083019} (\bibinfo {year} {2023})}\BibitemShut {NoStop}%
\bibitem [{\citenamefont {Heinzel}\ \emph {et~al.}(2024)\citenamefont
  {Heinzel}, \citenamefont {\'Alvarez-Vizoso}, \citenamefont
  {Dovale-\'Alvarez},\ and\ \citenamefont {Wiesner}}]{Heinzel2024}%
  \BibitemOpen
  \bibfield  {author} {\bibinfo {author} {\bibfnamefont {G.}~\bibnamefont
  {Heinzel}}, \bibinfo {author} {\bibfnamefont {J.}~\bibnamefont
  {\'Alvarez-Vizoso}}, \bibinfo {author} {\bibfnamefont {M.}~\bibnamefont
  {Dovale-\'Alvarez}},\ and\ \bibinfo {author} {\bibfnamefont {K.}~\bibnamefont
  {Wiesner}},\ }\bibfield  {title} {\bibinfo {title} {Frequency planning for
  lisa},\ }\href {https://doi.org/10.1103/PhysRevD.110.042002} {\bibfield
  {journal} {\bibinfo  {journal} {Phys. Rev. D}\ }\textbf {\bibinfo {volume}
  {110}},\ \bibinfo {pages} {042002} (\bibinfo {year} {2024})}\BibitemShut
  {NoStop}%
\bibitem [{\citenamefont {Inchausp\'e}\ \emph {et~al.}(2022)\citenamefont
  {Inchausp\'e}, \citenamefont {Hewitson}, \citenamefont {Sauter},\ and\
  \citenamefont {Wass}}]{Henri2022}%
  \BibitemOpen
  \bibfield  {author} {\bibinfo {author} {\bibfnamefont {H.}~\bibnamefont
  {Inchausp\'e}}, \bibinfo {author} {\bibfnamefont {M.}~\bibnamefont
  {Hewitson}}, \bibinfo {author} {\bibfnamefont {O.}~\bibnamefont {Sauter}},\
  and\ \bibinfo {author} {\bibfnamefont {P.}~\bibnamefont {Wass}},\ }\bibfield
  {title} {\bibinfo {title} {{New LISA dynamics feedback control scheme:
  Common-mode isolation of test mass control and probes of test-mass
  acceleration}},\ }\href {https://doi.org/10.1103/PhysRevD.106.022006}
  {\bibfield  {journal} {\bibinfo  {journal} {Phys. Rev. D}\ }\textbf {\bibinfo
  {volume} {106}},\ \bibinfo {pages} {022006} (\bibinfo {year}
  {2022})}\BibitemShut {NoStop}%
\bibitem [{\citenamefont {Tinto}\ \emph {et~al.}(2004)\citenamefont {Tinto},
  \citenamefont {Estabrook},\ and\ \citenamefont {Armstrong}}]{Tinto:2003vj}%
  \BibitemOpen
  \bibfield  {author} {\bibinfo {author} {\bibfnamefont {M.}~\bibnamefont
  {Tinto}}, \bibinfo {author} {\bibfnamefont {F.~B.}\ \bibnamefont
  {Estabrook}},\ and\ \bibinfo {author} {\bibfnamefont {J.~W.}\ \bibnamefont
  {Armstrong}},\ }\bibfield  {title} {\bibinfo {title} {{Time delay
  interferometry with moving spacecraft arrays}},\ }\href
  {https://doi.org/10.1103/PhysRevD.69.082001} {\bibfield  {journal} {\bibinfo
  {journal} {Phys. Rev. D}\ }\textbf {\bibinfo {volume} {69}},\ \bibinfo
  {pages} {082001} (\bibinfo {year} {2004})},\ \Eprint
  {https://arxiv.org/abs/gr-qc/0310017} {arXiv:gr-qc/0310017} \BibitemShut
  {NoStop}%
\bibitem [{\citenamefont {Muratore}\ \emph {et~al.}(2020)\citenamefont
  {Muratore}, \citenamefont {Vetrugno},\ and\ \citenamefont
  {Vitale}}]{Muratore:2020mdf}%
  \BibitemOpen
  \bibfield  {author} {\bibinfo {author} {\bibfnamefont {M.}~\bibnamefont
  {Muratore}}, \bibinfo {author} {\bibfnamefont {D.}~\bibnamefont {Vetrugno}},\
  and\ \bibinfo {author} {\bibfnamefont {S.}~\bibnamefont {Vitale}},\
  }\bibfield  {title} {\bibinfo {title} {{Revisitation of time delay
  interferometry combinations that suppress laser noise in LISA}},\ }\href
  {https://doi.org/10.1088/1361-6382/ab9d5b} {\bibfield  {journal} {\bibinfo
  {journal} {Class. Quant. Grav.}\ }\textbf {\bibinfo {volume} {37}},\ \bibinfo
  {pages} {185019} (\bibinfo {year} {2020})},\ \Eprint
  {https://arxiv.org/abs/2001.11221} {arXiv:2001.11221 [astro-ph.IM]}
  \BibitemShut {NoStop}%
\bibitem [{\citenamefont {Hartwig}\ and\ \citenamefont
  {Bayle}(2021)}]{Hartwig2021}%
  \BibitemOpen
  \bibfield  {author} {\bibinfo {author} {\bibfnamefont {O.}~\bibnamefont
  {Hartwig}}\ and\ \bibinfo {author} {\bibfnamefont {J.-B.}\ \bibnamefont
  {Bayle}},\ }\bibfield  {title} {\bibinfo {title} {{Clock-jitter reduction in
  LISA time-delay interferometry combinations}},\ }\href
  {https://doi.org/10.1103/PhysRevD.103.123027} {\bibfield  {journal} {\bibinfo
   {journal} {Phys. Rev. D}\ }\textbf {\bibinfo {volume} {103}},\ \bibinfo
  {pages} {123027} (\bibinfo {year} {2021})}\BibitemShut {NoStop}%
\bibitem [{\citenamefont {Yamamoto}\ \emph {et~al.}(2022)\citenamefont
  {Yamamoto}, \citenamefont {Vorndamme}, \citenamefont {Hartwig}, \citenamefont
  {Staab}, \citenamefont {Schwarze},\ and\ \citenamefont
  {Heinzel}}]{Yamamoto2022}%
  \BibitemOpen
  \bibfield  {author} {\bibinfo {author} {\bibfnamefont {K.}~\bibnamefont
  {Yamamoto}}, \bibinfo {author} {\bibfnamefont {C.}~\bibnamefont {Vorndamme}},
  \bibinfo {author} {\bibfnamefont {O.}~\bibnamefont {Hartwig}}, \bibinfo
  {author} {\bibfnamefont {M.}~\bibnamefont {Staab}}, \bibinfo {author}
  {\bibfnamefont {T.~S.}\ \bibnamefont {Schwarze}},\ and\ \bibinfo {author}
  {\bibfnamefont {G.}~\bibnamefont {Heinzel}},\ }\bibfield  {title} {\bibinfo
  {title} {Experimental verification of intersatellite clock synchronization at
  lisa performance levels},\ }\href
  {https://doi.org/10.1103/PhysRevD.105.042009} {\bibfield  {journal} {\bibinfo
   {journal} {Phys. Rev. D}\ }\textbf {\bibinfo {volume} {105}},\ \bibinfo
  {pages} {042009} (\bibinfo {year} {2022})}\BibitemShut {NoStop}%
\bibitem [{\citenamefont {Bayle}\ \emph {et~al.}(2022)\citenamefont {Bayle},
  \citenamefont {Hees}, \citenamefont {Lilley}, \citenamefont
  {Le~Poncin-Lafitte}, \citenamefont {Martens},\ and\ \citenamefont
  {Joffre}}]{LisaOrbits}%
  \BibitemOpen
  \bibfield  {author} {\bibinfo {author} {\bibfnamefont {J.-B.}\ \bibnamefont
  {Bayle}}, \bibinfo {author} {\bibfnamefont {A.}~\bibnamefont {Hees}},
  \bibinfo {author} {\bibfnamefont {M.}~\bibnamefont {Lilley}}, \bibinfo
  {author} {\bibfnamefont {C.}~\bibnamefont {Le~Poncin-Lafitte}}, \bibinfo
  {author} {\bibfnamefont {W.}~\bibnamefont {Martens}},\ and\ \bibinfo {author}
  {\bibfnamefont {E.}~\bibnamefont {Joffre}},\ }\href
  {https://doi.org/10.5281/zenodo.7700361} {\bibinfo {title} {L{ISA} {O}rbits}}
  (\bibinfo {year} {2022})\BibitemShut {NoStop}%
\bibitem [{\citenamefont {Roma-Dollase}\ \emph {et~al.}(2023)\citenamefont
  {Roma-Dollase}, \citenamefont {Gualani}, \citenamefont {Gohlke},
  \citenamefont {Abich}, \citenamefont {Morales}, \citenamefont {Gonzalvez},
  \citenamefont {Martín}, \citenamefont {Ramos-Castro}, \citenamefont
  {Sanjuan},\ and\ \citenamefont {Nofrarias}}]{Dollase2023}%
  \BibitemOpen
  \bibfield  {author} {\bibinfo {author} {\bibfnamefont {D.}~\bibnamefont
  {Roma-Dollase}}, \bibinfo {author} {\bibfnamefont {V.}~\bibnamefont
  {Gualani}}, \bibinfo {author} {\bibfnamefont {M.}~\bibnamefont {Gohlke}},
  \bibinfo {author} {\bibfnamefont {K.}~\bibnamefont {Abich}}, \bibinfo
  {author} {\bibfnamefont {J.}~\bibnamefont {Morales}}, \bibinfo {author}
  {\bibfnamefont {A.}~\bibnamefont {Gonzalvez}}, \bibinfo {author}
  {\bibfnamefont {V.}~\bibnamefont {Martín}}, \bibinfo {author} {\bibfnamefont
  {J.}~\bibnamefont {Ramos-Castro}}, \bibinfo {author} {\bibfnamefont
  {J.}~\bibnamefont {Sanjuan}},\ and\ \bibinfo {author} {\bibfnamefont
  {M.}~\bibnamefont {Nofrarias}},\ }\bibfield  {title} {\bibinfo {title}
  {Resistive-based micro-kelvin temperature resolution for ultra-stable space
  experiments},\ }\bibfield  {journal} {\bibinfo  {journal} {Sensors}\ }\textbf
  {\bibinfo {volume} {23}},\ \href {https://doi.org/10.3390/s23010145}
  {10.3390/s23010145} (\bibinfo {year} {2023})\BibitemShut {NoStop}%
\bibitem [{\citenamefont {Klioner}\ \emph {et~al.}(2017)\citenamefont
  {Klioner}, \citenamefont {Geyer}, \citenamefont {Steidelmüller},\ and\
  \citenamefont {Butkevich}}]{Klioner2017}%
  \BibitemOpen
  \bibfield  {author} {\bibinfo {author} {\bibfnamefont {S.~A.}\ \bibnamefont
  {Klioner}}, \bibinfo {author} {\bibfnamefont {R.}~\bibnamefont {Geyer}},
  \bibinfo {author} {\bibfnamefont {H.}~\bibnamefont {Steidelmüller}},\ and\
  \bibinfo {author} {\bibfnamefont {A.~G.}\ \bibnamefont {Butkevich}},\
  }\bibfield  {title} {\bibinfo {title} {{Data Timing, Time Transfer and
  On-board Clock Monitoring for Space Astrometry with Gaia}},\ }\href
  {https://doi.org/10.1007/s11214-017-0419-8} {\bibfield  {journal} {\bibinfo
  {journal} {Space Science Reviews}\ }\textbf {\bibinfo {volume} {212}},\
  \bibinfo {pages} {1423–1432} (\bibinfo {year} {2017})}\BibitemShut
  {NoStop}%
\bibitem [{\citenamefont {Wang}\ \emph {et~al.}(2024)\citenamefont {Wang},
  \citenamefont {Xu}, \citenamefont {Wu}, \citenamefont {Yang}, \citenamefont
  {Duan}, \citenamefont {Du}, \citenamefont {Zhang},\ and\ \citenamefont
  {Yeh}}]{Wang2024}%
  \BibitemOpen
  \bibfield  {author} {\bibinfo {author} {\bibfnamefont {Z.}~\bibnamefont
  {Wang}}, \bibinfo {author} {\bibfnamefont {W.}~\bibnamefont {Xu}}, \bibinfo
  {author} {\bibfnamefont {K.}~\bibnamefont {Wu}}, \bibinfo {author}
  {\bibfnamefont {S.}~\bibnamefont {Yang}}, \bibinfo {author} {\bibfnamefont
  {H.}~\bibnamefont {Duan}}, \bibinfo {author} {\bibfnamefont {Y.}~\bibnamefont
  {Du}}, \bibinfo {author} {\bibfnamefont {X.}~\bibnamefont {Zhang}},\ and\
  \bibinfo {author} {\bibfnamefont {H.-C.}\ \bibnamefont {Yeh}},\ }\bibfield
  {title} {\bibinfo {title} {{Time-delay interferometry with local delays for
  TianQin}},\ }\href {https://doi.org/10.1103/PhysRevD.110.082009} {\bibfield
  {journal} {\bibinfo  {journal} {Phys. Rev. D}\ }\textbf {\bibinfo {volume}
  {110}},\ \bibinfo {pages} {082009} (\bibinfo {year} {2024})}\BibitemShut
  {NoStop}%
\bibitem [{\citenamefont {Yamamoto}(2023)}]{YamamotoPhD}%
  \BibitemOpen
  \bibfield  {author} {\bibinfo {author} {\bibfnamefont {K.}~\bibnamefont
  {Yamamoto}},\ }\emph {\bibinfo {title} {Intersatellite clock synchronization
  and absolute ranging for gravitational wave detection in space}},\ \href
  {https://doi.org/10.15488/15682} {Ph.D. thesis},\ \bibinfo  {school}
  {Gottfried Wilhelm Leibniz Universit{\"a}t Hannover} (\bibinfo {year}
  {2023})\BibitemShut {NoStop}%
\bibitem [{\citenamefont {Baghi}\ \emph {et~al.}(2021)\citenamefont {Baghi},
  \citenamefont {Thorpe}, \citenamefont {Slutsky},\ and\ \citenamefont
  {Baker}}]{Baghi2021A}%
  \BibitemOpen
  \bibfield  {author} {\bibinfo {author} {\bibfnamefont {Q.}~\bibnamefont
  {Baghi}}, \bibinfo {author} {\bibfnamefont {J.~I.}\ \bibnamefont {Thorpe}},
  \bibinfo {author} {\bibfnamefont {J.}~\bibnamefont {Slutsky}},\ and\ \bibinfo
  {author} {\bibfnamefont {J.}~\bibnamefont {Baker}},\ }\bibfield  {title}
  {\bibinfo {title} {Statistical inference approach to time-delay
  interferometry for gravitational-wave detection},\ }\href
  {https://doi.org/10.1103/PhysRevD.103.042006} {\bibfield  {journal} {\bibinfo
   {journal} {Phys. Rev. D}\ }\textbf {\bibinfo {volume} {103}},\ \bibinfo
  {pages} {042006} (\bibinfo {year} {2021})}\BibitemShut {NoStop}%
\bibitem [{\citenamefont {Vallisneri}\ \emph {et~al.}(2021)\citenamefont
  {Vallisneri}, \citenamefont {Bayle}, \citenamefont {Babak},\ and\
  \citenamefont {Petiteau}}]{Vallisneri2021}%
  \BibitemOpen
  \bibfield  {author} {\bibinfo {author} {\bibfnamefont {M.}~\bibnamefont
  {Vallisneri}}, \bibinfo {author} {\bibfnamefont {J.-B.}\ \bibnamefont
  {Bayle}}, \bibinfo {author} {\bibfnamefont {S.}~\bibnamefont {Babak}},\ and\
  \bibinfo {author} {\bibfnamefont {A.}~\bibnamefont {Petiteau}},\ }\bibfield
  {title} {\bibinfo {title} {Time-delay interferometry without delays},\ }\href
  {https://doi.org/10.1103/PhysRevD.103.082001} {\bibfield  {journal} {\bibinfo
   {journal} {Phys. Rev. D}\ }\textbf {\bibinfo {volume} {103}},\ \bibinfo
  {pages} {082001} (\bibinfo {year} {2021})}\BibitemShut {NoStop}%
\bibitem [{\citenamefont {Vallisneri}(2005)}]{Vallisneri:2005}%
  \BibitemOpen
  \bibfield  {author} {\bibinfo {author} {\bibfnamefont {M.}~\bibnamefont
  {Vallisneri}},\ }\bibfield  {title} {\bibinfo {title} {{Geometric time delay
  interferometry}},\ }\href {https://doi.org/10.1103/PhysRevD.72.042003}
  {\bibfield  {journal} {\bibinfo  {journal} {Phys. Rev. D}\ }\textbf {\bibinfo
  {volume} {72}},\ \bibinfo {pages} {042003} (\bibinfo {year}
  {2005})}\BibitemShut {NoStop}%
\bibitem [{\citenamefont {Bayle}\ \emph {et~al.}(2021)\citenamefont {Bayle},
  \citenamefont {Hartwig},\ and\ \citenamefont {Staab}}]{Bayle2021}%
  \BibitemOpen
  \bibfield  {author} {\bibinfo {author} {\bibfnamefont {J.-B.}\ \bibnamefont
  {Bayle}}, \bibinfo {author} {\bibfnamefont {O.}~\bibnamefont {Hartwig}},\
  and\ \bibinfo {author} {\bibfnamefont {M.}~\bibnamefont {Staab}},\ }\bibfield
   {title} {\bibinfo {title} {{Adapting time-delay interferometry for LISA data
  in frequency}},\ }\href {https://doi.org/10.1103/PhysRevD.104.023006}
  {\bibfield  {journal} {\bibinfo  {journal} {Phys. Rev. D}\ }\textbf {\bibinfo
  {volume} {104}},\ \bibinfo {pages} {023006} (\bibinfo {year}
  {2021})}\BibitemShut {NoStop}%
\end{thebibliography}%

\end{document}